\documentclass[11pt, USenglish]{article}
\usepackage[letterpaper, top = 1 in, bottom = 1 in, left = 1 in, right = 1 in]{geometry}
\usepackage[singlespacing]{setspace}
\usepackage[utf8]{inputenc}
\usepackage{amsmath, amsfonts, amssymb, amsthm,mathtools,nicefrac} 
\usepackage[shortlabels]{enumitem}
\usepackage[linesnumbered,ruled,vlined]{algorithm2e}
\SetKwRepeat{DoWhile}{do}{while}
\usepackage{caption}
\usepackage{subcaption}
\usepackage{url}
\usepackage{hyperref}
\usepackage{tikz}
\usetikzlibrary{calc}
\usetikzlibrary{patterns}
\usetikzlibrary{decorations.pathreplacing}
\usetikzlibrary{fadings}
\title{The $2$-$3$-Set Packing problem and a $\nicefrac{4}{3}$-approximation for the Maximum Leaf Spanning Arborescence problem in rooted dags}
\author{Meike Neuwohner}
\date{}
 \theoremstyle{definition}
 \newtheorem{definition}{Definition}
 \theoremstyle{theorem}
 \newtheorem{theorem}[definition]{Theorem}
\newtheorem{proposition}[definition]{Proposition}
\newtheorem{lemma}[definition]{Lemma}
\newtheorem{corollary}[definition]{Corollary}
\newtheorem*{claim}{Claim}

\begin{document}

\maketitle
\begin{abstract}
(Weighted) $k$-Set Packing is a fundamental problem in theoretical computer science. The input consists of a collection $\mathcal{S}$ of sets, each of cardinality at most $k$ and equipped with a positive weight. The task is to find a disjoint sub-collection of maximum total weight. A straightforward reduction from 3-Dimensional Matching~\cite{karp1972reducibility} shows that already the special case where $k=3$ and all weights are $1$, the \emph{unweighted $3$-Set Packing problem}, is NP-hard. For this variant, the state-of-the-art are $\frac{4}{3}+\epsilon$-approximations by Cygan~\cite{Cygan} and F\"urer and Yu~\cite{FurerYu}. In contrast, for the weighted $3$-Set Packing problem, the long-standing guarantee of $2+\epsilon$ by Berman~\cite{Berman} has only recently been improved \cite{NeuwohnerLipics,ThieryWard}. The current state-of-the-art is a $1.786$-approximation by Thiery and Ward~\cite{ThieryWard}. The significant gap between the best known approximation guarantees for the unit weight setting and general weights raises the question for intermediate variants that strictly generalize the unit weight setting, but are easier to tackle than the general case.

In this paper, we study the $2$-$3$-Set Packing problem, a generalization of the unweighted $3$-Set Packing problem, where our set collection may contain sets of cardinality $3$ and weight $2$, as well as sets of cardinality $2$ and weight $1$. Building upon the state-of-the-art works on unweighted $3$-Set Packing, we manage to provide a $\frac{4}{3}+\epsilon$-approximation also for the more general $2$-$3$-Set Packing problem. We believe that this result can be a good starting point to identify classes of weight functions to which techniques from the unweighted setting can be generalized. Using a reduction by Fernandes and Lintzmayer~\cite{FernandesLintzmayer}, our result further implies a $\frac{4}{3}+\epsilon$-approximation for the Maximum Leaf Spanning Arborescence problem (MLSA) in rooted directed acyclic graphs, improving on the previously known $\frac{7}{5}$-approximation by Fernandes and Lintzmayer~\cite{FernandesLintzmayer}. By exploiting additional structural properties of the instance constructed in~\cite{FernandesLintzmayer}, we can further get the approximation guarantee for the MLSA down to $\frac{4}{3}$. The MLSA has applications in broadcasting where a message needs to be transferred from a source node to all other nodes along the arcs of an arborescence in a given network.
\end{abstract}
\newpage
\section{Introduction}
The (weighted) $k$-Set Packing problem is defined as follows: As input, we are given a collection $\mathcal{S}$ of sets, each of cardinality at most $k$, and strictly positive weights $w:\mathcal{S}\rightarrow\mathbb{Q}_{>0}$. The task is to find a sub-collection $A\subseteq \mathcal{S}$ consisting of pairwise disjoint sets, such that $w(A)$ is maximum. Note that we may equivalently assume all sets in $\mathcal{S}$ to be of cardinality \emph{exactly} $k$ by filling up sets of smaller size with pairwise distinct dummy elements. In particular, this shows that for $k\leq 2$, the weighted $k$-Set Packing problem can be solved in polynomial time via a reduction to the Maximum Weight Matching problem~\cite{edmonds1965maximum}.
In contrast, for $k\geq 3$, already the special case where $w\equiv 1$, the unweighted $k$-Set Packing problem, turns out to be NP-hard as a reduction from 3-Dimensional Matching~\cite{karp1972reducibility} shows. In fact, the best known approximation guarantee for the 3-Dimensional Matching problem stems from the reduction to unweighted $3$-Set Packing. 

As a consequence, there has been quite some research on approximation algorithms for both the weighted and the unweighted $k$-Set Packing problem for $k\geq 3$. It is well known that a simple greedy approach yields an approximation guarantee of $k$. On the other hand, Hazan, Safra and Schwartz~\cite{LowerBoundKSetPacking} have shown that unless $P=NP$, there is no $o\left(\frac{k}{\log k}\right)$-approximation even for unweighted $k$-Set Packing. 

 The technique that has proven most successful in designing approximation algorithms for both the weighted and the unweighted $k$-Set Packing problem is \emph{local search}. Given an instance $(\mathcal{S},w)$ of the (weighted) $k$-Set Packing problem and a feasible solution $A$, we call a disjoint set collection $X\subseteq \mathcal{S}$ a \emph{local improvement of $A$ of size $|X|$} if \[w(X)>w(\{a\in A:\exists x\in X: a\cap x\neq\emptyset\}),\] that is, if replacing the sets in $A$ that intersect with sets in $X$ by the sets in $X$ improves the total weight of the solution. A generic local search algorithm starts with any feasible solution, e.g.\ the empty one, and then iteratively applies local improvements from a certain class that can be searched for efficiently, until no more exist. At an arbitrarily small loss in the approximation guarantee, a polynomial number of iterations can be guaranteed even for general weights by scaling and truncating the weight function, see for example~\cite{ChandraHalldorsson}.
 
For the unweighted case, Hurkens and Schrijver~\cite{HurkensSchrijver} achieved the first improvement over the greedy algorithm by observing that local improvements of constant size suffice to obtain approximation guarantees arbitrarily close to $\frac{k}{2}$. Starting with a quasi-polynomial time $\frac{k+2}{3}$-approximation by Halld\'{o}rsson~\cite{Halldorsson}, there was a series of papers improving on the guarantee of $\frac{k}{2}+\epsilon$ in~\cite{HurkensSchrijver} by taking into account local improvements of up to logarithmic size~\cite{CyganGrandoniMastrolilli,SviridenkoWard,Cygan,FurerYu}. The state-of-the-art for unweighted $k$-Set Packing for $k\geq 3$ are $\frac{k+1}{3}+\epsilon$-approximations by Cygan~\cite{Cygan} and F\"urer and Yu~\cite{FurerYu}. They consider local improvements of logarithmically bounded size that bear enough structure to get down to a polynomial instead of quasi-polynomial running time by means of color coding.

For general weights, the situation appears to be more difficult. While in the unweighted setting, local improvements of constant size are enough to obtain an approximation ratio of $\frac{k}{2}+\epsilon$, for general weights, there are examples showing that such an approach cannot yield a guarantee below $k-1$~\cite{ArkinHassin}. A significant improvement over the greedy guarantee of $k$ was achieved by Chandra and Halld\'{o}rsson~\cite{ChandraHalldorsson}, who showed that considering, among certain local improvements of constant size those maximizing the ratio between the total weight of sets added to and removed from the current solution, leads to a $\frac{2(k+1)}{3}+\epsilon$-approximation. For twenty years, the state-of-the-art has been Berman's algorithm SquareImp~\cite{Berman}, which achieves an approximation ratio of $\frac{k+1}{2}+\epsilon$ by considering local improvements with respect to squared weights. Rather recently, first Neuwohner~\cite{NeuwohnerLipics,Neuwohner22,Neuwohner23} and then Thiery and Ward~\cite{ThieryWard} managed to improve on this. The best known guarantees for $k\geq 3$ are of the form $\min\{\frac{k+\tau_k}{2},0.4986\cdot(k+1)+0.0208\}$ with $\frac{1}{3}\leq \tau_k\leq 0.572$~\cite{ThieryWard,Neuwohner23}. In particular, the state-of-the-art for weighted $3$-Set Packing is a $1.786$-approximation~\cite{ThieryWard}.
Comparing the best known approximation guarantees for the weighted and the unweighted $k$-Set Packing, it is apparent that already for $k=3$, there is a significant gap between what can be achieved for unit weights and in general. This raises the question which properties of unit weights are needed to make the problem much easier to handle, or, phrased differently, whether there are ``intermediate'' classes of weight functions for which better, or even as good results as in the unweighted setting, can be obtained.

In this paper, we study the $2$-$3$-Set Packing problem, which is defined as follows:
\begin{definition}[$2$-$3$-Set Packing]
	\begin{description}
		\item[]
		\item[Input:] A collection $\mathcal{S}$ of non-empty sets of cardinality $2$ or $3$, $w:\mathcal{S}\rightarrow\{1,2\}$, $s\mapsto |s|-1$
		\item[Task:] Find a disjoint sub-collection $A\subseteq\mathcal{S}$ that maximizes $w(A)$.
	\end{description}	\label{Def23SetPacking}
\end{definition}
Note that every instance of the unweighted $3$-Set Packing problem can be interpreted as an instance of the $2$-$3$-Set Packing problem by adding dummy elements to ensure that all sets are of cardinality exactly $3$, and, thus, receive the same weight. 

We manage to adapt some of the key ideas from the works by Cygan~\cite{Cygan} and F\"urer and Yu~\cite{FurerYu} to obtain a polynomial time $\frac{4}{3}+\epsilon$-approximation for the $2$-$3$-Set Packing problem, matching the state-of-the-art guarantee for unweighted $3$-Set Packing.
\begin{theorem}
For any $\epsilon >0$, there is a polynomial time $\frac{4}{3}+\epsilon$-approximation algorithm for the $2$-$3$-Set Packing problem.\label{TheoGeneralProblem}
\end{theorem}
We believe that this result constitutes a first step towards identifying classes of weight functions that are ``easier to handle''. This in turn may lead to improvements for other problems whenever such weight functions arise.
To support the latter argument, we point out that by using a result by Fernandes and Lintzmayer~\cite{FernandesLintzmayer}, Theorem~\ref{TheoGeneralProblem} immediately implies a $\frac{4}{3}+\epsilon$-approximation for the Maximum Leaf Spanning Arborescence problem (MLSA) in rooted directed acyclic graphs (dags), improving on the previously best guarantee of $\frac{7}{5}$~\cite{FernandesLintzmayer}.
The MLSA is defined as follows:
\begin{definition}[Maximum Leaf Spanning Arborescence Problem (MLSA)]
	\begin{description}
		\item[]
		\item[Input:] A directed graph $G$, $r\in V(G)$ such that every vertex of $G$ is reachable from $r$.
		\item[Task:] Find a spanning $r$-arborescence with the maximum number of leaves possible.
	\end{description}
\end{definition}
It plays an important role in the context of broadcasting: Given a network consisting of a set of nodes containing one distinguished source and a set of available arcs, a message needs to be transferred from the source to all other nodes along a subset of the arcs, which forms (the edge set of) an arborescence rooted at the source. As internal nodes do not only need to be able to receive, but also to re-distribute messages, they are more expensive. Hence, it is desirable to have as few of them as possible, or equivalently, to maximize the number of leaves.

Already the special case where every arc may be used in both directions, the Maximum Leaves Spanning Tree problem, is known to be NP-hard, even if the input graph is $4$-regular or planar with maximum degree at most $4$ (see~\cite{GareyJohnson}, problem ND2). It has further been shown to be APX-hard~\cite{GALBIATI199445}\footnote{Note that MaxSNP-hardness implies APX-hardness, see~\cite{OnSyntacticVersusComputationalViewsOfApproximability}}, even when restricted to cubic graphs~\cite{BONSMA201214}. The best that is known for the Maximum Leaves Spanning Tree problem is an approximation guarantee of $2$~\cite{SolisOba2015A2A}. 

In contrast, for general digraphs, the state-of-the-art is a $\min\{\sqrt{\mathrm{OPT}},92\}$-approximation~\cite{Drescher2010AnAA,DaligaultThomasse}. Moreover, there is a line of research focusing on FPT-algorithms for the MLSA~\cite{SpanningDirectedTreesWithManyLeaves,KernelsForProblemsWithNoKernel,DaligaultThomasse}.

The special case where the graph $G$ is assumed to be a dag (directed acyclic graph) has been proven to be APX-hard by Schwartges, Spoerhase and Wolff~\cite{SchwartgesSpoerhaseWolff}. They further provided a $2$-approximation, which was then improved to $\frac{3}{2}$ by Fernandes and Lintzmayer~\cite{FERNANDES2022217}. Recently, they managed to enhance their approach to obtain a $\frac{7}{5}$-approximation~\cite{FernandesLintzmayer}, which has been unchallenged so far.

In~\cite{FernandesLintzmayer}, Fernandes and Lintzmayer tackle the MLSA in dags by reducing it, up to an approximation guarantee of $\frac{4}{3}$, to a special case of the $2$-$3$-Set Packing problem, which we call the \emph{hereditary $2$-$3$-Set Packing problem}\footnote{As we would like to distinguish between the hereditary and the non-hereditary case, we did not adapt the notation from~\cite{FernandesLintzmayer}}.
\begin{definition}[hereditary $2$-$3$-Set Packing]
An instance of the \emph{hereditary} $2$-$3$-Set Packing problem is an instance of the $2$-$3$-Set Packing problem with the additional property that for every set $s\in\mathcal{S}$ with $|s|=3$, all three two-element subsets of $s$ are contained in $\mathcal{S}$ as well. \label{DefHereditary23SetPacking}
\end{definition}
Fernandes and Lintzmayer~\cite{FernandesLintzmayer} prove the hereditary $2$-$3$-Set Packing problem to be NP-hard via a reduction from $3$-dimensional matching and provide a $\frac{7}{5}$-approximation algorithm. Via Theorem~\ref{TheoFernandesLintzmayer}, this results in a $\frac{7}{5}$-approximation for the MLSA in rooted dags, which has been unchallenged so far.
\begin{theorem}[\cite{FernandesLintzmayer}]
Let $\alpha\geq 1$ and assume that there is a polynomial time $\alpha$-approximation algorithm for the hereditary $2$-$3$-Set Packing problem.
Then there exists a polynomial-time $\max\{\alpha,\frac{4}{3}\}$-approximation for the MLSA in dags.\label{TheoFernandesLintzmayer}
\end{theorem}
By combining Theorem~\ref{TheoFernandesLintzmayer} with Theorem~\ref{TheoGeneralProblem}, we obtain an improved approximation guarantee of $\frac{4}{3}+\epsilon$ for the MLSA in dags. In particular, we get arbitrarily close to the lower bound of $\frac{4}{3}$ that is inherent to the reduction in~\cite{FernandesLintzmayer}, even though we consider a more general problem. 
To complete the picture, we further observe that we can actually exploit the additional structure of the hereditary problem to get rid of the additive $\epsilon$-term. As a consequence, we obtain a $\frac{4}{3}$-approximation for the MLSA in dags.
\section{Our algorithm}
In this section, we introduce our $\frac{4}{3}+\epsilon$-approximation algorithm for the $2$-$3$-Set Packing problem, which is inspired by the state-of-the-art algorithms for the unweighted $3$-Set Packing problem~\cite{Cygan,FurerYu}.
In order to phrase and analyze our algorithm, we introduce the notion of the \emph{conflict graph}, which enables us to talk about the $2$-$3$-Set Packing problem using graph terminology.

 Given an instance $(\mathcal{S},w)$ of the $2$-$3$-Set Packing problem, the conflict graph $G_{\mathcal{S}}$ of $\mathcal{S}$ is defined as follows: The vertices of $G_{\mathcal{S}}$ represent the sets in $\mathcal{S}$ and inherit their weights. Two vertices are connected by an edge if and only if the corresponding sets have a non-empty intersection. By definition, there is a weight-preserving one-to-one correspondence between disjoint sub-collections of $\mathcal{S}$ and independent sets in $G_\mathcal{S}$.
 
  While the Maximum Weight Independent Set problem in general graphs cannot even be approximated within $n^{1-\epsilon}$ unless $\mathrm{P}=\mathrm{NP}$~\cite{InapproxIndependentSet}, conflict graphs of instances of the $2$-$3$-Set Packing problem bear a certain structure that we can exploit. To describe it, we need to introduce the notion of a \emph{$d$-claw}.
\begin{definition}
	Let $d\geq 1$. A \emph{$d$-claw} is a star on $d+1$-vertices, one \emph{center node} and $d$ talons that are all connected to the center, but not to each other.
	
	A graph is said to be \emph{$d$-claw free} if it does not contain an induced sub-graph that constitutes a $d$-claw.\label{DefClawFree}
\end{definition}
Note that a graph is $d$-claw free if and only if every vertex has at most $d-1$ neighbors in any independent set.
\begin{proposition}
	Let $(\mathcal{S},w)$ be an instance of the $2$-$3$-Set Packing problem. Then $G_{\mathcal{S}}$ is $4$-claw free and every $3$-claw is centered at a vertex of weight $2$.\label{PropStructureConflictGraph}
\end{proposition}
\begin{proof}
	Note that every $4$-claw in particular contains a $3$-claw as an induced sub-graph. Thus, it suffices to prove that for $k\in\{2,3\}$, a vertex corresponding to a set $s$ of size $k$ cannot be the center of a $k+1$-claw. Assume towards a contradiction that this were the case. Then $s$ needs to intersect each of the $k+1$ pairwise disjoint sets corresponding to the talons in a different element. This contradicts the fact that the cardinality of $s$ is $k$.
\end{proof}
We introduce some further notation.
\begin{definition}
Let $G$ be a graph and let $U,W\subseteq V(G)$. The \emph{neighborhood} of $U$ in $W$ is \[N(U,W)\coloneqq U\cap W\cup\{w\in W:\exists u\in U:\{u,w\}\in E(G)\}.\] For $v\in V(G)$, we let $N(v,W)\coloneqq N(\{v\},W)$. We further write $E(U,W)$ to denote the set of edges with one endpoint in $U\setminus W$ and one endpoint in $W\setminus U$, and define $\delta(U)\coloneqq E(U,V(G)\setminus U)$. Again, for $v\in V(G)$, we let $\delta(v)\coloneqq\delta(\{v\})$.
\end{definition}
\begin{definition}
Let $V$ be a set and $w:V\rightarrow\{1,2\}$. For $X\subseteq V$, we define $X'\coloneqq\{x\in X: w(x)=1\}$ and $X''\coloneqq\{x\in X: w(x)=2\}$.
\end{definition}
Like the state-of-the-art algorithms for both the unweighted and the weighted $k$-Set Packing problem~\cite{Cygan,FurerYu,ThieryWard,Neuwohner23}, our algorithm is based on \emph{local search}. As a consequence, we need to introduce the notion of a \emph{local improvement}.
\begin{definition}
Let $G$ be a graph, $w:V(G)\rightarrow\{1,2\}$ and let $A,X\subseteq V(G)$ be independent. We say that $X$ is a \emph{local improvement of $A$} if
$w(X)>w(N(X,A))$, or $w(X)=N(X,A)$ and $X$ contains more vertices of weight $2$ than $N(X,A)$.
 We call $|X|$ the \emph{size} of $X$.\label{DefLocalImprovement}
\end{definition}
Similar to the algorithm in~\cite{FurerYu}, our algorithm will search for local improvements of constant size, as well as a certain type of local improvement of logarithmically bounded size that corresponds to a minimal binocular in an auxiliary graph. In the following, we will first define the auxiliary graph (Def.~\ref{DefEdgeInducingPair} and Def.~\ref{DefSearchGraph}) and show that we can construct it in polynomial time (Prop.~\ref{PropSearchGraphPolynomial}). Then, we introduce the notion of an improving binocular (Def.~\ref{DefImprovingBinocular}). Finally, we present our algorithm (Algorithm~\ref{OverallAlgorithm}). 
\begin{definition}[edge-inducing pair]
	Let $\tau > 0$, let $G$ be a graph, let $w:V(G)\rightarrow\{1,2\}$ and let \mbox{$A\subseteq V(G)$} be independent. An \emph{edge-inducing pair} is a pair $(U,W)$ where:
	\begin{enumerate}[(i)]
		\item \label{EdgeInducingPair1} $U\subseteq A$ and $W\subseteq V\setminus A$ is independent.
		\item\label{EdgeInducingPair2} $\max\{|U|,|W|\}\leq \tau$ and $w(U)+2=w(W)$.
		\item \label{EdgeInducingPair3} $N(W,A\setminus U)\subseteq A''$ and $1\leq |N(W,A\setminus U)|\leq 2$.
	\end{enumerate}
	The \emph{edge associated with $(U,W)$} is $e(U,W)\coloneqq N(W,A\setminus U)$. (If $|N(W,A\setminus U)|=1$, the edge is a loop.) For $e=e(U,W)$, we set $U(e)\coloneqq U$ and $W(e)\coloneqq W$.\label{DefEdgeInducingPair}
\end{definition}
\begin{definition}[search graph]
	Let $\tau > 0$, let $G$ be a graph, let $w:V(G)\rightarrow\{1,2\}$ and let $A\subseteq V(G)$ be independent. The \emph{search graph} $S\coloneqq S_\tau(G,w,A)$ is defined as follows: $V(S)\coloneqq A''$ and\\ $E(S)\coloneqq\{e(U,W): (U,W) \text{ is an edge-inducing pair}\}.$
		\label{DefSearchGraph}
\end{definition}
\begin{proposition}
Let $\tau > 0$ be a fixed constant, let $G$ be a graph, let $w:V(G)\rightarrow\{1,2\}$and let $A\subseteq V(G)$ be independent. We can construct $S_\tau(G,w,A)$ in polynomial time. In particular, the number of edges of $S_\tau(G,w,A)$ is polynomially bounded.\label{PropSearchGraphPolynomial}
\end{proposition}
\begin{proof}
There are  only polynomially many pairs $(U,W)$ with the properties that $U\subseteq A$, \mbox{$W\subseteq V\setminus A$} and $\max\{|U|,|W|\}\leq \tau$. For each of them, we can check whether they are edge-inducing in polynomial time, and, if they are, add the respective edge to the search graph.
\end{proof}
\begin{definition}
	Let $G$ be a graph. A \emph{binocular in $G$} is a sub-graph $B$ of $G$ with the property that $|E(B)|>|V(B)|$. A \emph{minimal binocular} is a minimal sub-graph of $G$ that is a binocular. The \emph{size} of a binocular is its number of edges.
\end{definition}

\begin{definition}[improving binocular]
		Let $\tau > 0$, let $G$ be a graph, let $w:V(G)\rightarrow\{1,2\}$ and let \mbox{$A\subseteq V(G)$} be independent.
	Let $\mathcal{B}$ be a binocular in $S_\tau(G,w,A)$ and let $E(\mathcal{B})\coloneqq E_1(\mathcal{B})\dot{\cup}E_2(\mathcal{B})$, where $E_1(\mathcal{B})$ is the set of edges of $\mathcal{B}$ with one endpoint, i.e.\ loops, and $E_2(\mathcal{B})$ is the set of edges of $\mathcal{B}$ with two endpoints. When $\mathcal{B}$ is clear from the context, we may just write $E_1$ and $E_2$ for short.
	
	 Let $U(\mathcal{B})\coloneqq\bigcup_{e\in E(\mathcal{B})} e\cup U(e)$ and $W(\mathcal{B})\coloneqq\bigcup_{e\in E(\mathcal{B})} W(e)$. We call $\mathcal{B}$ an \emph{improving binocular} if the following three conditions hold:
	\begin{enumerate}[(i)]
		\item \label{ImprovingBinocular1} The sets $W(e),e\in E_2$ are pairwise disjoint.
		\item \label{ImprovingBinocular2} We have $w\left(\bigcup_{e\in E_1} W(e)\setminus \bigcup_{e\in E_2} W(e)\right) \geq w\left(\bigcup_{e\in E_1} U(e)\setminus \bigcup_{e\in E_2} U(e)\right)+2\cdot|E_1|.$
		\item \label{ImprovingBinocular3} $W(\mathcal{B})$ is independent.
	\end{enumerate}
We further call $\mathcal{B}$ an \emph{improving minimal binocular} if $\mathcal{B}$ is a minimal binocular that is improving.\label{DefImprovingBinocular}
\end{definition}
\begin{lemma}
	Let $\mathcal{B}$ be an improving binocular. Then $N(W(\mathcal{B}),A)\subseteq U(\mathcal{B})$ and \\$w(W(\mathcal{B}))>w(U(\mathcal{B}))$.\label{LemBinocularYieldsLocalImprovement}
\end{lemma}
In particular, for any improving binocular $\mathcal{B}$, $W(\mathcal{B})$ constitutes a local improvement. 
\begin{proof}
	Let $\mathcal{B}$ be an improving binocular. By Def.~\ref{DefEdgeInducingPair}, we have $N(W(e),A)\subseteq U(e)\cup e$ for $e\in E(\mathcal{B})$. This implies $N(W(\mathcal{B}),A)\subseteq U(\mathcal{B})$. Moreover, $W(\mathcal{B})$ is independent by Def.~\ref{DefImprovingBinocular}~(\ref{ImprovingBinocular3}). Thus, it remains to show that $w(W(\mathcal{B}))>w(U(\mathcal{B}))$.
	We calculate\begingroup\allowdisplaybreaks
	\begin{align*}
	&\quad w(W(\mathcal{B}))=w\left(\bigcup_{e\in E_2} W(e)\right)+w\left(\bigcup_{e\in E_1} W(e)\setminus \bigcup_{e\in E_2} W(e)\right)\quad | \quad\text{Def.~\ref{DefImprovingBinocular}~(\ref{ImprovingBinocular1})}\\&=\sum_{e\in E_2} w(W(e))+w\left(\bigcup_{e\in E_1} W(e)\setminus \bigcup_{e\in E_2} W(e)\right)\quad | \quad\text{Def.~\ref{DefEdgeInducingPair}~(\ref{EdgeInducingPair2})}\\&=\sum_{e\in E_2} (w(U(e))+2)+w\left(\bigcup_{e\in E_1} W(e)\setminus \bigcup_{e\in E_2} W(e)\right)\quad | \quad\text{Def.~\ref{DefImprovingBinocular}~(\ref{ImprovingBinocular2})}\\
	&\geq\sum_{e\in E_2} w(U(e))+2\cdot |E_2|+w\left(\bigcup_{e\in E_1} U(e)\setminus \bigcup_{e\in E_2} U(e)\right)+2\cdot |E_1|\geq w\left(\bigcup_{e\in E(\mathcal{B})} U(e)\right)+2\cdot|E(\mathcal{B})|\\&>w\left(\bigcup_{e\in E(\mathcal{B})} U(e)\right)+2\cdot|V(\mathcal{B})|\geq w\left(\bigcup_{e\in E(\mathcal{B})} U(e)\right)+w\left(\bigcup_{e\in E(\mathcal{B})} e\right)\geq w(U(\mathcal{B})).
	\end{align*}\endgroup
\end{proof}
We are now ready to actually define the algorithm we would like to analyze. It is given by Algorithm~\ref{OverallAlgorithm}.
\begin{algorithm}[t]
	\DontPrintSemicolon
\KwIn{An instance $(\mathcal{S},w)$ of the $2$-$3$-Set Packing problem, $\tau > 0$}
\KwOut{An independent set in $G_{\mathcal{S}}$ (a.k.a.\ a disjoint sub-collection of $\mathcal{S}$)}
$G\gets G_{\mathcal{S}}$, $A\gets \emptyset$\;
\DoWhile{$\texttt{improvement\_found}$}{
	$\texttt{improvement\_found}\gets\texttt{false}$\;
	\If{$\exists$ a local improvement $X$ of size at most $\tau$}
	{
		$A\gets A\setminus N(X,A)\cup X$\;
		$\texttt{improvement\_found}\gets\texttt{true}$
	}
\If{$\exists$ improving min.\ binocular $\mathcal{B}_0$ in $S_\tau(G,w,A)$ with $|E(\mathcal{B}_0)|\leq \tau\cdot \log(|V(G)|)$\label{LineStartCheckForBinocular}}{
	Compute an improving binocular $\mathcal{B}$ in $S_\tau(G,w,A)$\;
$A\gets A\setminus N(W(\mathcal{B}),A)\cup W(\mathcal{B})$\;
$\texttt{improvement\_found}\gets\texttt{true}$\label{LineEndCheckForBinocular}
}
}
\textbf{Return} $A$\;
\caption{Local search algorithm for the $2$-$3$-Set Packing problem}\label{OverallAlgorithm}
\end{algorithm}
Our main result (Theorem~\ref{TheoGeneralProblem}) is implied by Theorem~\ref{TheoMain}.
\begin{theorem}
Let $\epsilon>0$. There exists a constant $\tau_\epsilon>0$ such that Algorithm~\ref{OverallAlgorithm} yields a $\frac{4}{3}+\epsilon$-approximation for the $2$-$3$-Set Packing problem.

For any fixed constant $\tau$, Algorithm~\ref{OverallAlgorithm} can be implemented to run in polynomial time.\label{TheoMain}
\end{theorem}

In the following section, we provide an overview of our analysis of Algorithm~\ref{OverallAlgorithm} and the key technical ideas. Sections~\ref{SecNormalize} to~\ref{SecFinalAnalysis} are dedicated to the proof of the approximation guarantee stated in Theorem~\ref{TheoMain}. In section~\ref{SecPolyTime}, we show how to implement Algorithm~\ref{OverallAlgorithm} in polynomial time. Finally, in section~\ref{Sec:Hereditary}, we point out how to exploit the additional structure of an instance of the hereditary $2$-$3$-Set Packing problem to obtain a guarantee of $\frac{4}{3}$ (instead of $\frac{4}{3}+\epsilon$) in this special case.
\section{Overview of the analysis\label{Sec:Overview}}
To prove the approximation guarantee claimed in Theorem~\ref{TheoMain}, we perform an analysis that uses similar ideas as the ones presented by Cygan~\cite{Cygan} and F\"urer and Yu~\cite{FurerYu} for the unweighted $3$-Set Packing problem. In the following, we provide some of the main ideas appearing in \cite{FurerYu} and comment on the difficulties we faced when trying to adapt them to our more general setting. This will motivate our approach and the statements we need to prove to make our analysis work.

In \cite{FurerYu}, the authors consider an algorithm that is very similar to Algorithm~\ref{OverallAlgorithm} and show that for any $\epsilon > 0$, there exists a constant $\tau_\epsilon$ such that Algorithm~\ref{OverallAlgorithm}, when run with parameter $\tau=\tau_\epsilon$, yields a $\frac{4}{3}+\epsilon$-approximation for the unweighted $3$-Set Packing problem\footnote{In fact, they show more generally that for any $k\geq 3$ and any $\epsilon>0$, there is $\tau_{\epsilon,k}$ such that Algorithm~\ref{OverallAlgorithm} with $\tau=\tau_{\epsilon,k}$ yields a $\frac{k+1}{3}+\epsilon$-approximation for the unweighted $k$-Set Packing problem.}.
We first sketch how to obtain a guarantee of $\frac{5}{3}+\epsilon$ for unweighted $3$-Set Packing via a relatively simple analysis, and then explain how F\"urer and Yu improve on this to obtain the desired guarantee of $\frac{4}{3}+\epsilon$. 
 \subsection{\texorpdfstring{Obtaining an approximation guarantee of $\nicefrac{5}{3}+\epsilon$}{Obtaining an approximation guarantee of 5/3+epsilon}}
Let $\epsilon\in (0,1)$, $G=G_{\mathcal{S}}$ be the conflict graph of a $3$-Set Packing instance, let $A$ be the solution returned by Algorithm~\ref{OverallAlgorithm} and let $B$ be an optimum solution. For the analysis, we may assume that $A\cap B=\emptyset$ because otherwise, we can work with $G[A\Delta B]$ instead. See section~\ref{SecNormalize} for more details. We choose $\tau_\epsilon \coloneqq 4\cdot \lceil\frac{2}{\epsilon}\rceil \geq 1.$ As there is no local improvement of constant size $\tau$, we can infer that every vertex in $B$ has a neighbor in $A$ (otherwise, it would constitute a local improvement of size $1$). In particular, we can partition $B$ into the three sets $B_1$, $B_2$ and $B_3$, where $B_i$ contains those vertices from $B$ with exactly $i$ neighbors in $A$. Note that no vertex from $B$ can have more than $3$ neighbors in the independent set $A$ since $G$ is $4$-claw free by Prop.~\ref{PropStructureConflictGraph}. Analogously, no vertex in $A$ can have more than $3$ neighbors in $B$. Thus, by counting the edges between $A$ and $B$ from both sides, we obtain \begin{equation}3\cdot |A|\geq |E(A,B)|=|B_1|+2\cdot|B_2|+3\cdot|B_3|.\label{EqCountDegrees}\end{equation}
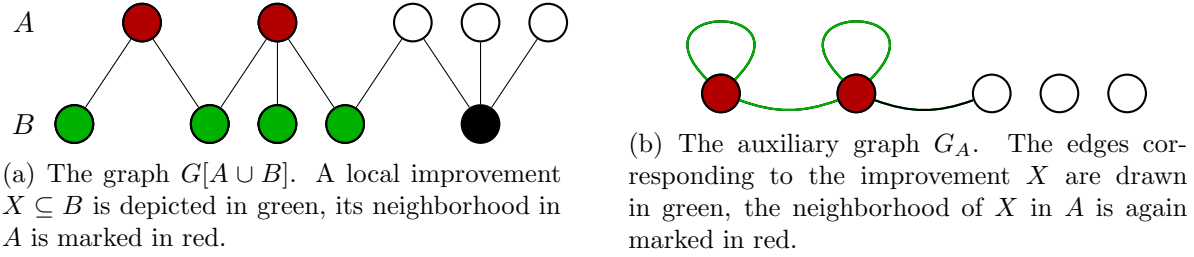
\begin{figure}
	\begin{subfigure}{0.45\textwidth}
		\begin{tikzpicture}[scale = 0.45,mynode/.style = {circle, draw = black, thick, fill = black, inner sep = 0mm, minimum size = 5mm}, solution/.style = {circle, draw = black, thick, fill = none, inner sep = 0mm, minimum size = 5mm},improvement/.style = {circle, draw = black, thick, fill = green!70!black, inner sep = 0mm, minimum size = 5mm}, loop/.style={min distance=10mm,in=45,out=135,looseness=10}]
		\node at (-1.5,0) {$B$};
		\node at (-1.5,3) {$A$};
		\node[mynode] (A) at (0,0){};
		\node[solution] (B) at (2,3){};
		\node[mynode] (C) at (4,0){};
		\node[solution] (D) at (6,3){};
		\node[mynode] (E) at (6,0){};
		\node[mynode] (F) at (8,0){};
		\node[solution] (G) at(10,3){};
		\node[mynode] (H) at (12,0){};
		\node[solution] (I) at (12,3){};
		\node[solution] (J) at (14,3){};
		\draw (A)--(B);
		\draw (B)--(C);
		\draw (C)--(D);
		\draw (D)--(E);
		\draw (D)--(F);
		\draw (F)--(G);
		\draw (G)--(H);
		\draw (H)--(I);
		\draw (H)--(J);
		{
			\node[improvement] (A) at (0,0){};
			\node[improvement] (E) at (6,0){};
		}
		{
			\node[improvement] (C) at (4,0){};
			\node[improvement] (F) at (8,0){};
		}
		{
			\node[mynode, fill = green!70!black] (A) at (0,0){};
			\node[solution, fill = red!70!black] (B) at (2,3){};
			\node[mynode, fill = green!70!black] (C) at (4,0){};
			\node[solution, fill = red!70!black] (D) at (6,3){};
			\node[mynode, fill = green!70!black] (E) at (6,0){};
		}
		\end{tikzpicture}
		\subcaption{The graph $G[A\cup B]$. A local improvement $X\subseteq B$ is depicted in green, its neighborhood in $A$ is marked in red.}
	\end{subfigure}
\qquad
	\begin{subfigure}{0.45\textwidth}
		\begin{tikzpicture}[scale = 0.45,mynode/.style = {circle, draw = black, thick, fill = black, inner sep = 0mm, minimum size = 5mm}, solution/.style = {circle, draw = black, thick, fill = none, inner sep = 0mm, minimum size = 5mm},improvement/.style = {circle, draw = green!70!black, thick, fill = green!70!black, inner sep = 0mm, minimum size = 5mm}, loop/.style={min distance=10mm,in=45,out=135,looseness=10}]
		\node[solution] (B) at (2,4){};
		\node[solution] (D) at (6,4){};
		\node[solution] (G) at(10,4){};
		\node[solution] (I) at (12,4){};
		\node[solution] (J) at (14,4){};
		{
			\draw[green!70!black, thick] (B) to[loop above] (B);
			\draw[green!70!black, thick] (D) to [loop above] (D);
		}
		{
			\draw[thick] (B) to[loop above] (B);
			\draw[thick] (D) to [loop above] (D);
		}
		{
			\draw[thick, green!70!black] (B) to[bend right = 20] (D);
			\draw[thick, green!70!black] (D) to[bend right = 20] (G);
		}
		{
			\draw[thick] (B) to[bend right = 20] (D);
			\draw[thick] (D) to[bend right = 20] (G);
		}
		{
			\node[solution, fill = red!70!black] (B) at (2,4){};
			\node[solution, fill = red!70!black] (D) at (6,4){};
			\draw[green!70!black, thick] (B) to[loop above] (B);
			\draw[green!70!black, thick] (D) to [loop above] (D);
			\draw[thick, green!70!black] (B) to[bend right = 20] (D);
		}
		
		\end{tikzpicture}
		\subcaption{The auxiliary graph $G_A$. The edges corresponding to the improvement $X$ are drawn in green, the neighborhood of $X$ in $A$ is again marked in red.}
	\end{subfigure}

	\caption{Construction of the auxiliary graph $G_A$. There is a one-to-one correspondence between local improvements featuring only vertices from $B_1\cup B_2$, and binoculars in $G_A$.\label{FigGA}}
\end{figure}
For vertices from $B_3$, this estimate is already sufficient for our purposes, but for vertices from $B_1\cup B_2$, we need to put in some further effort. To this end, we construct an auxiliary graph $G_A$ (see~\cite{FurerYu}) on the vertex set $A$ as follows: Every vertex from $B_1$ is represented by a loop on its unique neighbor in $A$, and every vertex in $B_2$ induces an edge between its two neighbors in $A$. See Figure~\ref{FigGA} for an example. By construction, we obtain a one-to-one correspondence between sub-graphs $H$ of $G_A$ with more edges than vertices and local improvements $X\subseteq B_1\cup B_2$ by mapping $H\subseteq G_A$ to the set of vertices from $B_1\cup B_2$ inducing its edges, and associating $X\subseteq B_1\cup B_2$ with the sub-graph containing precisely the edges induced by $X$. 
Now, a result by Berman and F\"urer~\cite{berman1994approximating} comes into play: It tells us that every graph whose number of edges is by a constant factor larger than its number of vertices contains a binocular (and in particular also a minimal binocular) of logarithmically bounded size (see Lemma~\ref{LemBinocular}).
\begin{lemma}[\cite{berman1994approximating}]
	Let $s\in\mathbb{N}$, $G=(V,E)$ be a graph s.t. $|E|\geq\frac{s+1}{s}\cdot|V|.$
	Then $G$ contains a binocular of size at most $4\cdot s\cdot \log(|V|)$.\label{LemBinocular}
\end{lemma}
We observe that  as $B$ is independent, a minimal binocular in $G_A$ yields an improving minimal binocular by choosing $U(e)=\emptyset$ and $W(e)=\{v\}$ for the edge $e$ induced by $v\in B_1\cup B_2$. (Recall that our reduction from the unweighted $3$-Set Packing problem to the $2$-$3$-Set Packing problem assigns a weight of $2$ to every set.)
By our choice of $\tau_\epsilon$, we know that \begin{equation}|B_1|+|B_2|=|E(G_A)|\leq \left(1+\frac{\epsilon}{2}\right)\cdot |V(G_A)|=\left(1+\frac{\epsilon}{2}\right)\cdot |A|.\label{EqBoundNoBinocular}\end{equation} Combining \eqref{EqCountDegrees} and \eqref{EqBoundNoBinocular} tells us that \[(5+\epsilon)\cdot |A|\geq 3\cdot|B_1|+4\cdot|B_2|+3\cdot|B_3|\geq 3\cdot|B|,\] which yields the desired approximation guarantee. 

\subsection{\texorpdfstring{Improving the guarantee to $\nicefrac{4}{3}+\epsilon$}{Improving the guarantee to 4/3+epsilon}}
Observe that the bottleneck in the previous analysis, which prevents us from showing a guarantee of $\frac{4}{3}+\epsilon$ instead of $\frac{5}{3}+\epsilon$, are the vertices in $B_1$ because they force us to add \eqref{EqBoundNoBinocular} twice instead of just once. Indeed, if we consider an instance where each vertex in $A$ has exactly one neighbor in $B_1$, and two neighbors in $B_3$, we end up with a ratio of $\frac{5}{3}$. (However, in such an instance, local improvements of constant size exist.)

In order to remedy this issue, F\"urer and Yu introduce the concept of a \emph{tail change}. Up to some technical details that we will discuss at a later point, a tail change consists of sets $U\subseteq A$ and $V\subseteq B$ with $|U|=|V|\leq \tau$ and $N(V,A)\subseteq U$. We remark that the sets $U$ and $W$ in our definition of an edge-inducing pair are used to represent tail changes. A family of tail changes is called \emph{consistent} if the corresponding subsets of $A$ and $B$ are pairwise disjoint. A nice property of a consistent family of tail changes $(U_i,V_i)_{i\in I}$ is that we can remove $A_t\coloneqq\bigcup_{i\in I} U_i$ and $B_t\coloneqq\bigcup_{i\in I} V_i$ from our instance and analyze $A\setminus A_t$ and $B\setminus B_t$ instead. As $|A_t|=|B_t|$, any bound $\alpha\geq 1$ on the ratio $\frac{|B\setminus B_t|}{|A\setminus A_t|}$ yields a bound on $\frac{|B|}{|A|}$. On the other hand, a local improvement $X\subseteq B\setminus B_t$ of constant size in the reduced instance translates into a local improvement of larger, but still constant size in the original instance by adding the sets $V_i$ for the (constant number of) tail changes $(U_i,V_i)$ for which $U_i$ is adjacent to $X$. With a little technical work, we can further translate (minimal) binoculars in the auxiliary graph $G_{A\setminus A_t}$ into improving binoculars in the search graph.

Hence, we can carry out essentially the same analysis as before after removing a consistent family of tail changes. The key insight of F\"urer and Yu is that it is possible to construct a consistent family of tail changes in such a way that after removing $A_t$ and $B_t$ from the instance, we can bound $|B_1|$ by $|A_t|+\epsilon\cdot |B|$\footnote{A similar idea also appears in~\cite{Cygan}}. Then, using \eqref{EqCountDegrees}, \eqref{EqBoundNoBinocular}, $|A_t|=|B_t|$ and $|B_1|\leq |A_t|+\epsilon\cdot|B|$ leads to an approximation guarantee of $\frac{4+\frac{\epsilon}{2}}{3-\epsilon}$.

\subsection{How to construct the tail changes}
The main question that remains is of course how to obtain a consistent family of tail changes with the desired property. To this end, we point out that a lot of the technicalities F\"urer and Yu have to deal with stem from the fact that they aim at a running time that is only singly exponential in $\epsilon^{-2}$, as opposed to the doubly exponential running time in~\cite{Cygan}. In this work, we do not care so much about the dependence of the running time on $\epsilon$, which is why we may settle for a simpler approach that is similar to the one in~\cite{Cygan}.

Our construction of tail changes proceeds in iterations. At the beginning of each iteration, we are given a consistent family $(U_i,V_i)_{i\in I}$ of tail changes, which will be empty initially. Let $A_t\coloneqq\bigcup_{i\in I} U_i$ and $B_t\coloneqq\bigcup_{i\in I} V_i$. We remove $A_t$ and $B_t$ from our instance and consider the set $B_1$. If $|B_1|\leq \epsilon \cdot |B|$, we stop. Otherwise, we observe that every vertex $v\in B_1$ has a neighbor $n(v)$ in $A$ and these neighbors are pairwise distinct because otherwise, we obtain a local improvement of constant size. Thus, we can obtain a new family of tail changes by adding, for each $v\in B_1$, a tail change exchanging $n(v)$ and the union of the sets $U_i$ such that $N(v,U_i)\neq \emptyset$ by $v$ and the corresponding $V_i$. We then remove all tail changes from the old family that have been subsumed. In order to ensure that this procedure preserves consistency, we need to perform the following technical adjustment: We associate each tail change with an incident edge $e_i$ of $U_i$ and restrict $B_1$ to those vertices that have degree at most $1$ after removing only those edges associated with the tail changes. In doing so, we make sure that for every tail change $(U_i,V_i)$ in our current family, there is at most one vertex in $B_1$ adjacent to $U_i$. This is enough to guarantee consistency. Moreover, we can show that each vertex in $A_t$ will have at most one incident edge connecting it to $B\setminus B_t$ that is not associated with any tail change. Hence, when the algorithm terminates, there can be at most $|A_t|+\epsilon\cdot |B|$ vertices in $B\setminus B_t$ with degree $1$ to $A\setminus A_t$ because either they are contained in $B_1$, or they have an incident edge to $A_t$ that is not associated with any tail change. Finally, it is not hard to see that our iterative tail change construction terminates after at most $\epsilon^{-1}$ iterations because $|B_t|$ increases by at least $\epsilon\cdot|B|$ in each iteration.
\subsection{Critical components pose a challenge}
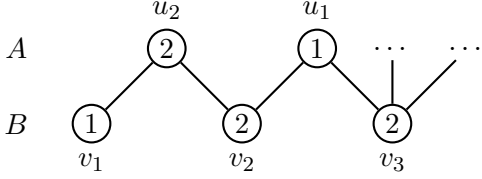
\begin{figure}
	\begin{tikzpicture}[mynode/.style = {circle, draw = black, thick, fill = none, inner sep = 0mm, minimum size = 5mm}]
	\node at (-1,1) {$A$};
	\node at (-1,0) {$B$};
	\node[mynode,label=below:$v_1$] (V1) at (0,0) {$1$};
	\node[mynode,label=below:$v_2$] (V2) at (2,0) {$2$};
	\node[mynode,label=below:$v_3$] (V3) at (4,0) {$2$};
	\node[mynode,label=above:$u_2$] (U2) at (1,1) {$2$};
	\node[mynode,label=above:$u_1$] (U1) at (3,1) {$1$};
	\node (D1) at (4,1) {$\dots$};
	\node (D2) at (5,1) {$\dots$};
	\draw[thick] (V1)--(U2)--(V2)--(U1)--(V3)--(D1);
	\draw[thick] (V3)--(D2);
	\end{tikzpicture}
	\caption{A critical component. \label{FigCritical}}
\end{figure}

The approach described above yields an approximation guarantee of $\frac{4}{3}+\epsilon$ for the unweighted $3$-Set Packing problem. Unfortunately, it is not quite sufficient for our purposes. The reason for this is that during the previous tail change construction, we can only remove vertices of weight $2$ from $B$. Indeed, if we would remove a vertex $v$ of weight $1$ with exactly one neighbor $u$ of weight $2$ in $A$, we would lose the crucial property that the sets of vertices we remove from $A$ and $B$ have the same weight. As a consequence, after constructing and removing our family of tail changes, we may run into the situation depicted in Figure~\ref{FigCritical}. There, we see three vertices from $B$, $v_1$ of weight $1$, and $v_2$ and $v_3$ of weight $2$, and their neighbors in $A$ (after removing the tail changes). $v_1$ is adjacent to a vertex $u_2\in A$ of weight $2$. $v_2$ is adjacent to $u_2$ and a vertex $u_1$ of weight $1$. Finally, the neighborhood of $v_3$ in $A$ consists of $u_1$ and two further vertices of unspecified weight. As in the previous analysis, for a vertex $v\in B$ with three neighbors in $A$, we would like each of these neighbors to ``pay'' for one third of the weight of $v$. On the other hand, we would like to bound the total weight of the vertices in $B$ with at most two neighbors in $A$ by $(1+\epsilon)$ times the weight of $A$. But now, we run into trouble at $u_1$: If we only take into account vertices from $B$ of degree $1$ or $2$ to $A$, we cannot find a local improvement in Figure~\ref{FigCritical}. In particular, $\{v_1,v_2\}$ does not constitute an improvement because both $\{v_1,v_2\}$ and $N(\{v_1,v_2\},A)=\{u_1,u_2\}$ contain exactly one vertex of weight $1$ and one vertex of weight $2$. However, we cannot make $u_1$ pay its full weight to cover for $v_1$ and $v_2$ because it might also have to pay $\frac{2}{3}$ for $v_3$, which results in a total of $\frac{5}{3}>\frac{4}{3}$. This is too much for the guarantee we are aiming at. 
To resolve this issue, we introduce the notion of a \emph{critical component}, which captures essentially the situation displayed in Figure~\ref{FigCritical}. Our idea is to simply transform small\footnote{If the component is sufficiently large, we can amortize what we need to pay for $v_3$ against all vertices in the component.} critical components into tail changes as well (in the given example, the tail change would consist of $\{u_1,u_2\}$ and $\{v_1,v_2\}$). However, in doing so, new vertices may enter the set $B_1$, which may thus grow in size. Hence, we launch another round of the tail change construction, removing all but an $\epsilon$-fraction of the new set $B_1$. We iterate this process until both $|B_1|$ and the number of critical components are bounded by $\epsilon\cdot |B|$. As in each iteration, we will add at least an $\epsilon$-fraction of the vertices in $B$ to the set of vertices contained in the tail changes, we can still bound the number of iterations by $\epsilon^{-1}$. This will ensure that we can further bound the sizes of the tail changes we construct by some constant (depending on $\epsilon$). The rest of the analysis then goes along a similar road as in \cite{FurerYu}.

The remainder of this paper is organized as follows: In section~\ref{SecNormalize}, we show that for the analysis of the approximation guarantee, we may restrict ourselves to a well-structured class of instances, which we call normalized. This will simplify the subsequent arguments. In section~\ref{SecConstructingTailChanges}, we then explain in more detail how to, given a consistent family of tail changes, obtain a new family of tail changes that removes all of $B_1$. Next, section~\ref{SecCritical} formally introduces the notion of a critical component and shows how to achieve a small number of vertices in $B_1$ and a small number of critical components via an iterative tail change construction. Finally, section~\ref{SecFinalAnalysis} puts everything together to prove the first part of Theorem~\ref{TheoMain}. In section~\ref{SecPolyTime}, we explain how to implement Algorithm~\ref{OverallAlgorithm} with a polynomial running time. This is done by applying the color coding technique, similar as in~\cite{FurerYu}.
Finally, in section~\ref{Sec:Hereditary}, we show that for the hereditary $2$-$3$-Set Packing problem, any solution that is locally optimal w.r.t.\ local improvements of size at most $10$ already is $\frac{4}{3}$-approximate. In particular, this shows that Algorithm~\ref{OverallAlgorithm} with $\tau=10$ yields a polynomial time $\frac{4}{3}$-approximation for the hereditary $2$-$3$-Set Packing problem. Moreover, the additional structure of the hereditary $2$-$3$-Set Packing problem even allows us to omit the (run-time expensive) search for improving binoculars.
\section{Normalizing the instance\label{SecNormalize}}
In this section, we first prove that for the analysis, we can restrict ourselves to a well-structured class of instances that we call \emph{normalized} (see Def.~\ref{DefNormalized}). Normalized instances feature some nice properties that simplify the presentation of the analysis.
\begin{definition}[normalized instance]
A \emph{normalized instance} is a tuple $(G,w,A,B)$ consisting of a graph $G$, $w:V(G)\rightarrow\{1\,2\}$ and two independent sets $A,B\subseteq V(G)$ with the following properties:
\begin{enumerate}[(i)]
	\item $G$ is $4$-claw free and every $3$-claw in $G$ is centered at a vertex of weight $2$.
	\item $G$ is bipartite with bipartitions $A$ and $B$ and $A'\cup B'$ forms an independent set in $G$.
\end{enumerate} \label{DefNormalized}
\end{definition}
Theorem~\ref{TheoNormalization} tells us that it suffices to analyze normalized instance to prove an approximation guarantee of $\frac{4}{3}+\epsilon$.
\begin{theorem}
	Let $\bar{\tau} > 0$ be a constant. Then there exists a constant $\tau > 0$ with the following property:
Let $(\mathcal{S},w)$ be an instance of the $2$-$3$-Set Packing problem and let $A$ and $B$ be two independent sets in $G=G_{\mathcal{S}}$. Then there exists a normalized instance $(\bar{G},\bar{w},\bar{A},\bar{B})$ such that:
\begin{itemize}
	\item For every constant $\alpha \geq \frac{4}{3}$, if we have $w(\bar{B})\leq \alpha\cdot w(\bar{A})$, then we also have $w(B)\leq w(A)$.
	\item If there exists a local improvement of $\bar{A}$ in $(\bar{G},\bar{w})$ of size at most $\bar{\tau}$, then there exists a local improvement of $A$ in $(G,w)$ of size at most $\tau$. 
	\item If there exists an improving minimal binocular in $S_{\bar{\tau}}(\bar{G},\bar{w},\bar{A})$ of size at most $\bar{\tau}\cdot\log|V(\bar{G})|$, then there exists an improving minimal binocular in $S_\tau(G,w,A)$ of size at most $\tau\cdot\log |V(G)|$. 
\end{itemize}\label{TheoNormalization}
\end{theorem}
In the remainder of this section, we briefly sketch the proof of Theorem~\ref{TheoNormalization}. A formal proof can be found in appendix~\ref{appendix:Normalization}.
In order to prove Theorem~\ref{TheoNormalization}, we first observe that we can replace $(G,w,A,B)$ by $(G[A\cup B],w\upharpoonright_{A\cup B},A,B)$ because the ratio between the weights of $A$ and $B$ remains the same and it is not hard to see that any local improvement of constant size that is a subset of $B$ also constitutes a local improvement of constant size in the original graph. Moreover, by carefully going through the respective definitions, one can show that an improving minimal binocular in $S_\tau(G[A\cup B], w\upharpoonright_{A\cup B},A)$ constitutes an improving minimal binocular in $S_\tau(G,w,A)$. As both $A$ and $B$ are independent, the vertices in $A\cap B$ are isolated in $G[A\cup B]$. Thus, we may further restrict ourselves to $G[A\Delta B]$. In particular, by replacing $A$ with $A\setminus B$ and $B$ with $B\setminus A$, we can assume that $G$ is bipartite with bipartitions $A$ and $B$. 
\begin{figure}
	\begin{subfigure}{\textwidth}
		\centering
		\begin{tikzpicture}[yscale = 0.8,mynode/.style = {circle, draw = black, thick, fill = none, inner sep = 0mm, minimum size = 5mm}, decoration = brace]
		\node[mynode] (A1) at (0,2){};
		\node[mynode, fill = black!30!white] (A2) at (2,2){};
		\node[mynode, fill = black!30!white] (A3) at (4,2){};
		\node[mynode, fill = black!30!white] (A4) at (6,2){};
		\node[mynode, fill = black!30!white] (B1) at (1,0){};
		\node[mynode, fill = black!30!white] (B2) at (3,0){};
		\node[mynode, fill = black!30!white] (B3) at (5,0){};
		\node[mynode, fill = black!30!white] (B4) at (7,0){};
		\draw[ultra thick, black!30!white] (B1)--(A2)--(B2)--(A3)--(B3)--(A4)--(B4);
		\draw[thick] (A1)--(B1)--(A2)--(B2)--(A3)--(B3)--(A4)--(B4);
		\node at (-1,2) {$A'$};
		\node at (-1,0) {$B'$};
		\draw[decorate, thick] (1.7,2.7) to node[midway, above = 5pt] {$\geq 3$ inner vertices from $A'$} (6.3,2.7);
		\end{tikzpicture}
		\caption{A connected component $P$ of $G[A'\cup B']$ that constitutes a path with at least three inner vertices from $A'$. $Q$ (marked in gray) arises from $P$ by deleting those endpoints of $P$ that are contained in $A'$.}\label{FigLongPathComponent}
	\end{subfigure}
	\begin{subfigure}{\textwidth}
		\centering
		\begin{tikzpicture}[yscale = 0.8,mynode/.style = {circle, draw = black, thick, fill = none, inner sep = 0mm, minimum size = 5mm}, decoration = brace]
		\node[mynode] (A2) at (2,2){$1$};
		\node[mynode] (A3) at (4,2){$1$};
		\node[mynode] (A4) at (6,2){$1$};
		\node[mynode, dashed] (B1) at (1,0){$2$};
		\node[mynode] (B2) at (3,0){$1$};
		\node[mynode] (B3) at (5,0){$1$};
		\node[mynode, dashed] (B4) at (7,0){$2$};
		\draw[thick] (A2)--(B2)--(A3)--(B3)--(A4);
		\draw[thick, dashed] (B1)--(A2);
		\draw[thick, dashed] (A4)--(B4);
		\node at (0,2) {$A$};
		\node at (0,0) {$B$};
		\draw[thick, ->] (7.5,1)--(8,1);
		\node[mynode] (A5) at (10,2) {$1$};
		\node[mynode, dashed] (B5) at (9,0) {$2$};
		\node[mynode,dashed] (B6) at (11,0) {$2$};
		\draw[thick, dashed] (B5)--(A5)--(B6);
		\end{tikzpicture}
		\caption{A connected component $P$ of $G[A'\cup B']$ that constitutes a short path with an odd number of vertices and both endpoints in $A'$. We contract it into a single vertex in $A'$. Dashed lines indicate neighbors of $P$ in $B''$ that may, but need not exist.}\label{FigOddPathComponentA}
	\end{subfigure}
	\begin{subfigure}{\textwidth}
		\centering
		\begin{tikzpicture}[yscale = 0.8,mynode/.style = {circle, draw = black, thick, fill = none, inner sep = 0mm, minimum size = 5mm}, decoration = brace]
		\node[mynode] (A2) at (2,0){$1$};
		\node[mynode] (A3) at (4,0){$1$};
		\node[mynode] (A4) at (6,0){$1$};
		\node[mynode, dashed] (B1) at (1,2){$2$};
		\node[mynode] (B2) at (3,2){$1$};
		\node[mynode] (B3) at (5,2){$1$};
		\node[mynode, dashed] (B4) at (7,2){$2$};
		\draw[thick] (A2)--(B2)--(A3)--(B3)--(A4);
		\draw[thick, dashed] (B1)--(A2);
		\draw[thick, dashed] (A4)--(B4);
		\node at (0,2) {$A$};
		\node at (0,0) {$B$};
		\draw[thick, ->] (7.5,1)--(8,1);
		\node[mynode] (A5) at (10,0) {$1$};
		\node[mynode, dashed] (B5) at (9,2) {$2$};
		\node[mynode,dashed] (B6) at (11,2) {$2$};
		\draw[thick, dashed] (B5)--(A5)--(B6);
		\end{tikzpicture}
		\caption{A connected component $P$ of $G[A'\cup B']$ that constitutes a short path with an odd number of vertices and both endpoints in $B'$. We contract it into a single vertex in $B'$. Dashed lines indicate neighbors of $P$ in $A''$ that may, but need not exist.}\label{FigOddComponentB}
	\end{subfigure}
	\begin{subfigure}{\textwidth}
		\centering
		\begin{tikzpicture}[yscale = 0.8,mynode/.style = {circle, draw = black, thick, fill = none, inner sep = 0mm, minimum size = 5mm}, decoration = brace]
		\node[mynode] (A2) at (2,2){$1$};
		\node[mynode] (A3) at (4,2){$1$};
		\node[mynode, dashed] (A4) at (6,2){$2$};
		\node[mynode, dashed] (B1) at (1,0){$2$};
		\node[mynode] (B2) at (3,0){$1$};
		\node[mynode] (B3) at (5,0){$1$};
		\draw[thick] (A2)--(B2)--(A3)--(B3);
		\draw[thick, dashed] (B1)--(A2);
		\draw[thick, dashed] (B3)--(A4);
		\node at (0,2) {$A$};
		\node at (0,0) {$B$};
		\draw[thick, ->] (7.5,1)--(8,1);
		\node[mynode, dashed] (A5) at (10,2) {$2$};
		\node[mynode, dashed] (B5) at (9,0) {$2$};
		\draw[thick, dashed] (B5)--(A5);
		\end{tikzpicture}
		\caption{A connected component $P$ of $G[A'\cup B']$ that constitutes a short path with an even number of vertices. We delete it and in case the endpoints of the path have neighbors in $A''$ and $B''$, respectively, we connect them by an edge.}\label{FigEvenComponent}
	\end{subfigure}
	\caption{Illustration of our instance reduction steps.}\label{FigInstanceReduction}
\end{figure}

It remains to show how to turn $A'\cup B'$ into an independent set. To this end, we first note that Proposition~\ref{PropStructureConflictGraph}, and the fact that $G$ is bipartite with bipartitions $A$ and $B$ implies that every vertex in $A'\cup B'$ has degree at most $2$ in $G$. Thus, every connected component of $G[A'\cup B']$ is an $A'$-$B'$-alternating path or cycle. In a cycle component, every vertex already attains its maximum possible degree of $2$, which implies that is also constitutes a connected component of $G$, and, hence, an independent sub-instance. As an $A'$-$B'$-alternating cycle further contains the same number of vertices from $A'$ and $B'$, we can simply delete it because adding it back will only improve the approximation guarantee. 

Hence, we are left with path components. For a path component $P$ that contains at least three inner vertices from $A'$, we consider the sub-path $Q$ that arises by deleting endpoints of $P$ that are contained in $A'$, and delete the vertices in $V(Q)$ (see Figure~\ref{FigLongPathComponent}). Then $V(P)\setminus V(Q)$ is independent in $G$. In addition, for every vertex in $V(Q)\cap A'$, both of its neighbors in $B$ are contained in $V(Q)$ as well. Thus, deleting $V(Q)$ preserves the property that every local improvement in the resulting instance also constitutes a local improvement in the original one. Finally, as $P$ contains at least three inner vertices from $A'$, we know that $|V(Q)\cap B'|\leq |V(Q)\cap A'|+1\leq \frac{4}{3}\cdot |V(Q)\cap A'|$. As a consequence, adding back $V(Q)$ preserves an approximation guarantee of up to $\frac{4}{3}$.
All of the remaining path components contain at most two inner vertices from $A'$, and, thus, at most $7$ vertices in total. We call such components \emph{short}.

 If a short path component $P$ features an odd number of vertices, we contract it into a single vertex of weight $1$, which is contained in $A'$ if both of the endpoints of $P$ are, and in $B'$ otherwise (see Figures~\ref{FigOddPathComponentA} and~\ref{FigOddComponentB}). In doing so, we reduce the number of vertices from $A'$ and $B'$ by the same amount. Thus, undoing the contraction preserves approximation guarantees and only blows up local improvements by a constant factor.
Finally, for a short path component $P$ with an even number of vertices, let $u_1$ be its endpoint in $A'$ and let $v_1$ be its endpoint in $B'$. If $u_1$ has a neighbor $v_2$ in $B''$ and $v_1$ has a neighbor $u_2$ in $A''$, we connect them by an edge (see Figure~\ref{FigEvenComponent}). Note that since every vertex in $A'\cup B'$ has degree at most two, there can be a most one such neighbor for each of $u_1$ and $v_1$. Then, we delete $V(P)$. Again, adding back $V(P)$ preserves approximation guarantees because $V(P)$ contains the same number of vertices from $A'$ and $B'$. Moreover, the edge we have added makes sure that whenever a local improvement in the remaining instance contains $v_2$ (if exists), $u_2$ (if exists) is contained in its neighborhood. Thus, we can add $V(P)\cap B'$ to obtain a local improvement in the original instance, only increasing the size by a constant factor.
 \begin{theorem}
 Let $\epsilon >0$. Then there exists a constant $\tau$ with the following property: For every normalized instance $(G,w,A,B)$ such that
 \begin{itemize}
 	\item there exists no local improvement of size at most $\tau$ w.r.t.\ $A$ and
 	\item there exists no improving minimal binocular of size at most $\tau\cdot\log(|V(G)|)$ in $S_\tau(G,w,A)$,
 \end{itemize}
we have $w(B)\leq \left(\frac{4}{3}+\epsilon\right)\cdot w(A)$. \label{TheoAnalysisNormalized}
 \end{theorem}
Sections~\ref{SecConstructingTailChanges} to \ref{SecFinalAnalysis} are dedicated to the proof of Theorem~\ref{TheoAnalysisNormalized}. We point out that Theorem~\ref{TheoAnalysisNormalized} implies that Algorithm~\ref{OverallAlgorithm} achieves the desired approximation guarantee.
\begin{corollary}
For every $\epsilon>0$, there exists a constant $\tau$ such that Algorithm~\ref{OverallAlgorithm} with parameter $\tau$ yields a $\frac{4}{3}+\epsilon$-approximation for the $2$-$3$-Set Packing problem.
\end{corollary}
\begin{proof}
Let $\epsilon >0$ and pick a constant $\bar{\tau}$ as provided by Theorem~\ref{TheoAnalysisNormalized}. Moreover, pick $\tau$ as given by Theorem~\ref{TheoNormalization}. We claim that $\tau$ meets the requirements of the corollary. Let $(\mathcal{S},w)$ be an instance of the $2$-$3$-Set Packing problem, let $B$ be an independent set in $G=G_{\mathcal{S}}$ of maximum weight (i.e.\ an optimum solution to $(\mathcal{S},w)$) and let $A$ be the output of Algorithm~\ref{OverallAlgorithm} with parameter $\tau$ and input $(\mathcal{S},w)$. Then there is no local improvement of size at most $\tau$ and no improving minimal binocular of size at most $\tau\cdot\log(|V(G)|)$ w.r.t.\ $A$. Let $(\bar{G},\bar{w},\bar{A},\bar{B})$ be as provided by Theorem~\ref{TheoNormalization}. Then there is no local improvement of size at most $\bar{\tau}$ and no improving minimal binocular of size at most $\bar{\tau}\cdot\log(|V(\bar{G}|)$ w.r.t.\ $\bar{A}$. Thus,Theorem~\ref{TheoAnalysisNormalized} tells us that $w(\bar{B})\leq \left(\frac{4}{3}+\epsilon\right)\cdot w(\bar{A})$. By the guarantees of Theorem~\ref{TheoNormalization}, this implies $w(B)\leq \left(\frac{4}{3}+\epsilon\right)\cdot w(A)$.
\end{proof}
\section{Constructing tail changes\label{SecConstructingTailChanges}}
In order to prove Theorem~\ref{TheoAnalysisNormalized}, we first need to formally introduce the notions of a tail change and a consistent family of tail changes.
\begin{definition}[tail change]
Let $(G,w,A,B)$ be normalized. A \emph{tail change} is a triple $(U,W,e)$, where $U\subseteq A$, $W\subseteq B$, $w(U)=w(W)$, $|U|=|W|$, $N(W,A)\subseteq U$ and $e$ is an incident edge of a vertex in $U$ (which can be a loop not contained in $G$)\footnote{For technical reasons, we will later add loops to vertices from $A''$. This is also done in~\cite{FurerYu}.}. We call $e$ the \emph{edge associated with the tail change}. The \emph{size} of the tail change is $|U|=|W|$.
\end{definition}
Observe that the definition of a tail change implies that $U$ and $W$ are non-empty because $e$ is incident to $U$ and $|W|=|U|$. Moreover, the two conditions $w(U)=w(W)$ and $|U|=|W|$ are equivalent to requiring that $U$ and $W$ feature the same amount of vertices of weight $1$ and weight $2$, respectively.
\begin{definition}[Family of consistent tail changes]
A family $(U_i,W_i, e_i)_{i\in I}$ of tail changes is called \emph{consistent} if $U_i\cap U_j=\emptyset$ and $W_i\cap W_j=\emptyset$ for all $i\neq j\in I$.
\end{definition}
Note that this notion of consistency is weaker than the one in~\cite{FurerYu}, where the authors additionally require that for every vertex $v\in B$ with $|\delta(v)\cap\{e_i,i\in I\}\neq \emptyset$, we have $|\delta(v)\setminus\{e_i,i\in I\}|\geq 2$. As this additional requirement makes the iterative construction of tail changes much more tedious and is not needed for our purposes, we omit it in our definition of consistency.

Our main result for this section is Theorem~\ref{TheoNewFamilyOfTailChanges}, which constitutes one of the two building blocks for our iterative tail change construction. Intuitively, it tells us, that, given a consistent family $\mathcal{F}$ of tail changes, we can construct a new consistent family of tail changes removing all of the vertices contained in tail changes in $\mathcal{F}$, as well as all of $B_1$.
\begin{theorem} Let $(G,w,A,B)$ be a normalized instance and let $\epsilon\in (0,1)$ and \mbox{$N\in\mathbb{N}$} be constants such that there is no local improvement of $A$ of size at most $6\cdot N+2$.
	
	Let further $(U_i,V_i,e_i)_{i\in\mathcal{I}}$ be a  consistent family of tail changes, each of size at most $N$.
Let $B_1$ be the set of vertices $v\in B''\setminus \bigcup_{i\in\mathcal{I}} V_i$ such that $|\delta(v)\setminus \{e_i,i\in \mathcal{I}\}|\leq 1$.
	Then there exists a consistent family $(U_k,V_k,e_k)_{k\in\mathcal{K}}$ of tail changes, each of size at most $2\cdot N+1$, with the following properties:
\begin{enumerate}[(a)]
	\item \label{NewTailChangesProp1} For every $i\in \mathcal{I}$, there exists $k\in\mathcal{K}$ with $U_i\subseteq U_k$ and $V_i\subseteq V_k$ and
	\begin{itemize}
		\item $(U_i,V_i,e_i)=(U_k,V_k,e_k)$, or
		\item $e_i\subseteq U_k\cup V_k$.
	\end{itemize} 
\item \label{NewTailChangesProp3}For every $b\in B_1$, there is $k\in\mathcal{K}$ with $b\in V_k$.
\item \label{NewTailChangesProp4} For every $u\in \bigcup_{k\in\mathcal{K}} U_k\setminus \bigcup_{i\in \mathcal{I}} U_i$, we have $u\in A''$ and $|\delta(u)\setminus (\delta(\bigcup_{k\in\mathcal{K}} V_k)\cup\{e_k,k\in\mathcal{K}\})|\leq 1$.		
\end{enumerate}
	 \label{TheoNewFamilyOfTailChanges}
\end{theorem}
The remainder of this section is dedicated to the proof of Theorem~\ref{TheoNewFamilyOfTailChanges}.

To this end, fix a normalized instance $(G,w,A,B)$, constants $\epsilon$ and $N$ and a family $(U_i,V_i,e_i)_{i\in\mathcal{I}}$ as stated in Theorem~\ref{TheoNewFamilyOfTailChanges}. Further assume that there is no local improvement of size at most $6\cdot N+2$.

We define $E_t\coloneqq\{e_i,i\in \mathcal{I}\}$ to be the set of edges associated with the tail changes indexed by $\mathcal{I}$ and let $A_t\coloneqq\bigcup_{i\in \mathcal{I}} U_i$, $A_r\coloneqq A\setminus A_t$, $B_t\coloneqq\bigcup_{i\in \mathcal{I}} V_i$ and $B_r\coloneqq B\setminus B_t$. Define $B_1$ as in Theorem~\ref{TheoNewFamilyOfTailChanges} and let $A_1\coloneqq N(B_1,A_r)$.

 We first prove the following very useful proposition, that tells us that we can translate local improvements in $G[A_r\cup B_r]$ into local improvements in $G$.
\begin{proposition}
	Let $X\subseteq B_r$ such that
	\begin{itemize}
		\item $w(X)> w(N(X,A_r))$, or
		\item $w(X)=w(N(X,A_r))$ and $|X\cap B''|>|N(X,A_r\cap A'')|$.
	\end{itemize}   Then there exists a local improvement of $A$ of size at most $|X|\cdot (3\cdot N +1)$.\label{PropLocalImprovement}
\end{proposition}
\begin{proof}
	Let $I\coloneqq\{i\in\mathcal{I}: N(X,U_i)\neq \emptyset\}$. As in a normalized instance, every vertex has degree at most $3$, and the sets $(U_i)_{i\in \mathcal{I}}$ are pairwise disjoint, we know that $|I|\leq 3\cdot |X|$. Define $Y\coloneqq X\cup\bigcup_{i\in I} V_i$. Then $|Y|\leq |X|\cdot (3\cdot N +1)$. We further have $N(Y,A)\setminus N(X,A_r)\subseteq \bigcup_{i\in I} U_i$ and, thus, \[w(Y\setminus X) = \sum_{i\in I} w(V_i)=\sum_{i\in I} w(U_i)\geq w(N(Y,A)\setminus N(X,A_r)).\] Moreover, \[|(Y\setminus X)\cap B''| = \sum_{i\in I} |V_i\cap B''| = \sum_{i\in I} |U_i\cap A''|\geq |(N(Y,A)\setminus N(X,A_r))\cap A''|.\] Hence, our assumptions on $X$ imply that $Y$ constitutes a local improvement.
\end{proof}
\begin{lemma}We have
	$|A_1|=|B_1|$ and $A_1\subseteq A''$. Moreover, each $u\in A_1$ has exactly one neighbor in $B_1$ and every $v\in B_1$ has exactly one neighbor in $A_1$. In particular, $E[A_1,B_1]$ is a matching.  \label{LemUniqueNeighbor}
\end{lemma}
\begin{proof}By definition of $A_1$ and $B_1$, every vertex in $A_1$ has at least one neighbor in $B_1$ and every vertex in $B_1$ has at most one neighbor in $A_1$. Thus, it suffices to verify that
	\begin{enumerate}
		\item Every vertex in $A_1$ has weight $2$.
		\item Every vertex in $B_1$ has at least one neighbor in $A_1$.
		\item Every vertex in $A_1$ has at most one neighbor in $B_1$.
	\end{enumerate}
	\begin{claim}
		If one of the above three conditions fails, then there exist $U\subseteq A_1$ and $V\subseteq B_1$ with the properties that
		\begin{enumerate}
			\item[(a)] $|U|\leq |V|\leq 2$,
			\item[(b)] $N(V,A_r)\subseteq U$ and
			\item[(c)] $w(U)<w(V)$.
		\end{enumerate}
	\end{claim}
Note that combining the claim with Prop.~\ref{PropLocalImprovement} yields a local improvement of size at most $6\cdot N+2$, a contradiction. To prove the claim, we explain how to define $U$ and $V$ in each of the three cases.
		\begin{enumerate}
			\item Pick a vertex $u\in A_1$ of weight $1$ and let $v\in N(u,B_1)$, which exists by definition of $A_1$. Let $U\coloneqq\{u\}$ and $V\coloneqq\{v\}$. Note that $v\in B''$ by definition of $B_1$, so $w(u)< w(v)$. Besides, $N(v,A_r)=\{u\}$  since $v$ has at most one neighbor in $A_r\ni u$. 
			\item Let $v\in B_1$ with $N(v,A_1)=\emptyset$. By definition of $A_1$, $N(v,A_r)=\emptyset$. Pick $U\coloneqq\emptyset$ and $V\coloneqq\{v\}$.
			\item Assume that $u\in A_1$ has two neighbors $v_1,v_2\in B_1$. Let $U\coloneqq\{u\}$ and $V\coloneqq\{v_1,v_2\}$.  As $B_1\subseteq B''$, we have $w(u)\leq 2<2+2=w(v_1)+w(v_2)$. Moreover, $N(\{v_1,v_2\},A_r)=\{u\}$  since $v_1$ and $v_2$ both have at most one neighbor in $A_r$.
		\end{enumerate}
\end{proof}
\begin{lemma}
Let $v\in B_1$. Then $E(\{v\},A_t)\subseteq E_t$.\label{LemSubsetOfEt}
\end{lemma}
\begin{proof}
By definition of $B_1$, we have $|\delta(v)\setminus E_t|\leq 1$. Moreover, Lemma~\ref{LemUniqueNeighbor} tells us that there is an edge connecting $v$ to a vertex in $A_1\subseteq A_r$. This edge cannot be contained in $E_t$ since all of these edges are incident to $A_t$. Thus, $E(\{v\},A_t)\subseteq E_t$.
\end{proof}
\begin{definition} For $v\in B_1$, denote the unique neighbor of $v$ in $A_1$ by $m(v)$.
	
	 Let $I_v\coloneqq\{i\in\mathcal{I}:e_i\in\delta(v)\}$ and define $U_v\coloneqq\bigcup_{i\in I_v} U_i\cup\{m(v)\}$ and $V_v\coloneqq\bigcup_{i\in I_v} V_i\cup\{v\}$. Moreover, if $E(\{m(v)\}, B_r\setminus B_1)\neq\emptyset$, pick $e_v\in E(\{m(v)\}, B_r\setminus B_1)$. Otherwise, let $e_v\coloneqq\{m(v)\}$ be a loop on $m(v)$.
 \end{definition}
\begin{lemma}
For $v\in B_1$, $(U_v,V_v,e_v)$ is a tail change of size at most $2\cdot N+1$.\label{LemTailChange}
\end{lemma}
\begin{proof}
As $v$ can have at most $3$ neighbors in $A$ and one of them is $m(v)\in A_r$, we have $|I_v|\leq 2$.
As $m(v)\in A_r\cap A''$ by Lemma~\ref{LemUniqueNeighbor} and $v\in B_r\cap B''$ by definition, we obtain
\[|U_v|=1+\sum_{i\in I_v} |U_i| = 1+\sum_{i\in I_v} |V_i| = |V_v|\leq 2\cdot N +1\] and \[w(U_v)=w(m(v))+\sum_{i\in I_v} w(U_i)=w(v)+\sum_{i\in I_v} w(V_i)=w(V_v)\] since $(U_i,V_i,e_i)_{i\in\mathcal{I}}$ is a consistent family of tail changes, each of size at most $N$. We have $e_v\in\delta(U_v)$ by definition. Hence, it remains to check that $N(V_v,A)\subseteq U_v$. For $w\in V_i$ with $i\in I_v$, we have $N(w,A)\subseteq U_i\subseteq U_v$ since $(U_i,V_i,e_i)_{i\in\mathcal{I}}$ is a consistent family of tail changes. By Lemma~\ref{LemSubsetOfEt}, we know that $E(\{v\},A_t)\subseteq \{e_i,i\in I_v\}$ and, thus, $N(v,A_t)\subseteq U_v$. By definition of $A_1$ and Lemma~\ref{LemUniqueNeighbor}, we further have $N(v,A_r)=\{m(v)\}\subseteq U_v$. This concludes the proof.
\end{proof}
Let $\mathcal{I'}\coloneqq\mathcal{I}\setminus\bigcup_{v\in B_1} I_v$ and let $\mathcal{K}\coloneqq B_1\cup\mathcal{I'}$.
\begin{lemma}
$(U_k,V_k,e_k)_{k\in\mathcal{K}}$ is a consistent family of tail changes, each of size at most $2\cdot N+1$.\label{LemFamilyTailChanges}
\end{lemma}
\begin{proof}
By Lemma~\ref{LemTailChange}, $(U_k,V_k,e_k)_{k\in\mathcal{K}}$ is a family of tail changes, each of size at most $2\cdot N+1$. Hence, it remains to verify consistency. As $(U_i,V_i,e_i)_{i\in\mathcal{I}}$ is consistent, we have $U_i\cap U_j=\emptyset$ and $V_i\cap V_j=\emptyset$ for $i,j\in \mathcal{I'}$ with $i\neq j$. 

For $i\in\mathcal{I'}$ and $v\in B_1$, we have $i\not\in I_v$, so $U_i\cap U_j=\emptyset$ and $V_i\cap V_j=\emptyset$ for all $j\in I_v$. As $U_i\subseteq A_t$ and $V_i\subseteq B_t$, but $m(v)\in A_r$ and $v\in B_r$, we get $U_i\cap U_v=\emptyset$ and $V_i\cap V_v=\emptyset$. 

Finally, let $v,w\in B_1$ with $v\neq w$. Then $I_v\cap I_w=\emptyset$ since each of the edges $e_l,l\in\mathcal{I}$ has at most one endpoint in $B$. Thus, $\bigcup_{i\in I_v} U_i\cap\bigcup_{i\in I_w} U_i =\emptyset$ and $\bigcup_{i\in I_v} V_i\cap\bigcup_{i\in I_w} V_i=\emptyset$. By Lemma~\ref{LemUniqueNeighbor}, we have $m(v)\neq m(w)$. As $\bigcup_{i\in\mathcal{I}} U_i\cup V_i\subseteq A_t\cup B_t$, but $\{m(v),m(w),v,w\}\subseteq A_r\cup B_r$, we obtain $U_v\cap U_w=\emptyset$ and $V_v\cap V_w=\emptyset$.
\end{proof}
\begin{lemma}
For every $i\in\mathcal{I}$, there is $k\in\mathcal{K}$ with $U_i\subseteq U_k$ and $V_i\subseteq V_k$ and
\begin{itemize}
	\item $(U_i,V_i,e_i)=(U_k,V_k,e_k)$, or
	\item $e_i\subseteq U_k\cup V_k$.
\end{itemize}\label{LemOldTailChanges}
\end{lemma}
\begin{proof}
If $i\in\mathcal{I'}$, then $i\in\mathcal{K}$ and the claim follows. Otherwise, there is $v\in B_1$ such that $i\in I_v$. Then $U_i\subseteq U_v$, $V_i\subseteq V_v$ and $e_i\subseteq U_i\cup \{v\}\subseteq U_v\cup V_v$.
\end{proof}
\begin{proposition}
For every $v\in B_1$, there is $k\in\mathcal{K}$ with $v\in V_k$.\label{PropVinVV}
\end{proposition}
\begin{proof}
We have $v\in V_v$.
\end{proof}
\begin{lemma}
For every $u\in\bigcup_{k\in\mathcal{K}} U_k\setminus\bigcup_{i\in\mathcal{I}} U_i$, we have $|\delta(u)\setminus (\delta(\bigcup_{k\in\mathcal{K}} V_k)\cup\{e_k,k\in\mathcal{K}\})|\leq 1$ and $u\in A''$.\label{LemRemainingDegree}
\end{lemma}
\begin{proof}
Let $u\in\bigcup_{k\in\mathcal{K}} U_k\setminus\bigcup_{i\in\mathcal{I}} U_i$. Then there is $v\in B_1$ such that $u=m(v)$. By Lemma~\ref{LemUniqueNeighbor}, $u\in A''$. If $E(\{u\},B_r\setminus B_1)=\emptyset$, then $\delta(u)\subseteq \delta(\bigcup_{k\in\mathcal{K} }V_k)$ because $\bigcup_{k\in\mathcal{K}} V_k = \bigcup_{i\in\mathcal{I}} V_i\cup B_1$. Otherwise, we know that $\{u,v\}\in \delta(u)\cap \delta(\bigcup_{k\in\mathcal{K} }V_k)$, and $\{u,v\}\neq e_v\in \{e_k,k\in\mathcal{K}\}$. The fact that $|\delta(u)|\leq 3$ since the instance is normalized concludes the proof.
\end{proof}
Combining Lemma~\ref{LemFamilyTailChanges}, Lemma~\ref{LemOldTailChanges}, Proposition~\ref{PropVinVV} and Lemma~\ref{LemRemainingDegree} proves Theorem~\ref{TheoNewFamilyOfTailChanges}.
\section{Handling critical components via iterative tail change construction\label{SecCritical}}
In this section, we first introduce the notion of a critical component. Then, we state our main theorem for this section, telling us that we can construct a family of consistent tail changes removing all current critical components.
\subsection{The auxiliary graph and critical components}
Let again $(G,w,A,B)$ be normalized and let $(U_i,V_i,e_i)_{i\in\mathcal{I}}$ be a consistent family of tail changes, each of size at most $N$. As in the previous section, we define $A_t\coloneqq\bigcup_{i\in\mathcal{I}} U_i$, $A_r\coloneqq A\setminus A_t$, $B_t\coloneqq\bigcup_{i\in\mathcal{I}} V_i$ and $B_r\coloneqq B\setminus B_t$.

 We further define special subsets of $B_r$ that we need in order to construct an auxiliary graph in which we will search for local improvements.
\begin{definition}
	We define the following subsets of $ B_r$: 
	\begin{description}
		\item[$B'_1$] consists of all $v\in B'\cap B_r$ such that $E(\{v\},A_t)\subseteq \{e_i,i\in\mathcal{I}\}$ and such that $v$ has exactly one neighbor in $A_r$. Note that by the definition of normalized instances, this neighbor is contained in $A_r\cap A''$.
		\item[$B^{2,1}$] consists of all $v\in B''\cap B_r$ such that $E(\{v\},A_t)\subseteq \{e_i,i\in \mathcal{I}\}$ and such that $v$ has exactly one neighbor $n_1(v)$ in $A'\cap A_r$ and exactly one neighbor $n_2(v)$ in $A''\cap A_r$.
		\item[$B^{2,2}$] consists of all $v\in B''\cap B_r$ such that $E(\{v\},A_t)\subseteq \{e_i,i\in \mathcal{I}\}$ and such that $v$ has exactly two neighbors in $A''\cap A_r$ and no neighbor in $A'\cap A_r$.
		\item[$B^1_0$] consists of all $v\in B''\cap B_r$ such that $E(\{v\},A_t)\subseteq \{e_i,i\in \mathcal{I}\}$ and such that $v$ has exactly one neighbor in $A''\cap A_r$ and no neighbor in $A'\cap A_r$.
		\item[$B^1_1$] consists of all $v\in B''\cap B_r$ such that $|E(\{v\},A_t)\setminus \{e_i,i\in \mathcal{I}\}|= 1$ and such that $v$ has exactly one neighbor in $A''\cap A_r$ and no neighbor in $A'\cap A_r$.
	\end{description}\label{DefSubsetsB}
\end{definition}

Similar as in~\cite{FurerYu}, we define an auxiliary graph $G^{\mathcal{I}}_A$ on the vertex set $A''\cap A_r$. See Figure~\ref{FigAuxiliaryGraph} for an illustration.

\begin{definition} The auxiliary graph $G^{\mathcal{I}}_A$ with vertex set $A''\cap A_r$ is defined as follows:
	\begin{itemize}
		\item Each vertex from $B^1_1$ induces a loop on its neighbor in $A''\cap A_r$.
		\item Each vertex from $B^{2,2}$ induces an edge between its two neighbors in $A''\cap A_r$.
	\end{itemize}
	In addition, we define a set $F$ of virtual edges, which we call \emph{dashed} or \emph{dotted}. They are not part of the edge set of $G^{\mathcal{I}}_A$.
	\begin{itemize}
		\item Every vertex from $B'_1$ induces a dotted loop on its neighbor in $A''\cap A_r$.
		\item Every vertex from $B^{2,1}$ induces a dashed loop on its neighbor in $A''\cap A_r$.
	\end{itemize}
\label{DefGA}
\end{definition}
\begin{proposition}
Let $v\in B^{2,2}\cup B^{2,1}\cup B^1_1\cup B'_1$ induce an edge $e\in E(G^{\mathcal{I}}_A)\cup F$.\\ Then $w(v)> w(N(v,A_r\setminus e))$. \label{PropAddingEdgeHelps}
\end{proposition}
\begin{proof}
If $v\in B^{2,2}\cup B^1_1\cup B'_1$, then $N(v,A_r)\subseteq e$ and $w(v)>0$ yields the desired statement. If $v\in B^{2,1}$, then $N(v,A_r\setminus e)$ consists of one vertex of weight $1$, whereas $v$ has weight $2$.
\end{proof}
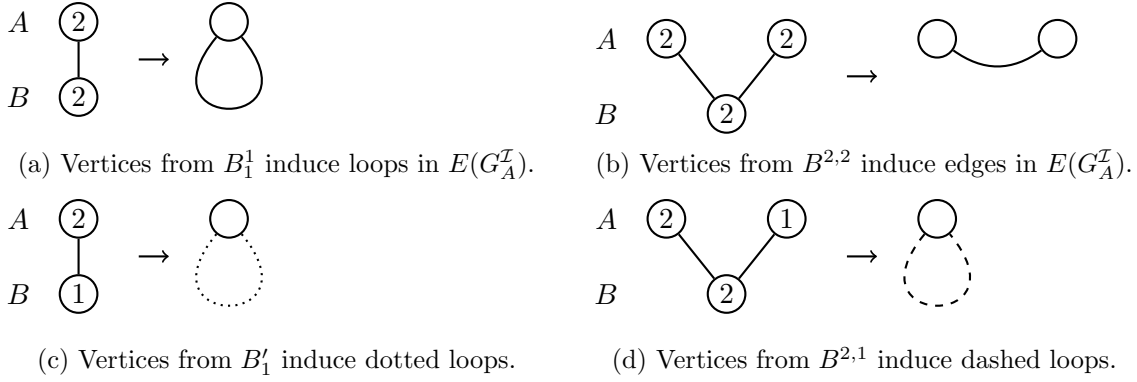
\begin{figure}
\begin{subfigure}[t]{0.45\textwidth}
\begin{tikzpicture}[mynode/.style = {circle, draw = black, thick, fill = none, inner sep = 0mm, minimum size = 5mm}, loop/.style={min distance=10mm,in=-45,out=-135,looseness=10},xscale = 0.8]
\node[mynode] (A) at (0,1) {$2$};
\node[mynode] (B) at (0,0) {$2$};
\node at (-1,1) {$A$};
\node at (-1,0) {$B$};
\draw[thick] (A)--(B);
\draw[thick,->] (1,0.5)--(1.5,0.5);
\node[mynode] (U) at (2.5,1){};
\draw[thick] (U) to[loop] (U);
\end{tikzpicture}
\subcaption{Vertices from $B_1^1$ induce loops in $E(G^{\mathcal{I}}_A)$.}
\end{subfigure}\quad
\begin{subfigure}[t]{0.45\textwidth}
	\begin{tikzpicture}[mynode/.style = {circle, draw = black, thick, fill = none, inner sep = 0mm, minimum size = 5mm}, loop/.style={min distance=10mm,in=-45,out=-135,looseness=10},xscale = 0.8]
	\node[mynode] (A) at (0,1) {$2$};
	\node[mynode] (B) at (1,0) {$2$};
	\node[mynode] (A2) at (2,1) {$2$};
	\node at (-1,1) {$A$};
	\node at (-1,0) {$B$};
	\draw[thick] (A)--(B)--(A2);
	\draw[thick,->] (3,0.5)--(3.5,0.5);
	\node[mynode] (U) at (4.5,1){};
	\node[mynode] (U2) at (6.5,1){};
	\draw[thick] (U) to[bend right = 30] (U2);
	\end{tikzpicture}
	\subcaption{Vertices from $B^{2,2}$ induce edges in $E(G^{\mathcal{I}}_A)$.}
\end{subfigure}
\vspace{0.2cm}

\begin{subfigure}[t]{0.45\textwidth}
	\begin{tikzpicture}[mynode/.style = {circle, draw = black, thick, fill = none, inner sep = 0mm, minimum size = 5mm}, loop/.style={min distance=10mm,in=-45,out=-135,looseness=10},xscale = 0.8]
	\node[mynode] (A) at (0,1) {$2$};
	\node[mynode] (B) at (0,0) {$1$};
	\node at (-1,1) {$A$};
	\node at (-1,0) {$B$};
	\draw[thick] (A)--(B);
	\draw[thick,->] (1,0.5)--(1.5,0.5);
	\node[mynode] (U) at (2.5,1){};
	\draw[thick,dotted] (U) to[loop] (U);
	\end{tikzpicture}
	\subcaption{Vertices from $B'_1$ induce dotted loops.}
\end{subfigure}\quad
\begin{subfigure}[t]{0.45\textwidth}
	\begin{tikzpicture}[mynode/.style = {circle, draw = black, thick, fill = none, inner sep = 0mm, minimum size = 5mm}, loop/.style={min distance=10mm,in=-45,out=-135,looseness=10},xscale = 0.8]
	\node[mynode] (A) at (0,1) {$2$};
	\node[mynode] (B) at (1,0) {$2$};
	\node[mynode] (A2) at (2,1) {$1$};
	\node at (-1,1) {$A$};
	\node at (-1,0) {$B$};
	\draw[thick] (A)--(B)--(A2);
	\draw[thick,->] (3,0.5)--(3.5,0.5);
	\node[mynode] (U) at (4.5,1){};
	\draw[thick,dashed] (U) to[loop] (U);
	\end{tikzpicture}
	\subcaption{Vertices from $B^{2,1}$ induce dashed loops.}
\end{subfigure}
\caption{Construction of $G^{\mathcal{I}}_A$ and $F$ (see Definition~\ref{DefGA}).\label{FigAuxiliaryGraph}}
\end{figure}
We are now ready to introduce the notion of a critical component. See Figure~\ref{FigCritical2} for an illustration.
\begin{definition}
	Let $\alpha$ be a constant.
	We call a connected component of $G^{\mathcal{I}}_A$ \emph{$\alpha$-small} if it contains less than $\alpha$ vertices, and \emph{$\alpha$-large} otherwise.

	We call an $\alpha$-small component $C$ \emph{$\alpha$-critical} if it is a tree with exactly one dotted loop (induced by $v_1^C$) and one dashed loop (induced by $v_2^C$) attached to it, and additionally, $n_1(v_2^C)$ has a neighbor in $B_r$ other than $v_2^C$. Call this neighbor $w_C$.\label{DefCritical}
 \end{definition}
Note that as a neighbor of the vertex $n_1(v_2^C)$ of weight $1$, $w_C$ must be of weight $2$ by definition of a normalized instance.
\begin{figure}
\begin{subfigure}[t]{0.5\textwidth}
\begin{tikzpicture}[mynode/.style = {circle, draw = black, thick, fill = none, inner sep = 0mm, minimum size = 5mm}, loop/.style={min distance=10mm,in=70,out=160,looseness=10}]
\node[mynode] (A) at (-1,1) {};
\node[mynode,label= below :{$n_2(v_2^C)$}] (B) at (-1,-1) {};
\node[mynode] (C) at (0,0) {};
\node[mynode] (D) at (2,0) {};
\node[mynode] (E) at (3,1) {};
\node[mynode] (F) at (3,-1) {};
\draw[thick] (A)--(C);
\draw[thick, red!70!black] (B)--(C);
\draw[thick, red!70!black] (D)--(C);
\draw[thick] (D)--(E);
\draw[thick] (D)--(F);
\draw[thick, dotted, red!70!black] (D) to [loop](D);
\draw[thick, dashed, red!70!black] (B) to [loop](B);
\end{tikzpicture}
\subcaption{An $\alpha$-critical componenent $C$ in $G_A^{\mathcal{I}}$. The dashed and the dotted loop attached to it are drawn in the respective style and in red. The edges of the path $P_C$ (see Def.~\ref{DefCriticalTailChange}) are drawn in red as well.}
\end{subfigure}\quad
\begin{subfigure}[t]{0.5\textwidth}
\begin{tikzpicture}[mynode/.style = {circle, draw = black, thick, fill = none, inner sep = 0mm, minimum size = 5mm}, loop/.style={min distance=10mm,in=70,out=160,looseness=10}]
\node[mynode,label= above :{$n_2(v_2^C)$}] (N2) at (0,1) {$2$};
\node[mynode,label= below :{$v_2^C$}] (V2) at (-1,0) {$2$};
\node[mynode,label= above :{$n_1(v_2^C)$}] (N1) at (-2,1) {$1$};
\node[mynode,label= below :{$w_C$}] (X) at (-3,0) {$2$};
\draw[thick] (N2)--(V2)--(N1)--(X);
\end{tikzpicture}
\subcaption{The situation at $v_2^C$.}
\end{subfigure}
\begin{subfigure}[t]{\textwidth}
	\begin{tikzpicture}[mynode/.style = {circle, draw = black, thick, fill = none, inner sep = 0mm, minimum size = 5mm}, loop/.style={min distance=10mm,in=70,out=160,looseness=10}]
	\node[mynode,label= below :{$v_1^C$}] (V1) at (5,0) {$1$};
	\node[mynode] (U2) at (4,1){$2$};
	\node[mynode] (E2) at (3,0){$2$};
	\node[mynode] (U1) at (2,1){$2$};
	\node[mynode] (E1) at (1,0){$2$};
	\node[mynode,label= above :{$n_2(v_2^C)$}] (N2) at (0,1) {$2$};
	\node[mynode,label= below :{$v_2^C$}] (V2) at (-1,0) {$2$};
	\node[mynode,label= above :{$n_1(v_2^C)$}] (N1) at (-2,1) {$1$};
	\node[mynode,label= below :{$w_C$}] (X) at (-3,0) {$2$};
	\draw[thick] (V1)--(U2)--(E2)--(U1)--(E1)--(N2)--(V2)--(N1);
	\draw[thick,blue] (N1) to node[midway,below=4pt, right=1pt] {$\color{blue}e_C$} (X);
	\draw[rounded corners, red!70!black] (-1.4,-0.3) rectangle (5.4,0.3);
	\node at (6,0){$\color{red!70!black}W'_C$};
		\draw[rounded corners, green!70!black] (-2.4,0.7) rectangle (4.4,1.3);
	\node at (5,1){$\color{green!70!black}U'_C$};
	\end{tikzpicture}
	\subcaption{Illustration of Definition~\ref{DefCriticalTailChange}. The set $U'_C$ is marked in green, $W'_C$ is indicated in red. The edge $e_C$ is drawn in blue.}
\end{subfigure}
\caption{Illustration of Definition~\ref{DefCritical} and Definition~\ref{DefCriticalTailChange}.\label{FigCritical2}.}
\end{figure}
\subsection{A nice family of tail changes}
Our main result for this section is given by the following theorem:
\begin{theorem}
Let $\delta\in (0,1)$ be a constant. Then there exist constants $N_\delta$ and $\pi_\delta$ with the following properties: Let $(G,w,A,B)$ be a normalized instance such that there is no local improvement of size at most $\pi_\delta$. Then there exists a consistent family $(U_i,V_i,e_i)_{i\in\mathcal{I}}$ of tail changes, each of size at most $N_\delta$, such that:
\begin{enumerate}[(i)]
	\item \label{NiceTailChanges1} For every $u\in \bigcup_{i\in\mathcal{I}} U_i$, we have $|\delta(u)\setminus (\delta(\bigcup_{i\in\mathcal{I}} V_i)\cup\{e_i,i\in\mathcal{I}\})|\leq 1$. If $u\in A'\cap\bigcup_{i\in\mathcal{I}} U_i$, then $\delta(u)\subseteq \delta(\bigcup_{i\in\mathcal{I}} V_i)\cup\{e_i,i\in\mathcal{I}\}$.
	\item  \label{NiceTailChanges2} The number of $\delta^{-1}$-critical components in $G_A^\mathcal{I}$ is bounded by $\delta\cdot |B|$.
	\item \label{NiceTailChanges3} $|B^1_0|\leq \delta\cdot |B|$, where $B^1_0$ is defined according to Def.~\ref{DefSubsetsB}.
\end{enumerate}\label{TheoNiceFamilyofTailChanges}
\end{theorem}

The remainder of this section is dedicated to the proof of Theorem~\ref{TheoNiceFamilyofTailChanges}. We first show how to obtain a consistent family of tail changes that remove all critical components that are currently present. However, this might result in new vertices entering the set $B_0^1$ and new critical components may arise. Thus, as long as the number of vertices in $B_0^1$ is at least $\delta\cdot |B|$, or there are least $\delta\cdot|B|$ many critical components, we first apply Theorem~\ref{TheoNewFamilyOfTailChanges} to remove the vertices in $B_0^1$, and then obtain a new family of consistent tail changes that removes all critical components. In each of these iterations, the cardinality of $B_t$ will increase by at least $\delta\cdot |B|$. In particular, the total number of iterations will be bounded by $\delta^{-1}$, which is a constant. Thus, we also obtain a constant bound (depending on $\delta$, of course) on the sizes of the tail changes we construct. When the algorithm terminates, the requirements of Theorem~\ref{TheoNiceFamilyofTailChanges} will be met.
\subsection{Obtaining consistent tail changes that remove critical components}
In this section, we show how to obtain a consistent family of tail changes that remove all critical components. We consider the following scenario, which we denote by ($*$) so that we do not have to write the same set of assumptions for each of the following lemmata:
Let $(G,w,A,B)$ be a normalized instance, let constants $\alpha$ and $N$ be given an let $(U_i,V_i,e_i)_{i\in\mathcal{I}}$ be a consistent family of tail changes, each of size at most $N$. Assume that there is no local improvement of size at most $2\cdot(\alpha+1)\cdot (N+1)$. 

\begin{definition}
	Let $C$ be an $\alpha$-critical component. Let $U'_C$ consist of the vertices of $C$ that lie on the path $P_C$ from $n_2(v_2^C)$ to the neighbor of $v_1^C$, and the vertex $n_1(v_2^C)$. Let $W'_C$ be the set of vertices from $B_r$ that induce an edge of $P_C$ or the dotted or dashed loop attached to it (see Figure~\ref{FigCritical2}).
	
	 Let $I_C\coloneqq\{i\in \mathcal{I}:e_i\in \delta(W'_C)\}$ and define $U_C\coloneqq U'_C\cup\bigcup_{i\in\mathcal{I}} U_i$ and $V_C\coloneqq W'_C\cup\bigcup_{i\in\mathcal{I}} V_i$. 
	
	The \emph{tail change corresponding to $C$} is $(U_C,V_C,e_C)$, where $e_C\coloneqq\{n_1(v_2^C),w_C\}$. \label{DefCriticalTailChange}
\end{definition}
\begin{lemma}
	Assume ($*$) and let $C$ be an $\alpha$-critical component. Then $(U_C,V_C,e_C)$ is a tail change of size at most $(\alpha+1)\cdot (N+1)$.\label{LemCriticalTailChange}
\end{lemma}
\begin{proof}
	First of all, we have $U_C\subseteq A$ and $V_C\subseteq B$. By definition, $U'_C$ consists of $|V(P_C)|$ vertices of weight $2$ and one vertex of weight $1$. As $P_C$ is a path, $W'_C$ consists of $|V(P_C)|-1$ vertices of weight $2$ inducing the edges of $C$, the vertex $v_1^C$ of weight $1$ inducing the dotted loop, and the vertex $v_2^C$ of weight $2$ inducing the dashed loop. Thus, $|U'_C|=|W'_C|$ and $w(U'_C)=w(W'_C)$. The fact that $(U_i,V_i,e_i)_{i\in\mathcal{I}}$ is a consistent family of tail changes implies $|U_C|=|V_C|$ and $w(U_C)=w(V_C)$. We have $|W'
	_C|= |V(P_C)|+1\leq \alpha+1$. Moreover, every vertex from $B'$ has at most $2$ neighbors in $A$, every vertex from $B''$ has at most $3$ neighbors in $A$, one neighbor of $v_1^C$ is contained in $A_r$, and two neighbors of each vertex in $W'_C\setminus \{v_1^C\}$ are contained in $A_r$. Hence, $|I_C|\leq \alpha+1$. This yields $|V_C|\leq (\alpha+1)\cdot (N+1)$.  Moreover, $e_C$ is incident to $U_C$. Finally, by construction of $G^{\mathcal{I}}_A$ and $F$, we have $N(W'_C,A_r)\subseteq U'_C$. Moreover, all incident edges of the vertices in $W'_C$ that go to $A_t$ are of the form $e_i$ with $i\in\mathcal{I}$ because this holds for all vertices in $B'_1\cup B^{2,1}\cup B^{2,2}$. Thus, $N(W'_C,A\setminus A_r)\subseteq \bigcup_{i\in I_C} U_i\subseteq U_C$. As $N(V_i,A)\subseteq U_i$ for all $i\in I_C$, we can conclude that $N(V_C,A)\subseteq U_C$.
\end{proof}
We denote the set of $\alpha$-critical components by $\mathcal{C}$. We show that we can obtain a consistent family of tail changes incorporating the tail changes corresponding to the $\alpha$-critical components (see Lemma~\ref{LemConsistentFamily}).
\begin{lemma}
	Assume ($*$) and let $\mathcal{K}\coloneqq\mathcal{C}\cup\mathcal{I}\setminus\bigcup_{C\in\mathcal{C}} I_C$. Then $(U_k,V_k,e_k)_{k\in\mathcal{K}}$ is a consistent family of tail changes, each of size at most $(\alpha+1)\cdot (N+1)$, with the following properties:
	\begin{enumerate}[(i)]
		\item \label{CriticalFamilyProp1} For every $i\in\mathcal{I}$, there is $k\in\mathcal{K}$ such that 
		\begin{itemize}
			\item $(U_i,V_i,e_i)=(U_k,V_k,e_k)$, or
			\item $U_i\subseteq U_k$, $V_i\subseteq V_k$ and $e_i\subseteq U_k\cup V_k$.
		\end{itemize}
		\item \label{CriticalFamilyProp2}For every $u\in \bigcup_{k\in\mathcal{K}} U_k\setminus\bigcup_{i\in\mathcal{I}} U_i$, we have $|\delta(u)\setminus(\delta(\bigcup_{k\in\mathcal{K}} V_k)\cup\{e_k,k\in\mathcal{K}\})|\leq 1$. Moreover, if $u\in A'\cap \bigcup_{k\in\mathcal{K}} U_k\setminus\bigcup_{i\in\mathcal{I}} U_i$, then $\delta(u)\subseteq \delta(\bigcup_{k\in\mathcal{K}} V_k)\cup\{e_k,k\in\mathcal{K}\}$.
	\end{enumerate}
	
	\label{LemConsistentFamily}
\end{lemma}
\begin{proof}
	By ($*$), Lemma~\ref{LemCriticalTailChange} and $\alpha\geq 1$, $(U_k,V_k,e_k)_{k\in\mathcal{K}}$ is a family of tail changes meeting the required size bound.
	\begin{claim}
	$(U_k,V_k,e_k)_{k\in\mathcal{K}}$ is consistent.
	\end{claim}
\begin{proof}
	As $(U_i,V_i,e_i)_{i\in \mathcal{I}}$ is consistent, it suffices to show the following two statements: \begin{itemize}
		\item For $i\in \mathcal{I}\setminus\bigcup_{C\in\mathcal{C}} I_C$ and $C\in\mathcal{C}$, we have $U_C\cap U_i=\emptyset$ and $V_C\cap V_i=\emptyset$.
		\item For two distinct $\alpha$-critical components $C$ and $C'$, we have $U_C\cap U_{C'}=\emptyset$ and $V_C\cap V_{C'}=\emptyset$.
	\end{itemize}
	For the first item, pick $i\in \mathcal{I}\setminus\bigcup_{C\in\mathcal{C}} I_C$ and $C\in\mathcal{C}$. Then $U_i\subseteq A_t$, $U'_C\subseteq A_r=A\setminus A_t$, $V_i\subseteq B_t$ and $W'_C\subseteq B_r=B\setminus B_t$, so $U_i\cap U'_C=\emptyset$ and $V_i\cap W'_C=\emptyset$. Let $j\in I_C$. Then $j\neq i$, so $U_i\cap U_j=\emptyset$ and $V_i\cap V_j=\emptyset$ by consistency of $(U_l,V_l,e_l)_{l\in\mathcal{I}}$. Hence, $U_i\cap U_C=\emptyset$ and $V_i\cap V_C=\emptyset$.
	
	For the second item, let $C$ and $C'$ be two distinct $\alpha$-critical components. Then the vertex sets $V(P_C)\subseteq V(C)$ and $V(P_{C'})\subseteq V(C')$ are disjoint. As every vertex from $B_r$ induces at most one edge in $E(G^{\mathcal{I}}_A)\cup F$, and every vertex from $W'_C$ induces an edge incident to $V(C)$, whereas every vertex from $W'_{C'}$ induces an edge incident to $V(C')$, we must have $W'_C\cap W'_{C'}=\emptyset$. As every edge $e_i$ with $i\in\mathcal{I}$ has at most one endpoint in $B_r$, it can only be incident to one of the two sets $W'_C$ or $W'_{C'}$, implying $I_C\cap I_{C'}=\emptyset$. Thus, consistency of $(U_l,V_l,e_l)_{l\in\mathcal{I}}$ tells us that $V_i\cap V_j=\emptyset$ for all $i\in I_C$ and $j\in I_{C'}$. Finally, as $\bigcup_{i\in \mathcal{I}} V_i\subseteq B_t$, but $W'_C\cup W'_{C'}\subseteq B_r=B\setminus B_t$, we obtain \[V_C\cap V_{C'}=\left(W'_C\cup\bigcup_{i\in I_C} V_i\right)\cap \left(W'_{C'}\cup\bigcup_{i\in I_{C'}} V_i\right)=\emptyset.\] By Lemma~\ref{LemCriticalTailChange}, we know that $N(V_C\cup W_{C'},A)\subseteq U_C\cup U_{C'}$ and $|V_C\cup W_{C'}|\leq 2\cdot (\alpha+1)\cdot (N+1)$. As ($*$) tells us that there is no local improvement meeting this size bound, we must have $w(V_C\cup V_{C'})\leq w(U_C\cup U_{C'})$. Together with Lemma~\ref{LemCriticalTailChange} and the fact that $V_C\cap V_{C'}=\emptyset$, this results in
	\[w(U_C)+w(U_{C'})=w(V_C)+w(V_{C'})=w(V_C\cup V_{C'})\leq w(U_C\cup U_{C'}).\] As all vertex weights are positive, this yields $U_C\cap U_{C'}=\emptyset$. This concludes the proof of consistency.
\end{proof}
\begin{claim}
	(\ref{CriticalFamilyProp1}) holds.
\end{claim}
\begin{proof}
Let $i\in\mathcal{I}$. If $i\in\mathcal{I}\setminus\bigcup_{C\in\mathcal{C}} I_C$, then $i\in\mathcal{K}$ and the claim follows. Otherwise, let $C\in\mathcal{C}$ such that $i\in I_C$. Then $U_i\subseteq U_C$, $V_i\subseteq V_C$ and $e_i$ has one endpoint in $U_i$ and one endpoint in $W'_C$, which yields $e_i\subseteq U_C\cup V_C$.
\end{proof}
\begin{claim}
	(\ref{CriticalFamilyProp2}) holds.
\end{claim}
\begin{proof}
By construction, we have $\bigcup_{k\in\mathcal{K}} U_k\setminus \bigcup_{i\in\mathcal{I}} U_i\subseteq \bigcup_{C\in\mathcal{C}} U'_C$. Pick $C\in\mathcal{C}$ and let $u\in U'_C$. We first show that $|\delta(u)\cap(\delta(\bigcup_{k\in\mathcal{K}} V_k)\cup\{e_k,k\in\mathcal{K}\})|\geq 2$. If $u\in V(P_C)$, then $u$ has two neighbors in $W'_C\subseteq V_C$ because
\begin{itemize}
	\item $u$ is an inner vertex of $P_C$ and has two neighbors in $W'_C$ inducing the incident edges of $u$ in $P_C$, or
	\item $|V(P_C)|\geq 2$ and $u$ is an endpoint of $P_C$ that has one neighbor in $W'_C$ inducing its incident edge and another one that is $v_1^C$ or $v_2^C$, or
	\item $|V(P_C)|=1$ and $u$ is adjacent to both $v_1^C$ and $v_2^C$.
\end{itemize}
If $u\not\in V(P_C)$, then $u=n_1(v_2^C)$. $u$ is adjacent to $v_2^C\in W'_C$ and $w_C$, which is not contained in $W'_C$ because the only vertex in $W'_C$ with a neighbor in $B_r\cap B'$ is $v_2^C$, but $w_C\neq v_2^C$ by definition. We have $e_C=\{u,w_C\}$. Thus, $u$ has two incident edges in $\delta(\bigcup_{k\in\mathcal{K}} V_k)\cup\{e_k,k\in\mathcal{K}\}$.

By definition of a normalized instance, we have $|\delta(u)|\leq 3$, which yields \[|\delta(u)\setminus(\delta(\bigcup_{k\in\mathcal{K}} V_k)\cup\{e_k,k\in\mathcal{K}\})|\leq 1.\] If $u\in A'$, then $|\delta(u)|\leq 2$, so $\delta(u)\subseteq \delta(\bigcup_{k\in\mathcal{K}} V_k)\cup\{e_k,k\in\mathcal{K}\}$.
\end{proof}
\end{proof}
\subsection{Iterative tail change construction}
\subsubsection{How to construct the tail changes}
Now, we are ready to show how to apply Theorem~\ref{TheoNewFamilyOfTailChanges} and Lemma~\ref{LemConsistentFamily} in order to prove Theorem~\ref{TheoNiceFamilyofTailChanges}. Let $\delta\in (0,1)$ and choose $\alpha\coloneqq\delta^{-1}$. Let $\gamma\coloneqq 2\cdot(\alpha+1) > 3$, $\pi_\delta\coloneqq \max\{4\cdot(\alpha+1),8\}\cdot\gamma^{\delta^{-1}}$ and $N_\delta\coloneqq\gamma^{\delta^{-1}}$.
 In the remainder of this section, we would like to show that the family of tail changes that Algorithm~\ref{AlgIterativeTailChanges} produces meets the requirements of Theorem~\ref{TheoNiceFamilyofTailChanges}.

\begin{algorithm}[t]
	\caption{Iterative tail change construction}\label{AlgIterativeTailChanges}
\DontPrintSemicolon
$\mathcal{I}^0\gets \emptyset$\;
$N_0\gets 0$\;
$j\gets 0$ \tcp{iteration index}
\While{$\mathrm{true}$\label{LineWhileLoop}}
{
	$j\gets j+1$\label{LineUpdateIndex}\;
	$B_t^j\gets \bigcup_{i\in\mathcal{I}^{j-1}} V_i$, $B_r^j\gets B\setminus B_t^j$\;
$B_1^j\gets \{v\in B''\cap B_r^j: |\delta(v)\setminus\{e_i,i\in\mathcal{I}^{j-1}\}|\leq 1\}$\;
$\mathcal{C}^j\gets$ $\alpha$-critical components in $G_A^{\mathcal{I}^{j-1}}$ \tcp{see Def.~\ref{DefCritical}}
\If{$|B_1^j|> \delta\cdot|B|$}{
	Apply Theorem~\ref{TheoNewFamilyOfTailChanges} with $\mathcal{I}=\mathcal{I}^{j-1}$ and $N=N_{j-1}$ to obtain a new consistent family $(U_i,V_i,e_i)_{i\in\mathcal{I}^{j}}$ of tail changes\label{LineApplyThm}\;
	$N_j\gets 2\cdot N_{j-1}+1$\;
	\textbf{goto} line~\ref{LineWhileLoop}\;
}
\If{$|\mathcal{C}^j|> \delta\cdot |B|$}
{
	$(U_i,V_i,e_i)_{i\in\mathcal{I}^j}\gets (U_i,V_i,e_i)_{i\in\mathcal{C}^{j}\cup \mathcal{I}^{j-1}\setminus\bigcup_{C\in\mathcal{C}^j} I_C}$\label{LineApplyLemma}\;
	$N_j\gets ( \alpha +1)\cdot (N_{j-1}+1)$\;
	\textbf{goto} line~\ref{LineWhileLoop}\;
}
\textbf{return} $(U_i,V_i,e_i)_{i\in\mathcal{I}^{j-1}}$ \tcp{$|B_1^j|\leq \delta\cdot|B|$ and $|\mathcal{C}^j|\leq \delta\cdot |B|$}\;
}
\end{algorithm}
The \emph{index of an iteration} of the while-loop in line~\ref{LineWhileLoop} of Algorithm~\ref{AlgIterativeTailChanges} is the value to which $j$ is set in line~\ref{LineUpdateIndex}.
\begin{lemma}
Let $k\in\mathbb{N}_{>0}$. If there is an iteration with index $k$, then the following properties hold:
\begin{enumerate}[(i)]
	\item $(U_i,V_i,e_i)_{i\in\mathcal{I}^{k-1}}$ is a consistent family of tail changes, each of size at most $N_{k-1}\leq \gamma^{k-1}$.\label{ItAlgProp1}
	\item $|B_t^k|\geq (k-1)\cdot\delta\cdot|B|$ and in particular, $k\leq \delta^{-1}+1$.\label{ItAlgProp2}
	\item There is no local improvement of size at most $\max\{(6\cdot N_{k-1}+2,2\cdot(\alpha+1)\cdot ( N_{k-1}+1)\}$.\label{ItAlgProp3}
\end{enumerate}
In particular, we can indeed apply Theorem~\ref{TheoNewFamilyOfTailChanges} in line~\ref{LineApplyThm} and Lemma~\ref{LemConsistentFamily} in line~\ref{LineApplyLemma}.\label{LemIterativeAlg1}
\end{lemma}
\begin{proof}
	First, observe that $(k-1)\cdot\delta\cdot |B|\leq |B_t^k|\leq |B|$ yields $k-1\leq \delta^{-1}$ since either $k=1$, or $k\geq 2$ and the fact that we have reached the second iteration implies $B\neq\emptyset$.
Next, we show that (\ref{ItAlgProp1}) and (\ref{ItAlgProp2}) imply (\ref{ItAlgProp3}) and then prove (\ref{ItAlgProp1}) and (\ref{ItAlgProp2}) by induction on $k$. By (\ref{ItAlgProp1}) and (\ref{ItAlgProp2}), we obtain
\begin{align*}\max\{6\cdot N_{k-1}+2,2\cdot(\alpha+1)\cdot ( N_{k-1}+1)\}&\leq \max\{6\cdot \gamma^{k-1}+2,2\cdot(\alpha+1)\cdot (\gamma^{k-1}+1)\}\\&\leq \max\{8,4\cdot(\alpha+1)\}\cdot\gamma^{\delta^{-1}}=\pi_\delta,\end{align*} which shows (\ref{ItAlgProp3}.

 To inductively prove (\ref{ItAlgProp1}) and (\ref{ItAlgProp2}), let $k\in\mathbb{N}_{>0}$ and assume that the statement of the lemma holds for all smaller values of $k$. (Observe that for $k=1$, this is a void claim.) 
	 If $k=1$, then (\ref{ItAlgProp1}) and (\ref{ItAlgProp2}) are clear since $\mathcal{I}^0=\emptyset$ and $N_0=0\leq \gamma^0=1$.
	 
	  Now, let $k\geq 2$. We distinguish two cases:\\
	  \textbf{Case 1:} $(U_i,V_i,e_i)_{i\in\mathcal{I}^{k-1}}$ is defined in line~\ref{LineApplyThm} in iteration $k-1$. Then the induction hypothesis and Theorem~\ref{TheoNewFamilyOfTailChanges} tell us that $(U_i,V_i,e_i)_{i\in\mathcal{I}^{k-1}}$ is a consistent family of tail changes, each of size at most \[N_{k-1}=2\cdot N_{k-2}+1\leq 2\cdot\gamma^{k-2}+1\leq \gamma^{k-2}\cdot 3\leq \gamma^{k-1}.\] Theorem~\ref{TheoNewFamilyOfTailChanges}~(\ref{NewTailChangesProp1}) tells us that $B_t^{k-1}\subseteq B_t^k$, and Theorem~\ref{TheoNewFamilyOfTailChanges}~(\ref{NewTailChangesProp3}), together with the definition of $B^{k-1}_1\subseteq B\setminus B_t^{k-1}$, yields $B_1^{k-1}\subseteq B_t^k\setminus B_t^{k-1}$. By the induction hypothesis, we obtain \[|B_t^k|\geq |B_t^{k-1}|+|B_1^{k-1}|\geq (k-2)\cdot \delta\cdot |B|+\delta\cdot|B|=(k-1)\cdot\delta\cdot|B|.\]
	  \textbf{Case 2:} $(U_i,V_i,e_i)_{i\in\mathcal{I}^{k-1}}$ is defined in line~\ref{LineApplyLemma} in iteration $k-1$. By the induction hypothesis, we may apply Lemma~\ref{LemConsistentFamily} to conclude that $(U_i,V_i,e_i)_{i\in\mathcal{I}^{k-1}}$ is a consistent family of tail changes, each of size at most
	  \[N_{k-1}=(\alpha+1)\cdot (N_{k-2}+1)\leq (\alpha +1)\cdot (\gamma^{k-2}+1)\leq 2\cdot(\alpha+1)\cdot \gamma^{k-2}\leq \gamma^{k-1}.\] By Lemma~\ref{LemConsistentFamily}~(\ref{CriticalFamilyProp1}), we have $B_t^{k-1}\subseteq B_t^{k}$.
	  Moreover, we have \[\{v_2^C,C\in\mathcal{C}^{k-1}\}\subseteq \bigcup_{C\in\mathcal{C}^{k-1}} V_C\setminus B_t^{k-1}\subseteq B_t^k\setminus B_t^{k-1}\] (see Def.~\ref{DefCritical} and Def.~\ref{DefCriticalTailChange}) and the vertices $v_2^C,C\in\mathcal{C}^{k-1}$ are pairwise distinct since their neighbors $n_2(v_2^C)$ belong to distinct critical components. Thus, by the induction hypothesis, we obtain
	  \[|B_t^{k}|\geq |B_t^{k-1}|+|\mathcal{C}^{k-1}|\geq \delta\cdot(k-2)\cdot|B|+\delta \cdot|B|=(k-1)\cdot\delta\cdot|B|.\]
\end{proof}
\begin{lemma}
Let $k\in\mathbb{N}_{>0}$ such that there exists an iteration with index $k$.

 Then for every $u\in\bigcup_{i\in\mathcal{I}^{k-1}} U_i$, we have $|\delta(u)\setminus(\delta(\bigcup_{i\in\mathcal{I}^{k-1}} V_i)\cup \{e_i,i\in\mathcal{I}^{k-1}\})|\leq 1$. Moreover, if $u\in A'\cap \bigcup_{i\in\mathcal{I}^{k-1}} U_i$, then $\delta(u)\subseteq \delta(\bigcup_{i\in\mathcal{I}^{k-1}} V_i)\cup \{e_i,i\in\mathcal{I}^{k-1}\}$.\label{LemIterativeAlg2}
\end{lemma}
\begin{proof}
By induction on $k$. We have $\mathcal{I}^0=\emptyset$, so there is nothing to show for $k=1$. Now, let $k\geq 2$ and assume that the statement is true for $k-1$. Then for $u\in\bigcup_{i\in\mathcal{I}^{k-2}} U_i$, the induction hypothesis and Theorem~\ref{TheoNewFamilyOfTailChanges}~(\ref{NewTailChangesProp1}) or Lemma~\ref{LemConsistentFamily}~(\ref{CriticalFamilyProp1}), respectively, yield the desired statement. On the other hand, if $u\in \bigcup_{i\in\mathcal{I}^{k-1}} U_i\setminus\bigcup_{i\in\mathcal{I}^{k-2}} U_i$, then Theorem~\ref{TheoNewFamilyOfTailChanges}~(\ref{NewTailChangesProp4}) or Lemma~\ref{LemConsistentFamily}~(\ref{CriticalFamilyProp2}) proves the claim.
\end{proof}
Now, we are ready to prove Theorem~\ref{TheoNiceFamilyofTailChanges}.
\begin{proof}[Proof of Theorem~\ref{TheoNiceFamilyofTailChanges}]
By Lemma~\ref{LemIterativeAlg1}~(\ref{ItAlgProp2}, we know that Algorithm~\ref{AlgIterativeTailChanges} terminates after at most $\delta^{-1}+1$ iterations. Let $k$ be the index of the last iteration and let $\mathcal{I}\coloneqq\mathcal{I}^{k-1}$. We claim that $(U_i,V_i,e_i)_{i\in\mathcal{I}}$ meets the requirements of Theorem~\ref{TheoNiceFamilyofTailChanges}. By Lemma~\ref{LemIterativeAlg1}, it constitutes a consistent family of tail changes, each of size at most $N_\delta$. The termination criterion of our algorithm tells us that we have $|B^1_0|\leq |B_1|\leq \delta\cdot |B|$, and that the number of $\alpha$-critical components is bounded by $\delta\cdot|B|$. So Theorem~\ref{TheoNiceFamilyofTailChanges}~(\ref{NiceTailChanges2}) and (\ref{NiceTailChanges3}) are satisfied. Finally, Lemma~\ref{LemIterativeAlg2} yields the Theorem~\ref{TheoNiceFamilyofTailChanges}~(\ref{NiceTailChanges1}).	
\end{proof}

\section{Final analysis\label{SecFinalAnalysis}}
In this section, we can finally prove Theorem~\ref{TheoAnalysisNormalized}. Let $\epsilon > 0$ and pick $\delta$ with $\delta^{-1}\in\mathbb{N}$ such that $\frac{4+6\cdot\delta}{3-7\cdot\delta}\leq\frac{4}{3}+\epsilon$. Let $\pi_\delta$ and $N_\delta$ as implied by Theorem~\ref{TheoNiceFamilyofTailChanges}.

 Define $\tau\coloneqq\max\{\pi_\delta,4\cdot\delta^{-1},(3\cdot\delta^{-1}+3)\cdot (3\cdot N_\delta +1)\}$. Let $(G,w,A,B)$ be a normalized instance with the property that there is neither a local improvement of size at most $\tau$, nor an improving binocular of size at most $\tau\cdot\log(|V(G)|)$ in $S_{\tau}(G,w,A)$. As $\pi_\delta\leq \tau$, let  $(U_i,V_i,e_i)_{i\in \mathcal{I}}$ be a consistent family of tail changes as guaranteed by Theorem~\ref{TheoNiceFamilyofTailChanges}.

As before, we set $A_t\coloneqq\bigcup_{i\in\mathcal{I}} U_i$, $A_r\coloneqq A\setminus A_t$, $B_t\coloneqq\bigcup_{i\in\mathcal{I}} V_i$ and $B_r\coloneqq B\setminus B_t$. 
We start by proving a corollary of Proposition~\ref{PropLocalImprovement}, which tells us that certain configurations can not occur.
\begin{corollary}
	Let the \emph{size} of a sub-graph $H$ of $G_A^{\mathcal{I}}$ be its number of edges.
	None of the following situations can occur:
	\begin{enumerate}
		\item a vertex $v\in B_r$ with $N(v,A_r)=\emptyset$
		\item a vertex $v\in B_r$ such that $N(v,A_r)\subseteq A'$ and $|N(v,A_r)|\leq 2$
		\item a connected sub-graph of $G_A^{\mathcal{I}}$ of size at most $\delta^{-1}$ that contains more edges than vertices
		\item a connected sub-graph of $G_A^{\mathcal{I}}$ of size at most $3\cdot \delta^{-1}$ that has three attached virtual edges
		\item a connected sub-graph of $G_A^{\mathcal{I}}$ of size at most $\delta^{-1}$ that contains a cycle and has an attached virtual edge
		\item a connected sub-graph of $G_A^{\mathcal{I}}$ of size at most $\delta^{-1}$ that has two attached dashed loops
	\end{enumerate}\label{CorImpossibleSituations}
\end{corollary}
\begin{proof}
	We show that each of these situations implies the existence of $X\subseteq B_r$ with $|X|\leq 3\cdot\delta^{-1}+3$ and $w(X)>w(N(X,A_r))$ or $w(X)=w(N(X,A_r))$ and $|X\cap B''|>|N(X,A_r)\cap A''|$. By Prop.~\ref{PropLocalImprovement}, this yields the existence of a local improvement of size at most $(3\cdot\delta^{-1}+3)\cdot (3\cdot N_\delta+1)\leq \tau$, a contradiction.
	\begin{enumerate}
		\item Clear since $N(\{v\},A_r)=\emptyset$.
		\item Clear since $w(N(v,A_r))\leq 2\leq w(v)$ and $|N(v,A_r)\cap A''|=0<|\{v\}\cap B''|$.
		\item Pick such a sub-graph $H$ and let $X$ be the set of vertices from $B_r$ inducing the edges of $H$. Then \[w(X)=2\cdot |E(H)|> 2\cdot |V(H)|=w(V(H))=w(N(X,A_r)).\] Thus, $X$ is as required.
		\item Pick such a sub-graph $H$, let $v_1$, $v_2$ and $v_3$ be the vertices inducing the virtual edges and let $u_1$, $u_2$ and $u_3$ be the vertices they are attached to (which need not be distinct). Then $w(N(\{v_1,v_2,v_3\},A_r\setminus\{u_1,u_2,u_3\}))\leq w(\{v_1,v_2,v_3\})-3$ because if $v_i$ induces a dotted loop, then $w(v_i)=1$ and $u_i$ is the only neighbor of $v_i$ in $A_r$, and if $v_i$ induces a dashed loop, then $w(v_i)=2$ and the only neighbor of $v_i$ in $A_r$ other than $u_i$ has weight $1$. Let $X'$ be the set of vertices inducing the edges of $H$. Then $N(X',A_r)=V(H)\supseteq \{u_1,u_2,u_3\}$ and 
		\[w(X')=2\cdot |E(H)|\geq 2\cdot(|V(H)|-1) = w(V(H))-2.\] Thus, \begin{align*}w(X'\cup \{v_1,v_2,v_3\})&\geq w(V(H))-2+w(N(\{v_1,v_2,v_3\},A_r\setminus\{u_1,u_2,u_3\}))+3\\& > w(N(X'\cup\{v_1,v_2,v_3\},A_r)).\end{align*} Hence, $X\coloneqq X'\cup\{v_1,v_2,v_3\}$ has the required properties.
		\item Pick such a sub-graph $H$, let $v$ induce the virtual loop and let $u\in V(H)$ be the vertex it is attached to. Then $w(v)>w(N(v,A_r\setminus V(H)))$ as in the previous case. Let $X'\subseteq B_r$ induce the edges of $H$.
		Then $N(X',A_r)=V(H)$ and we obtain
		\begin{align*}
		w(X'\cup \{v\})&=2\cdot|E(H)|+w(v)> 2\cdot |V(H)|+w(N(v,A_r\setminus V(H)))\\&=w(N(X'\cup\{v\},A_r)).
		\end{align*}
		Again, $X\coloneqq X'\cup\{v\}$ has the desired properties.
		\item Pick such a sub-graph $H$, let $v_1$ and $v_2$ induce the two dashed loops and let $u_1$ and $u_2$ be the vertices they are attached to. As before, we get $w(\{v_1,v_2\})\geq w(N(\{v_1,v_2\},A_r\setminus V(H)))+2$. Let $X'$ be the set of vertices inducing the edges of $H$. Then
		\begin{align*}
		w(X'\cup\{v_1,v_2\})&\geq 2\cdot |E(H)|+w(N(\{v_1,v_2\},A_r\setminus V(H)))+2\\&\geq 2\cdot|V(H)|+w(N(\{v_1,v_2\},A_r\setminus V(H)))\\&=w(N(X'\cup\{v_1,v_2\},A_r)).
		\end{align*}
		In addition, as $N(\{v_1,v_2\},A_r\setminus V(H))$ only contains vertices of weight $1$, we have \begin{align*}|(X'\cup\{v_1,v_2\})\cap B''|&=|X'\cup\{v_1,v_2\}|=|E(H)|+2\geq |V(H)|+1\\&> |V(H)|=|N(X'\cup\{v_1,v_2\},A_r)\cap A''|.\end{align*} Thus, $X\coloneqq X'\cup\{v_1,v_2\}$ is as desired.
	\end{enumerate}
\end{proof}  As a consequence, we can partition the vertices in $B_r\cap B'$ into the following three sets:
\begin{itemize}
	\item $B'_1$ consists of those vertices $b\in B_r\cap B'$ such that $E(\{b\},A_t)\subseteq \{e_i,i\in\mathcal{I}\}$ and $b$ has exactly one neighbor in $A_r$.
	\item $B'_2$ consists of those vertices $b\in B_r\cap B'$ such that $b$ has two neighbors in $A_r$.
	\item $\tilde{B}'_{1}$ consists of those vertices $b\in B_r\cap B'$ such that $E(\{b\},A_t)\setminus\{e_i,i\in\mathcal{I}\}\neq\emptyset$ and $b$ has exactly one neighbor in $A_r$.
\end{itemize} 
In addition, we can partition the vertices in $B_r\cap B''$ into the following sets:
	\begin{description}	
		\item[$B^1_0$] consists of all $v\in B''\cap B_r$ such that $E(\{v\},A_t)\subseteq \{e_i,i\in \mathcal{I}\}$ and such that $v$ has exactly one neighbor in $A''\cap A_r$ and no neighbor in $A'\cap A_r$. 
	\item[$B^{2,1}$] consists of all $v\in B''\cap B_r$ such that $E(\{v\},A_t)\subseteq \{e_i,i\in \mathcal{I}\}$ and such that $v$ has exactly one neighbor $n_1(v)$ in $A'\cap A_r$ and exactly one neighbor $n_2(v)$ in $A''\cap A_r$.
	\item[$B^{2,2}$] consists of all $v\in B''\cap B_r$ such that $E(\{v\},A_t)\subseteq \{e_i,i\in \mathcal{I}\}$ and such that $v$ has exactly two neighbors in $A''\cap A_r$ and no neighbor in $A'\cap A_r$.
	\item[$\tilde{B}^2$] consists of all $v\in B''\cap B_r$ such that $E(\{v\},A_t)\setminus\{e_i,i\in \mathcal{I}\}\neq\emptyset$ and such that $v$ has exactly two neighbors in $A_r$.
	\item[$B^1_1$] consists of all $v\in B''\cap B_r$ such that $|E(\{v\},A_t)\setminus \{e_i,i\in \mathcal{I}\}|= 1$ and such that $v$ has exactly one neighbor in $A''\cap A_r$ and no neighbor in $A'\cap A_r$.

	\item[$\tilde{B}^1$] consists of all $v\in B''\cap B_r$ such that $|E(\{v\},A_t)\setminus \{e_i,i\in \mathcal{I}\}|= 2$ and such that $v$ has exactly one neighbor in $A''\cap A_r$ and no neighbor in $A'\cap A_r$.
	\item[$B^3$] consists of all $v\in B''\cap B_r$ such that $v$ has three neighbors in $A_r$.
\end{description}Note that this definition is consistent with Def.~\ref{DefSubsetsB}.
We further partition $B'_1=B'_{1,c}\dot{\cup}B'_{1,s}\dot{\cup}B'_{1,l}$, as well as $B^{2,1}=B^{2,1}_c\dot{\cup}B^{2,1}_s\dot{\cup}B^{2,1}_l$, where the sets with lower index $c$/$s$/$l$ contain those vertices for which the respective virtual loop is incident to a $\delta^{-1}$-critical/$\delta^{-1}$-small, but non-critical/$\delta^{-1}$-large connected component of $G_A^{\mathcal{I}}$.
\begin{lemma}
$|A_t\cap A''|\geq |\tilde{B}'_1|+|\tilde{B}^2|+|B^1_1|+2|\tilde{B}^1|$\label{LemBound1}
\end{lemma} 
\begin{proof}
By Theorem~\ref{TheoNiceFamilyofTailChanges}~(\ref{NiceTailChanges1}), we know that $|E(A_t,B_r)\setminus\{e_i,i\in\mathcal{I}\}|\leq |A''\cap A_t|$. As each vertex in $\tilde{B}'_1\cup\tilde{B}^2\cup B^1_1$ has at least one incident edge in  $E(A_t,B_r)\setminus\{e_i,i\in\mathcal{I}\}$, and each vertex from $\tilde{B}^1$ has two such incident edges, the claim follows.
\end{proof}
\begin{lemma}
$3|A_r\cap A''|+\sum_{u\in A'\cap A_r} |N(u,B_r)|\geq |B'_1|+2|B'_2|+|\tilde{B}'_1|+2|B^{2,1}|+2|B^{2,2}|+2|\tilde{B}^2|+|B^1_1|+|B^1_0|+|\tilde{B}^1|+3|B^3|$. \label{LemBound2}
\end{lemma}
\begin{proof}
We count the edges between $A_r$ and $B_r$ by counting degrees from $A_r$ to $B_r$ and vice versa.
\end{proof}
\begin{lemma}
There is no binocular of size at most $4\cdot\delta^{-1}\cdot\log(|A|)$ in $G^{\mathcal{I}}_A$.\label{LemNoBinocular}
\end{lemma}
\begin{proof}
	Assume that there were a binocular of size at most $4\cdot\delta^{-1}\cdot\log(|A|)$ in $G^{\mathcal{I}}_A$. In particular, there would also be a \emph{minimal} binocular in $G^{\mathcal{I}}_A$ meeting this size bound.
We show that such a minimal binocular would give rise to an improving minimal binocular of the same size. This concludes the proof since $4\cdot\delta^{-1}\leq \tau$. Let $\mathcal{B}$ be a minimal binocular in $G_A^{\mathcal{I}}$ of size at most $4\cdot\delta^{-1}\cdot\log(|A|)$. We construct a binocular in $S_\tau(G,w,A)$ with edge set $\mathcal{E}$ as follows:
\begin{itemize}
	\item For every edge $e$ of $\mathcal{B}$ that is induced by a vertex $v\in B^{2,2}$, let $I_v\coloneqq\{i: e_i\in \delta(v)\}$. Then $|I_v|\leq 1$ since $v$ has degree at most $3$ in total. Let $U_v\coloneqq\bigcup_{i\in I_v} U_i$ and $W_v\coloneqq\bigcup_{i\in I_v} V_i\cup\{v\}$. Then $U_v\subseteq A$ and $W_v\subseteq B = V\setminus A$ is independent. We further have $\max\{|U_v|,|W_v|\}\leq N_\delta+1\leq \tau$ and $w(U_v)+2=w(W_v)$. Finally, by definition of $B^{2,2}$, $N(W_v,A\setminus U_v)\subseteq A''$ and $|N(W_v,A\setminus U_v)|=2$. Thus, $(U_v,W_v)$ is an edge-inducing pair. We add $e(U_v,W_v)$ to $\mathcal{E}$. Observe that $e(U_v,W_v)$ is a two-vertex edge between the same vertices as $e$.
	\item For every edge $e$ of $\mathcal{B}$ that is induced by a vertex $v\in B^1_1$, let $I_v\coloneqq\{i: E(\{v\},U_i)\neq\emptyset\}$. Then $|I_v|\leq 2$ since $v$ has degree at most $3$ in total and possesses one incident edge to $B_r$. Let $U_v\coloneqq\bigcup_{i\in I_v} U_i$ and $W_v\coloneqq\bigcup_{i\in I_v} V_i\cup\{v\}$. Then $U_v\subseteq A$ and $W_v\subseteq B=V\setminus A$ is independent. Furthermore, $\max\{|U_v|,|W_v|\}\leq 2\cdot N_\delta+1\leq \tau$ and $w(U_v)+2=w(W_v)$. Finally, by definition of $B_1^1$, we have $N(W_v,A\setminus U_v)\subseteq A''$ and $|N(W_v,A\setminus U_v)|=1$. Thus, $(U_v,W_v)$ is an edge-inducing pair. We add $e(U_v,W_v)$ to $\mathcal{E}$. Note that $e(U_v,W_v)$ is a loop attached to the same vertex as $e$.
\end{itemize}
By construction, the sub-graph of $S_\tau(G,w,A)$ with edge set $\mathcal{E}$ is isomorphic to $\mathcal{B}$. In particular, it constitutes a minimal binocular of size at most $4\cdot\delta^{-1}$. It remains to check that this binocular is improving. Let $V_1$ be the set of vertices from $B_1^1$ inducing loops of $\mathcal{B}$, and let $V_2$ be the set of vertices from $B^{2,2}$ inducing (two-endpoint) edges of $\mathcal{B}$. Denote the sets of two-endpoint edges and loops in $\mathcal{E}$ by $\mathcal{E}_2$ and $\mathcal{E}_1$, respectively. Then $\mathcal{E}_i=\{e(U_v,W_v),v\in V_i\}$ for $i=1,2$. We check the conditions on an improving binocular (see Def.~\ref{DefImprovingBinocular}) one by one:
\begin{enumerate}[(i)]
	\item For $v\in V_2$ and $i\in I_v$, we have $e_i\in\delta(v)$, so the sets $I_v,v\in V_2$ are pairwise disjoint. By consistency of $(U_i,V_i,e_i)_{i\in\mathcal{I}}$, the sets $\bigcup_{i\in I_v} V_i,v\in V_2$ are pairwise disjoint. As $\bigcup_{i\in\mathcal{I}} V_i=B_t$, and $V_2\subseteq B_r=B\setminus B_t$, the sets $W_v,v\in V_2$ are pairwise disjoint. They agree with the sets $W(e),e\in\mathcal{E}_2$.
	\item As $\bigcup_{i\in\mathcal{I}} V_i\subseteq B_t$, $V_2\subseteq B^{2,2}$ and $V_1\subseteq B_1^1$, we have \begin{equation}V_1\cap \bigcup_{v\in V_2} W_v=V_1\cap \bigcup_{e\in \mathcal{E}_2} W(e)=\emptyset.\label{EqNothingFromV1Removed}\end{equation} Let $J\coloneqq\bigcup_{v\in V_1} I_v\setminus\bigcup_{v\in V_2} I_v$. By consistency of $(U_i,V_i,e_i)_{i\in\mathcal{I}}$, we have
	\[\bigcup_{e\in \mathcal{E}_1} U(e)\setminus\bigcup_{e\in\mathcal{E}_2} U(e)=\bigcup_{v\in V_1} U_v\setminus\bigcup_{v\in V_2}U_v = \bigcup_{j\in J} U_j \text{ and }\]
		\[\bigcup_{e\in \mathcal{E}_1} W(e)\setminus\bigcup_{e\in\mathcal{E}_2} W(e)=\bigcup_{v\in V_1} W_v\setminus\bigcup_{v\in V_2}W_v = V_1\cup\bigcup_{j\in J} V_j.\]
		As every vertex in $V_1$ is of weight $2$ and $|V_1|=|\mathcal{E}_1|$, this yields \[w\left(\bigcup_{e\in \mathcal{E}_1} W(e)\setminus\bigcup_{e\in\mathcal{E}_2} W(e)\right)=w\left(\bigcup_{e\in \mathcal{E}_1} U(e)\setminus\bigcup_{e\in\mathcal{E}_2} U(e)\right)+2\cdot|\mathcal{E}_1|.\]
		\item $\bigcup_{e\in\mathcal{E}_1\cup\mathcal{E}_2} W(e)\subseteq B$ is independent.
\end{enumerate}
\end{proof}
\begin{lemma}
$(1+\delta)\cdot |A_r\cap A''|+0.5|\{u\in A'\cap A_r: |N(u,B_r)|=1\}|\geq |B^1_1| + |B^{2,2}| + |B^{2,1}_s|+0.5|B^{'}_{1,s}|$.\label{LemBound3}
\end{lemma}
\begin{proof}
It suffices to show that for each connected component of $G^{\mathcal{I}}_A$, the respective inequality holds when restricting the left-hand-side to the vertices of the component, plus the vertices of weight $1$ that are neighbors of vertices inducing dashed loops in the components and have degree $1$ to $B_r$ (so that they are counted for at most one component), and the right-hand-side to those vertices inducing edges of the component or dashed or dotted loops attached to it.
Pick such a connected component $C$.

\textbf{Case 1: $C$ is $\delta^{-1}$-large.} By definition of these sets, no vertex from $B^{2,1}_s\cup B^{'}_{1,s}$ induces a dashed respectively dotted loop incident to $V(C)$. By Lemma~\ref{LemNoBinocular}, $C$ contains no binocular of size at most $4\cdot\delta^{-1}\cdot\log(|A|)\geq 4\cdot \delta^{-1}\cdot \log(|V(C)|)$. Thus, Lemma~\ref{LemBinocular} tells us that $|E(C)|\leq (1+\delta)\cdot |V(C)|$. This yields the desired inequality.

\textbf{Case 2: $C$ is $\delta^{-1}$-small, but non-critical.} \\
First of all, we can conclude that $|E(C)|\leq |V(C)|$ because if not, we could pick a spanning tree in $C$ and then add two further edges, which yields a connected sub-graph of $G_A^{\mathcal{I}}$ of size $|V(C)|+1\leq \delta^{-1}$ and with more edges than vertices, a contradiction to Corollary~\ref{CorImpossibleSituations}.

\textbf{Case 2.1}: $|E(C)|=|V(C)|$.\\ Then $C$ contains a cycle (or a non-virtual loop), so $C$ cannot have any attached dotted or dashed loop by Corollary~\ref{CorImpossibleSituations} and since $|V(C)|\leq\delta^{-1}-1$.

\textbf{Case 2.2}: $|E(C)|\leq |V(C)|-1$.\\As $C$ is connected, this means $C$ is a tree and $|E(C)|=|V(C)|-1$.

 If $C$ has at least one attached dashed loop (induced by $v$) and one attached dotted loop, then there cannot be a further virtual edge by Corollary~\ref{CorImpossibleSituations}. As $C$ is non-critical, we know that $n_1(v)$ has degree $1$ to $B_r$ and we can count it with value $0.5$. Thus, on the left-hand-side, we get $(1+\delta)\cdot|V(C)|+0.5$, and on the right-hand-side, we obtain $|E(C)|=|V(C)|-1$ for the edges of $C$, $1$ for the vertex inducing the dashed loop and $0.5$ for the vertex inducing the dotted loop. Thus, the left-hand-side is greater or equal than the right-hand-side.

We are left with the case where $C$ can have an attached dashed loop or an attached dotted loop, but not both. If $C$ has an attached dashed loop, $C$ cannot have another attached dashed loop by Corollary~\ref{CorImpossibleSituations}. Thus, $C$ is a tree with one attached dashed loop, and we have $|V(C)|\geq |E(C)|+1$, which yields the desired inequality.

If $C$ only has attached dotted loops, there can be at most two of them by Corollary~\ref{CorImpossibleSituations}. Moreover, we have $|V(C)|=|E(C)|+1=|E(C)|+0.5+0.5$. Again, the desired inequality follows.

\textbf{Case 3: $C$ is $\delta^{-1}$-critical.} Then $C$ does not have any attached virtual edges that are induced by vertices from $B^{2,1}_s\cup B^{'}_{1,s}$. Moreover, $C$ is a tree, so $|V(C)|=|E(C)|+1$. This gives the desired statement.
\end{proof}

\begin{lemma}
$|B^{2,1}_c|+0.5|B^{'}_{1,c}|\leq 1.5\cdot \delta\cdot |B|.$\label{LemBound4}
\end{lemma}
\begin{proof}
As every $\delta^{-1}$-critical component has precisely one attached dashed and one attached dotted loop, this follows from the fact that the number of $\delta^{-1}$-critical components is bounded by $\delta\cdot |B|$.
\end{proof}
\begin{lemma}
$0.5\cdot |B'_{1,l}|+|B^{2,1}_l|\leq |B'_{1,l}|+|B^{2,1}_l|\leq 2\cdot\delta\cdot |A|$.\label{LemBound5}
\end{lemma}
\begin{proof}
Consider a $\delta^{-1}$-large component. For every vertex $v$ in that component, there are at most $2$ dotted or dashed loops at distance at most $\delta^{-1}$ to it by Corollary~\ref{CorImpossibleSituations}. On the other hand, as our component is large, for every dotted or dashed loop, there are at least $\delta^{-1}$ vertices of the component at distance at most $\delta^{-1}$ to the loop. This yields the desired bound.
\end{proof}
Putting everything together, we obtain
\begin{align*}
&\quad\underbrace{|\tilde{B}'_1|+|\tilde{B}^2|+|B^1_1|+2|\tilde{B}^1|}_{\text{Lemma~\ref{LemBound1}}}\\&\quad+\underbrace{|B'_1|+2\cdot|B'_2|+|\tilde{B}'_1|+2\cdot|B^{2,1}|+2\cdot|B^{2,2}|+2\cdot|\tilde{B}^2|+|B^1_1|+|B^1_0|+|\tilde{B}^1|+3\cdot|B^3|}_{\text{Lemma~\ref{LemBound2}}}\\&\quad+\underbrace{|B^1_1| + |B^{2,2}| + |B^{2,1}_s|+0.5\cdot|B^{'}_{1,s}}_{\text{Lemma~\ref{LemBound3}}}|+\underbrace{2\cdot|B^1_0|}_{\text{assumption}}\\&\quad+\underbrace{|B^{2,1}_c|+0.5\cdot|B^{'}_{1,c}|}_{\text{Lemma~\ref{LemBound4}}}+\underbrace{0.5\cdot |B'_{1,l}|+|B^{2,1}_l|}_{\text{Lemma~\ref{LemBound5}}} \\&\leq \underbrace{|A_t\cap A''|}_{\text{Lemma~\ref{LemBound1}}}+\underbrace{3\cdot|A_r\cap A''|+\sum_{u\in A'\cap A_r} |N(u,B_r)|}_{\text{Lemma~\ref{LemBound2}}}\\&\quad+\underbrace{(1+\delta)\cdot |A_r\cap A''|+0.5\cdot|\{u\in A'\cap A_r: |N(u,B_r)|=1\}|}_{\text{Lemma~\ref{LemBound3}}}\\&+\underbrace{2\cdot\delta\cdot |B|}_{\text{assumption}}+\underbrace{1.5\cdot \delta\cdot |B|}_{\text{Lemma~\ref{LemBound4}}}+\underbrace{2\cdot\delta\cdot |A|}_{\text{Lemma~\ref{LemBound5}}}.
\end{align*}
Simplification yields
\begin{align*}
&\quad 1.5\cdot|B'\cap B_r| + 3\cdot|B''\cap B_r|\\&\leq 1.5\cdot |B'_1| + 2\cdot|B'_2|+2\cdot|\tilde{B}'_1|+3\cdot|B^{2,1}|+3\cdot|B^{2,2}|+3\cdot|\tilde{B}_2|+3\cdot|B^1_1|+3\cdot|B^1_0|+3\cdot|\tilde{B}^1|+3\cdot|B^3|\\&\leq |A_t\cap A''|+(4+\delta)\cdot|A_r\cap A''|+2\cdot|\{u\in A'\cap A_r: |N(u,B_r)|=2\}|\\&\quad +1.5\cdot|\{u\in A'\cap A_r: |N(u,B_r)|=1\}|+2\cdot\delta\cdot |A|+3.5\cdot\delta\cdot |B|\\
&\leq |A_t\cap A''| + 4\cdot|A''\cap A_r| + 2\cdot|A'\cap A_r| + 3\cdot\delta\cdot |A|+3.5\cdot\delta\cdot |B|.
\end{align*}
By definition of tail changes, we know that $|B_t\cap B''|=|A_t\cap A''|$ and $|B'\cap B_t|=|A'\cap A_t|$. This gives
\begin{align*}&\quad 1.5\cdot w(B)= 1.5\cdot|B'|+3\cdot|B''|=1.5\cdot|B'\cap B_t| + 3\cdot|B''\cap B_t|+1.5\cdot|B'\cap B_r| + 3\cdot|B''\cap B_r|\\&\leq  1.5\cdot|A_t\cap A'| + 4\cdot|A_t\cap A''| + 4\cdot|A''\cap A_r| + 2\cdot|A'\cap A_r| + 3\cdot\delta\cdot |A|+3.5\cdot\delta\cdot |B|\\
&\leq 2\cdot|A'|+4\cdot|A''|+ 3\cdot\delta\cdot |A|+3.5\cdot\delta\cdot |B| \leq 2\cdot w(A)+3\cdot\delta\cdot w(A)+3.5\cdot\delta\cdot w(B).\end{align*} Thus,
\[w(B)\leq \frac{4+6\cdot\delta}{3-7\cdot\delta}\cdot w(A)\leq \left(\frac{4}{3}+\epsilon\right)\cdot w(A).\]
This concludes the analysis.
\section{A polynomial running time \label{SecPolyTime}}
In this section, we prove that Algorithm~\ref{OverallAlgorithm} can be implemented to run in polynomial time for any fixed constant $\tau$. In doing so, Lemma~\ref{LemNumIterations} tells us that the number of iterations of the while-loop Algorithm~\ref{OverallAlgorithm} performs is polynomially bounded in the input size. In order to bound the running time of a single iteration, we employ Proposition~\ref{PropImprovementConstantSize} and Theorem~\ref{TheoSearchforBinocular}. Proposition~\ref{PropImprovementConstantSize} tells us that we can, in polynomial time, find a local improvement of constant size, or decide that none exists. Theorem~\ref{TheoSearchforBinocular} guarantees that we can search for an improving binocular in the search graph in polynomial time.
\begin{lemma}
Let $(\mathcal{S},w)$ be the input to Algorithm~\ref{OverallAlgorithm} and let $G=G_{\mathcal{S}}$.	The number of iterations of the while-loop that Algorithm~\ref{OverallAlgorithm} performs is bounded by $\mathcal{O}(|V(G)|^2)=\mathcal{O}(|\mathcal{S}|^2)$.\label{LemNumIterations}
\end{lemma}
\begin{proof}
	By Lemma~\ref{LemBinocularYieldsLocalImprovement}, we know that for an improving binocular $\mathcal{B}$, $W(\mathcal{B})$ constitutes a local improvement. By definition of a local improvement, we know that throughout the algorithm, the total weight of $A$ cannot decrease. As all weights are integral, whenever it increases, it increases by at least $1$. As $w(A)=w(\emptyset)=0$ initially and $w(A)\leq w(V(G))\leq 2\cdot |V(G)|$ throughout, there can be at most $2\cdot |V(G)|$ iterations in which the weight of $A$ strictly increases. Note that the first iteration falls into this category since all weights are positive. In between two consecutive iterations where $w(A)$ strictly increases, or after the last such iteration, the weight of $A$ remains constant, but the number of vertices of weight $2$ contained in $A$ increases. Thus, there can be at most $|V(G)|+1$ consecutive iterations where the weight remains the same. Hence, the total number of iterations can be bounded by $\mathcal{O}(|V(G)|^2)$. Note that $|V(G)|=|\mathcal{S}|$ by definition.
\end{proof}
\begin{proposition}
Let $\tau>0$ be a fixed constant. Let $(\mathcal{S},w)$ be an instance of the $2$-$3$-Set Packing problem, let $G=G_{\mathcal{S}}$ and let $A\subseteq V(G)$ be independent. We can, in polynomial time, either return a local improvement of $A$ of size at most $\tau$, or decide that none exists.\label{PropImprovementConstantSize}
\end{proposition}
\begin{proof}
	We can simply try every single one of the $\mathcal{O}(|V(G)|^\tau)$ many subsets of $V(G)$ of size at most $\tau$ and check whether it constitutes a local improvement in polynomial time.
\end{proof}
\begin{theorem}
Let $\tau>0$ be a fixed constant. Let $(\mathcal{S},w)$ be an instance of the $2$-$3$-Set Packing problem, let $G=G_{\mathcal{S}}$ and let $A\subseteq V(G)$ be independent. We can, in polynomial time, either return an improving binocular $\mathcal{B}$ in $S_\tau(G,w,A)$, or decide that there is no improving minimal binocular of size at most $\tau\cdot\log(|V(G)|)$.\label{TheoSearchforBinocular}
\end{theorem}
The remainder of this section is dedicated to the proof of Theorem~\ref{TheoSearchforBinocular}. For this purpose, we employ the color coding technique, similar as in~\cite{FurerYu}. This requires the following terminology:
\begin{definition}[$t$-perfect family of hash functions, \cite{ColorCoding}]
	For $t,m\in\mathbb{N}$ with $t\leq m$, a family \mbox{$\mathcal{F}\subseteq {}^{\{1,\dots,m\}}\{1,\dots,t\}$} of functions mapping $\{1,\dots,m\}$ to $\{1,\dots,t\}$ is called a \emph{$t$-perfect family of hash functions} if for all $I\subseteq \{1,\dots,m\}$ of size at most $t$, there is $f\in\mathcal{F}$ with $f\upharpoonright I$ injective.
\end{definition}
\begin{theorem}[stated in \cite{ColorCoding} referring to \cite{SchmidtSiegel}]
	For $t,m\in\mathbb{N}$ with $t\leq m$, a $t$-perfect family $\mathcal{F}$ of hash functions of cardinality $\mathcal{O}(2^{\mathcal{O}(t)}\cdot (\log (m))^2)$, where each function is encoded using $\mathcal{O}(t)+2\log\log m$ many bits, can be explicitly constructed such that the query time is constant.\label{TheoFamilyHashFunctions}
\end{theorem}
Let $G\coloneqq(V,E)\coloneqq G_{\mathcal{S}}$ be the conflict graph of an instance $(\mathcal{S},w)$ of the $2$-$3$-Set Packing problem and let $\mathcal{U}\coloneqq\bigcup \mathcal{S}$ be the underlying universe. Let $t\coloneqq3\cdot\tau^2\cdot\log(|V|)$ and let $\mathcal{F}$ be a $t$-perfect family of hash functions $\mathcal{U}\rightarrow\{1,\dots,t\}$ of cardinality $\mathcal{O}(2^{\mathcal{O}(t)}\cdot (\log (|\mathcal{U}|))^2)$ as stated in Theorem~\ref{TheoFamilyHashFunctions}. We have $|\mathcal{U}|\leq 3\cdot|\mathcal{S}|=3\cdot |V|$, and, thus, \[|\mathcal{F}|\in \mathcal{O}\left(2^{\mathcal{O}(t)}\cdot (\log (|\mathcal{U}|))^2\right)=2^{\mathcal{O}(3\cdot \tau^2\cdot\log(|V|))}\cdot (\log (|V|))^2=|V|^{\mathcal{O}(1)},\] which is polynomial.
\begin{definition}[colors]
	Let $f\in\mathcal{F}$ and $v\in V$. We interpret $v$ as the corresponding set in $\mathcal{S}$ and let $\mathrm{col}_f(v)\coloneqq f(v)$ be the set of colors assigned to the elements of $v$ by $f$.
	For a subset $W\subseteq V$, we define $\mathrm{col}_f(W)\coloneqq\bigcup_{v\in W} \mathrm{col}_f(v)$.
\end{definition}
\begin{definition}[colorful binocular]
	Let $f\in\mathcal{F}$ and let $\mathcal{B}$ be a binocular in $S_\tau(G,w,A)$ and let $E_1$, $E_2$ and $W(\mathcal{B})$ be as in Def.~\ref{DefImprovingBinocular}. We call $\mathcal{B}$ \emph{colorful}
	if the following properties hold:
	\begin{enumerate}[(i)]
		\item\label{ColorfulBinocular1} Each two among the sets $\mathrm{col}_f(W(e)),e\in E_2$ and $\mathrm{col}_f(\bigcup_{e\in E_1} W(e)\setminus \bigcup_{e\in E_2} W(e))$ are disjoint.
		\item\label{ColorfulBinocular2} We have $w\left(\bigcup_{e\in E_1} W(e)\setminus \bigcup_{e\in E_2} W(e)\right)\geq w\left(\bigcup_{e\in E_1} U(e)\setminus \bigcup_{e\in E_2} U(e)\right)+2\cdot |E_1|.$
		\item\label{ColorfulBinocular3} For every $e\in E(\mathcal{B})$, the sets $\mathrm{col}_f(v)$, $v\in W(e)$ are pairwise disjoint.
		\item \label{ColorfulBinocular4} The sets $\mathrm{col}_f(v)$, $v\in \bigcup_{e\in E_1} W(e)\setminus \bigcup_{e\in E_2} W(e)$ are pairwise disjoint.
	\end{enumerate}\label{DefColorfulBinocular}
\end{definition}
\begin{lemma}
	Every colorful binocular is improving. For every improving binocular of size at most $\tau\cdot\log(|V|)$, there exists $f\in\mathcal{F}$ for which it is colorful.
\end{lemma}
\begin{proof}
	Let $\mathcal{B}$ be a colorful binocular. Def.~\ref{DefColorfulBinocular}~(\ref{ColorfulBinocular1}) yields Def.~\ref{DefImprovingBinocular}~(\ref{ImprovingBinocular1}). Def.~\ref{DefColorfulBinocular}~(\ref{ColorfulBinocular2}) is the same as Def.~\ref{DefImprovingBinocular}~(\ref{ImprovingBinocular2}). Combining Def.~\ref{DefColorfulBinocular}~(\ref{ColorfulBinocular1}),~(\ref{ColorfulBinocular3}) and (\ref{ColorfulBinocular4}) tells us that the color sets $\mathrm{col}_f(v),v\in W(\mathcal{B})$ are pairwise disjoint because (\ref{ColorfulBinocular3}) and (\ref{ColorfulBinocular4}) imply that for $v\neq w$ contained in the same among the sets $W(e),e\in E_2$ or $\bigcup_{e\in E_1} W(e)\setminus \bigcup_{e\in E_2} W(e)$, $\mathrm{col}_f(v)$ and $\mathrm{col}_f(w)$ are disjoint, whereas (\ref{ColorfulBinocular1}) ensures that for $v\neq w$ coming from different ones of these sets, $\mathrm{col}_f(v)$ and $\mathrm{col}_f(w)$ do not intersect. In particular, we can infer that the underlying sets of the vertices in $W(\mathcal{B})$ are pairwise disjoint. Hence, they form an independent set in the conflict graph, which gives Def.~\ref{DefImprovingBinocular}~(\ref{ImprovingBinocular3}).
	
	Now, let $\mathcal{B}$ be an improving binocular of size at most $\tau\cdot\log(|V|)$. Then \[|W(\mathcal{B})|\leq \sum_{e\in E(\mathcal{B})} |W(e)|\leq \tau^2\cdot\log(|V|).\] In particular, the underlying collection of sets contains at most $3\cdot \tau^2\cdot\log(|V|)=t$ elements in total. Thus, there is $f\in\mathcal{F}$ that assigns a different color to each element in $\bigcup W(\mathcal{B})$. As $W(\mathcal{B})$ is independent in the conflict graph $G$ by Def.~\ref{DefImprovingBinocular}~(\ref{ImprovingBinocular3}), the sets corresponding to the vertices in $W(\mathcal{B})$ are pairwise disjoint. Thus, the sets $\mathrm{col}_f(v),v\in W(\mathcal{B})$ are pairwise disjoint. In particular, Def.~\ref{DefColorfulBinocular}~(\ref{ColorfulBinocular3}) and (\ref{ColorfulBinocular4}) are satisfied. By disjointness of the sets $W(e),e\in E_2$ by Def.~\ref{DefImprovingBinocular}~(\ref{ImprovingBinocular1}), we can further infer Def.~\ref{DefColorfulBinocular}~(\ref{ColorfulBinocular1}). Finally, Def.~\ref{DefImprovingBinocular}~(\ref{ImprovingBinocular2}) agrees with Def.~\ref{DefColorfulBinocular}~(\ref{ColorfulBinocular2}).
\end{proof}
Thus, the task of finding an improving binocular of size at most $\tau\cdot\log(|V|)$ reduces to the task of finding a colorful binocular obeying the latter size bound. Note that Def.~\ref{DefColorfulBinocular}~(\ref{ColorfulBinocular3}) only depends on the individual edges, and can be checked in polynomial time for each edge of the search graph. Thus, by, for a fixed function $f\in\mathcal{F}$, throwing away all edges that do not meet Def.~\ref{DefColorfulBinocular}~(\ref{ColorfulBinocular3}), we can restrict ourselves to checking the remaining three conditions. Formally, we introduce the \emph{colorful sub-graph} of $S_\tau(G,w,A)$ and observe Proposition~\ref{PropColorfulSubgraph}.
\begin{definition}
	Let $f\in\mathcal{F}$. The \emph{colorful sub-graph} $S^f_\tau(G,w,A)$ of $S_\tau(G,w,A)$ contains all vertices of $S_\tau(G,w,A)$ and precisely those edges $e$ for which the sets $\mathrm{col}_f(v)$, $v\in W(e)$ are pairwise disjoint.\label{DefColorfulSubgraph}
\end{definition}
\begin{proposition}
	Let $f\in\mathcal{F}$ and let $\mathcal{B}$ be a binocular in $S_\tau(G,w,A)$. The following are equivalent:
	\begin{itemize}
		\item $\mathcal{B}$ is a colorful binocular.
		\item $\mathcal{B}$ is a sub-graph of $S^f_\tau(G,w,A)$ and satisfies properties (\ref{ColorfulBinocular1}), (\ref{ColorfulBinocular2}) and (\ref{ColorfulBinocular4}) from Definition~\ref{DefColorfulBinocular}.
	\end{itemize}\label{PropColorfulSubgraph}
\end{proposition}
Note that Proposition~\ref{PropSearchGraphPolynomial} tells us that we can construct $S^f_\tau(G,w,A)$ in polynomial time.
\begin{proposition}
	We can construct $S^f_\tau(G,w,A)$ in polynomial time. In particular, the number of edges of $S^f_\tau(G,w,A)$ is polynomially bounded.\label{PropColorfulSearchGraphPolynomial}
\end{proposition}
It is a known a fact that every minimal binocular either consists of two cycles joined by a path, or of three paths between two vertices $u$ and $v$. For completeness, we nevertheless prove this statement again in appendix~\ref{appendix:binoculars}.
By employing a standard dynamic programming approach, we can, for $u,v\in A''$ and $C\subseteq \{1,\dots,t\}$, easily check for the existence of a $u$-$v$-walk with the property that the color sets of the edges are pairwise disjoint and their union is $C$. We will then stitch the information about sets of at most three walks together to search for a colorful binocular.

In the following, we first introduce the notion of a \emph{colorful walk} and show how to search for colorful walks efficiently (see Lemma~\ref{LemWALK}). We then employ our knowledge about the structure of minimal binoculars to prove a structural result about colorful minimal binoculars (Lemma~\ref{LemStructureColorfulBinocular}). This enables us to, in polynomial time, check for the existence of a colorful binocular of size at most $\tau\cdot\log(|V|)$, and return one if existent (see Lemma~\ref{LemNiceStructureGivesBinocular} and Lemma~\ref{LemSearchForNiceStructure}).
\begin{definition}[colorful walk]
	Let $f\in\mathcal{F}$ and let $P$ be a walk in $S^f_\tau(G,w,A)$ that does \emph{not contain any loop} and let $U(P)\coloneqq\bigcup_{e\in E(P)} U(e)$ and $W(P)\coloneqq\bigcup_{e\in E(P)} W(e)$. We call $P$ \emph{colorful} if the sets $\mathrm{col}_f(W(e)),e\in E(P)$ are pairwise disjoint.
	
	We define the \emph{color set} of $P$ to be $\mathrm{col}_f(P)\coloneqq\bigcup_{e\in E(P)} \mathrm{col}_f(W(e))$.
\end{definition}
\begin{definition}
	Let $f\in\mathcal{F}$. For 
	\begin{itemize}
		\item $u,v\in A''$,
		\item $1\leq l\leq \tau\cdot\log(|V|)$,
		\item $C\subseteq \{1,\dots,t\}$,
		\item $X\subseteq U\subseteq A$ and $Y\subseteq W\subseteq V\setminus A$ with $\max\{|U|,|W|\}\leq 2\cdot\tau$,
	\end{itemize} we define the boolean values $\mathrm{WALK}_f(u,v,C,U,W,X,Y,l)$ by setting
	\[\mathrm{WALK}_f(u,v,C,U,W,X,Y,l)=\begin{cases}
	\mathrm{true} &,\substack{\text{if there exists a colorful $u$-$v$-walk $P$ of length $l$ with $\mathrm{col}_f(P)=C$}\\\text{ s.t.\ $U(P)\cap U=X$ and $W(P)\cap W=Y$}}\\
	\mathrm{false} &, \text{otherwise}
	\end{cases}.\]\label{DefWalkValues}
\end{definition}
The sets $U$ and $W$ are used to represent $\bigcup_{e\in E_1(\mathcal{B})}U(e)$ and $\bigcup_{e\in E_1(\mathcal{B})}W(e)$, respectively. We will see later that any minimal binocular contains at most two loops, which yields the size bound of $2\cdot\tau$. $X$ and $Y$ allow us to compute $\bigcup_{e\in E_1(\mathcal{B})}U(e)\cap \bigcup_{e\in E_2(\mathcal{B})}U(e)$ and $\bigcup_{e\in E_1(\mathcal{B})}W(e)\cap \bigcup_{e\in E_2(\mathcal{B})}W(e)$.
\begin{lemma}
	Le $f\in\mathcal{F}$. We can, in polynomial time, compute all of the values \\$\mathrm{WALK}_f(u,v,C,U,W,X,Y,l)$ for tuples $(u,v,C,U,W,X,Y,l)$ as in Def.~\ref{DefWalkValues}, and store, for each tuple $(u,v,C,U,W,X,Y,l)$ where $\mathrm{WALK}_f(u,v,C,U,W,X,Y,l)=\mathrm{true}$, a colorful $u$-$v$-walk $P$ of length $l$ with $\mathrm{col}_f(P)=C$, $U(P)\cap U=X$ and $W(P)\cap W=Y$.\label{LemWALK}
\end{lemma}
\begin{proof}
	We first observe that the number of values we need to compute is polynomially bounded because the number of possibilities for $C$ equals $2^t=2^{3\cdot\tau^2\cdot\log(|V|)}=|V|^{3\cdot\tau^2}$, and the number of possible choices for $U$, $W$, $X$ and $Y$ is in $|V|^{\mathcal{O}(\tau)}$ each.\\
	 We employ dynamic programming to iteratively compute the values $\mathrm{WALK}_f(u,v,C,U,W,X,Y,l)$ in order of increasing $l$. As the total number of values we need to compute is polynomially bounded, we only need to show how to compute $\mathrm{WALK}_f(u,v,C,U,W,X,Y,l)$ in polynomial time, provided we already know all of the values $\mathrm{WALK}_f(u',v',C',U',W',X',Y',l')$ for $l'<l$.
	
	We first consider the base case $l=0$. We have $\mathrm{WALK}_f(u,v,C,U,W,X,Y,0)=\mathrm{true}$ if and only if $u=v$, $C=\emptyset$, $X=\emptyset$ and $Y=\emptyset$. We can store the empty walk to witness this.
	
	Now, pick a tuple $(u,v,C,U,W,X,Y,l)$ as in Def.~\ref{DefWalkValues} with $l>0$. Then \[\mathrm{WALK}_f(u,v,C,U,W,X,Y,l)=\mathrm{true}\] if and only if there exist $C'\subseteq C$, $X'\subseteq X$, $Y'\subseteq Y$ and an incident non-loop edge $e=\{v,x\}$ of $v$ such that $\mathrm{col}_f(e)=C\setminus C'$, $U(e)\cap U=X\setminus X'$, $W(e)\cap W=Y\setminus Y'$ and \[\mathrm{WALK}(u,x,C',U,W,X',Y',l-1)=\mathrm{true}.\] As there are $2^{|C|}\leq 2^t=|V|^{3\cdot\tau^2}$ many choices for $C'$ and at most $2^{2\cdot\tau}$ choices for each of $X'$ and $Y'$, this can be checked in polynomial time. Moreover, if there exist $x$, $e$, $C'$, $X'$ and $Y'$ for which the above conditions are satisfied, we can obtain a colorful $u$-$v$-walk $P$ of length $l$ with color set $C$, $U(P)\cap U=X$ and $W(P)\cap W=Y$ by, for the first such choice of $x$, $e$, $C'$, $X'$ and $Y'$ we encounter, taking the walk stored for $(u,x,C',U, W, X', Y', l-1)$ and appending $e$. Note that the disjointness of color sets ensures that we indeed find a walk and not an edge sequence containing some edge multiple times.
\end{proof}
\begin{lemma}
Assume that $S^f_\tau(G,w,A)$ contains a colorful minimal binocular of size at most \mbox{$\tau\cdot\log(|V|)$}. Then there exists $L$ consisting of at most two loops in $S^f_\tau(G,w,A)$ and a set $F$ consisting of two-endpoint edges in $S^f_\tau(G,w,A)$ with the following properties: Let $C\coloneqq\mathrm{col}_f(\bigcup_{e\in F} W(e))$, let $X\coloneqq\bigcup_{l\in L} U(l)\cap \bigcup_{e\in F} U(e)$ and let $Y\coloneqq\bigcup_{l\in L} W(l)\cap\bigcup_{e\in F} W(e)$.
\begin{enumerate}[(i)]
	\item \label{StructureLemma1} $C$ and $\mathrm{col}_f(\bigcup_{l\in L} W(l)\setminus Y)$ are disjoint.
	\item \label{StructureLemma2} We have $w\left(\bigcup_{l\in L} W(l)\setminus Y\right)\geq w\left(\bigcup_{l\in L} U(l)\setminus X\right)+2\cdot |L|.$
	\item \label{StructureLemma3}The sets $\mathrm{col}_f(v),v\in \bigcup_{l\in L} W(l)\setminus Y$ are pairwise disjoint.
	\item \label{StructureLemma4}One of the following cases applies:
	\begin{itemize}
		\item $L=\emptyset$ and there exist $u,v\in A''$ (not necessarily distinct) such that $F$ is the union of the edge sets of a colorful $u$-$u$-walk of length at least $2$ and at most $\tau\cdot\log(|V|)$, a colorful $v$-$v$-walk of length at least $2$ and at most $\tau\cdot\log(|V|)$ and a colorful $u$-$v$-walk of length at most $\tau\cdot\log(|V|)$, such that the color sets of these three walks are pairwise disjoint.
		\item $L=\emptyset$ and there exist $u\neq v\in A''$ such that $F$ is the union of the edge sets of three colorful $u$-$v$-walks with pairwise disjoint color sets.
		\item There is $u\in A''$ such that $L$ consists of a single loop incident to $u$, and there is $v\in A''$ (not necessarily distinct from $u$) such that $F$ is the union of the edge sets of a colorful $u$-$v$-walk of length at most $\tau\cdot\log(|V|)$ and a colorful $v$-$v$-walk of length at least $2$ and at most $\tau\cdot\log(|V|)$, such that the color sets of these two walks are pairwise disjoint.
		\item There are $u, v\in A''$ (not necessarily distinct) such that $L$ consists of a loop incident to $u$ and a loop incident to $v$ and such that $F$ is the edge set of a colorful $u$-$v$ walk of length at most $\tau\cdot\log(|V|)$. 
	\end{itemize}
\end{enumerate}\label{LemStructureColorfulBinocular}
\end{lemma}
\begin{proof}
Let $\mathcal{B}$ be a colorful minimal binocular. Let $L\coloneqq E_1(\mathcal{B})$ and $F\coloneqq E_2(\mathcal{B})$. By Lemma~\ref{LemStructureBinoculars}, we have $|L|\leq 2$. Moreover, the fact that $\mathcal{B}$ is colorful, together with Lemma~\ref{LemStructureBinoculars}, yields (\ref{StructureLemma4}). Def.~\ref{DefColorfulBinocular}~(\ref{ColorfulBinocular1}) yields (\ref{StructureLemma1}). Def.~\ref{DefColorfulBinocular}~(\ref{ColorfulBinocular2}) implies (\ref{StructureLemma2}). Def.~\ref{DefColorfulBinocular}~(\ref{ColorfulBinocular4}) agrees with (\ref{StructureLemma3}).
\end{proof}
\begin{lemma}
For $L$ and $F$ as described in Lemma~\ref{LemStructureColorfulBinocular}, $L\cup F$ is the edge set of a colorful binocular.\label{LemNiceStructureGivesBinocular}
\end{lemma}
\begin{proof}
Lemma~\ref{LemStructureColorfulBinocular}~(\ref{StructureLemma4}) tells us that $L\cup F$ is the edge set of a binocular. The definition of a colorful walk, Lemma~\ref{LemStructureColorfulBinocular}~(\ref{StructureLemma4}) and Lemma~\ref{LemStructureColorfulBinocular}~(\ref{StructureLemma1}) yield Def.~\ref{DefColorfulBinocular}~(\ref{ColorfulBinocular1}). Lemma~\ref{LemStructureColorfulBinocular}~(\ref{StructureLemma2}) and (\ref{StructureLemma3}) directly imply Def.~\ref{DefColorfulBinocular}~(\ref{ColorfulBinocular2}) and (\ref{ColorfulBinocular4}). Prop.~\ref{PropColorfulSubgraph} concludes the proof.
\end{proof}
\begin{lemma}
We can, in polynomial time, decide whether sets $L$ and $F$ as described in Lemma~\ref{LemStructureColorfulBinocular} exist, and, if this is the case, return such sets $L$ and $F$.\label{LemSearchForNiceStructure}
\end{lemma}
\begin{proof}
By Prop.~\ref{PropColorfulSearchGraphPolynomial}, there are only polynomially many choices for $L$ (which we can loop over in polynomial time). For every fixed choice of $L$, we let $U\coloneqq\bigcup_{l\in L} U(l)$ and $W\coloneqq\bigcup_{l\in L} W(l)$. Then $\max\{|U|,|W|\}\leq 2\cdot\tau$. We can further determine all triples $(C,X,Y)$ with $C\subseteq \{1,\dots,t\}$, $X\subseteq U$ and $Y\subseteq W$ for which (\ref{StructureLemma1}), (\ref{StructureLemma2}) and (\ref{StructureLemma3}) hold in polynomial time because the number of such triples is polynomially bounded and we can check each of the three conditions in polynomial time. For each such triple, we can check whether $F$ as in Lemma~\ref{LemStructureColorfulBinocular}~(\ref{StructureLemma4}) with $C=\mathrm{col}_f(\bigcup_{e\in F} W(e))$, $X=\bigcup_{l\in L} U(l)\cap \bigcup_{e\in F} U(e)$ and $Y=\bigcup_{l\in L} W(l)\cap\bigcup_{e\in F} W(e)$  exists by combining the information from at most three of the $\mathrm{WALK}$-values we have computed. Moreover, if such a set $F$ exist, we can obtain one by uniting the edge sets of the stored walks.
\end{proof}
\section{\texorpdfstring{A guarantee of $\nicefrac{4}{3}$ for the hereditary $2$-$3$-Set Packing problem}{A guarantee of 4/3 for the hereditary 2-3-Set Packing problem}\label{Sec:Hereditary}}
In this section, we show that Algorithm~\ref{OverallAlgorithm} with $\tau\geq 10$ yields a $\frac{4}{3}$-approximation for the hereditary $2$-$3$-Set Packing problem. More precisely, we prove Theorem~\ref{TheoMainHereditary}, which tells us that every solution that is locally optimum with respect to improvements of size at most $10$ is a $\frac{4}{3}$-approximation of the optimum. In particular, this implies that Algorithm~\ref{OverallAlgorithm} is a $\frac{4}{3}$-approximation for the hereditary $2$-$3$-Set Packing problem, even if we omit lines~\ref{LineStartCheckForBinocular}-\ref{LineEndCheckForBinocular}. Note that the arguments in section~\ref{SecPolyTime} show that both variants run in polynomial time.
\begin{theorem}
Let $(\mathcal{S},w)$ be an instance of the hereditary $2$-$3$-Set Packing problem and let $A\subseteq \mathcal{S}$ be a feasible solution such that there is no local improvement (in the sense of Def.~\ref{DefLocalImprovement} with $G=G_{\mathcal{S}}$) of size at most $10$. Let further $B\subseteq \mathcal{S}$ be an optimum solution. Then $w(B)\leq \frac{4}{3}\cdot w(A)$.\label{TheoMainHereditary}
\end{theorem}
The remainder of this section is dedicated to the proof of Theorem~\ref{TheoMainHereditary}. Let $\mathcal{S}$, $w$, $A$ and $B$ be as in the statement of the theorem. Our goal is to distribute the weights of the sets in $B$ among the sets in $A$ they intersect in such a way that no set in $A$ receives more than $\frac{4}{3}$ times its own weight. We remark that each set in $B$ must intersect at least one set in $A$ because otherwise, it would constitute a local improvement of size $1$. 

For the analysis, we can assume without loss of generality that $A\cap B=\emptyset$. If not, we can just restrict $\mathcal{S}$ and $w$ to the sets in $A\Delta B=A\setminus B\cup B\setminus A$ and their two-element subsets and then prove that $w(B\setminus A)\leq \frac{4}{3}\cdot w(A\setminus B)$.

In order to present our weight distribution, it is more convenient to work with a multi-graph version of the conflict graph. More precisely, for two sets $s_1,s_2\in\mathcal{S}$ that intersect, we add $|s_1\cap s_2|$ many parallel edges instead of just one edge connecting them. We denote the resulting graph by $G$. See Figure~\ref{FigConflictGraph} for an illustration. In the following, we will simultaneously interpret sets from $A$ and $B$ as the corresponding vertices in $G$ and talk about their degree, their incident edges and their neighbors.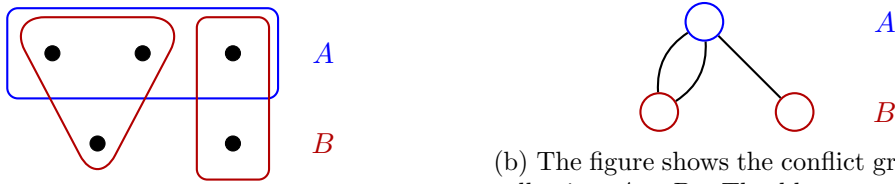
\begin{figure}
	\begin{subfigure}{0.45\textwidth}
		\centering
		\begin{tikzpicture}[scale = 1.2,elem/.style = {circle, draw = black, fill, inner sep = 0.5mm, minimum size = 2mm}]
		\node[elem] (A) at (0,1) {};
		\node[elem] (B) at (1,1) {};
		\node[elem] (C) at (2,1) {};
		\node[elem] (D) at (0.5,0) {};
		\node[elem] (E) at (2,0) {};
		\draw[blue, thick, rounded corners] (-0.5,0.5) rectangle (2.5,1.5);
		\draw[red!70!black, thick, rounded corners] (1.6,-0.4) rectangle (2.4,1.4);
		\draw[rounded corners = 5mm, red!70!black, thick] (-0.5,1.4)--(1.5,1.4)--(0.5,-0.5)--cycle;
		\node at (3,1) {\color{blue}$A$};
		\node at (3,0) {\color{red!70!black}$B$};
		\end{tikzpicture}
		\caption{The figure displays two collections $A$ (blue) and $B$ (red) consisting of pairwise disjoint sets of cardinality $2$ or $3$. Black dots represent set elements. $A$ contains the set drawn horizontally at the top, whereas $B$ contains the other two sets.}
	\end{subfigure}
	\quad
	\begin{subfigure}{0.45\textwidth}
		\centering
		\begin{tikzpicture}[scale = 1.2,anode/.style = {circle, draw = blue, thick, fill = none, inner sep = 0mm, minimum size = 5mm},bnode/.style = {circle, draw = red!70!black, thick, fill = none, inner sep = 0mm, minimum size = 5mm}]
		\node[anode] (A2) at (1,1){};
		\node[bnode] (B2) at (0.5,0){};
		\node[bnode] (B3) at (2,0){};
		\node at (3,1) {\color{blue}$A$};
		\node at (3,0) {\color{red!70!black}$B$};
		\draw[thick] (A2) to[bend left = 30] (B2);
		\draw[thick] (A2) to[bend right = 30] (B2);
		\draw[thick] (A2)--(B3);
		\end{tikzpicture}
		\caption{The figure shows the conflict graph of the set collection $A\cup B$. The blue vertex on the top represents the set contained in $A$. Two parallel edges connect it to the vertex representing the set of cardinality $3$ from $B$, a single edge connects it to the vertex representing the set of cardinality $2$ from $B$.}
	\end{subfigure}
	\caption{The multi-graph version of the conflict graph.\label{FigConflictGraph}}
\end{figure}

Our weight distribution proceeds in two steps. Let $B_1$ consist of all sets $v\in B$ with exactly one neighbor in $A$ and let $B_2$ consist of those $v\in B$ with $w(v)=2$ and exactly two incident edges, with the additional property that they connect to two distinct sets from $A$. In the first step, each $v\in B_1$ sends its full weight to its neighbor in $A$. Moreover, each $v\in B_2$ sends half of its weight (i.e.\ $1$) along each of its edges. See Figure~\ref{FigStep1} for an illustration.
\begin{figure}[t]
	\begin{subfigure}[t]{0.45\textwidth}
		\centering
		\begin{tikzpicture}[scale = 1.5,anode/.style = {circle, draw = blue, thick, fill = none, inner sep = 0mm, minimum size = 5mm},bnode/.style = {circle, draw = red!70!black, thick, fill = none, inner sep = 0mm, minimum size = 5mm}]
		\node[anode] (A1) at (0,1){};
		\node[anode] (A2) at (1,1){};
		\node[bnode] (B1) at (0,0){$1$};
		\node[bnode] (B2) at (1,0){$2$};
		\node at (2,1) {\color{blue}$A$};
		\node at (2,0) {\color{red!70!black}$B$};
		\draw[thick] (A1) to node[midway, left] {$1$}(B1);
		\draw[thick, dashed] (A1) to[bend left = 30] (B1);
		\draw[thick, dashed] (A2) to[bend left = 30] (B2);
		\draw[thick] (A2) to node[midway, right=10pt] {$2$}(B2);
		\end{tikzpicture}
		\caption*{Sets in $B_1$ send their whole weight to their unique neighbor in $A$ (to which they may be connected via multiple edges).}
	\end{subfigure}
\quad
	\begin{subfigure}[t]{0.45\textwidth}
		\centering
		\begin{tikzpicture}[scale = 1.5,anode/.style = {circle, draw = blue, thick, fill = none, inner sep = 0mm, minimum size = 5mm},bnode/.style = {circle, draw = red!70!black, thick, fill = none, inner sep = 0mm, minimum size = 5mm}]
		\node[anode] (A1) at (0,1){};
		\node[anode] (A2) at (1,1){};
		\node[bnode] (B1) at (0.5,0){$2$};
		\node at (2,1) {\color{blue}$A$};
		\node at (2,0) {\color{red!70!black}$B$};
		\draw[thick] (A1) to node[midway, left=10pt] {$1$}(B1);
		\draw[thick] (A2) to node[midway, right=10pt] {$1$}(B1);
		\end{tikzpicture}
		\caption*{Sets in $B_2$ send one unit of weight to each of their neighbors in $A$.}
	\end{subfigure}
	\caption{First step of the weight distribution.\label{FigStep1}}
\end{figure}

We first prove Lemma~\ref{LemFirstStep}, which tells us that we can represent the total amount of weight a collection $U\subseteq A$ receives in the first step as the weight of a set collection $X$ with $N(X,A)\subseteq U$. In particular, this implies that no set can receive more than its own weight in the first step, see Corollary~\ref{CorNoMoreThanOwnWeight}. We further show that we can obtain $X$ by picking a sub-collection of $B_1\cup B_2$ and replacing some of the contained sets of cardinality $3$ by a subset of size $2$. This will allow us to combine $X$ with sub-collections of $B\setminus(B_1\cup B_2)$ to construct local improvements.
\begin{lemma}
	For each $U\subseteq A$, there is a collection $X\subseteq \mathcal{S}$ with the following properties:
	\begin{enumerate}[(i)]
		\item \label{FirstStepProp1} $N(X,A)\subseteq U$ and every set in $X$ intersects at least one set in $U$. In particular, $|X|\leq 3\cdot |U|$.
		\item \label{FirstStepProp2} There is an injective map $f:X\rightarrow B_1\cup B_2$ such that $x\subseteq f(x)$ for all $x\in X$.
		\item \label{FirstStepProp3} $w(X)$ equals the total amount that $U$ receives in the first step.
		
	\end{enumerate} \label{LemFirstStep}
\end{lemma}
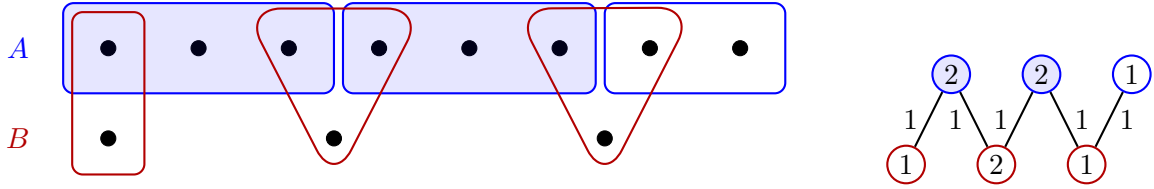
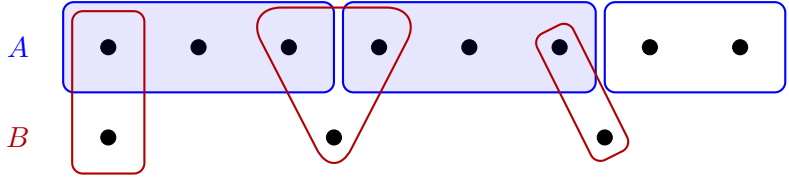
\begin{figure}[t]
	\begin{subfigure}[t]{0.65\textwidth}
		\centering
		\begin{tikzpicture}[scale = 1.2,elem/.style = {circle, draw = black, fill, inner sep = 0.5mm, minimum size = 2mm}]
		\node[elem] (A) at (0,1) {};
		\node[elem] (B) at (1,1) {};
		\node[elem] (C) at (2,1) {};
		\node[elem] at (3,1) {};
		\node[elem] at (4,1) {};
		\node[elem] (I) at (-1,1) {};
		\node[elem] (J) at (-2,1) {};
		\node[elem] (H) at (-3,1) {};
		\node[elem] (K) at (-3,0) {};
		\node[elem] (D) at (-0.5,0) {};
		\node[elem] (E) at (2.5,0) {};
		\draw[blue, thick, rounded corners, fill = blue, fill opacity = 0.1] (-0.4,0.5) rectangle (2.4,1.5);
		\draw[blue, thick, rounded corners, fill = blue, fill opacity = 0.1] (-3.5,0.5) rectangle (-0.5,1.5);
		\draw[blue, thick, rounded corners] (2.5,0.5) rectangle (4.5,1.5);
		\draw[rounded corners = 5mm, red!70!black, thick] (-1.5,1.44)--(0.5,1.44)--(-0.5,-0.5)--cycle;
		\draw[rounded corners = 5mm, red!70!black, thick] (1.5,1.44)--(3.5,1.45)--(2.5,-0.5)--cycle;
		\draw[rounded corners, red!70!black, thick] (-3.4,-0.4) rectangle (-2.6,1.4);
		\node at (-4,1) {\color{blue}$A$};
		\node at (-4,0) {\color{red!70!black}$B$};
		\end{tikzpicture}
		\caption{The left red set is contained in $B_1$ and sends its whole weight to the unique set from $A$ it intersects. It is added to $Y$ in the proof of Lemma~\ref{LemFirstStep}. The two triangular red sets are contained in $B_2$. The left one is contained in $Y$ as well because it only intersects sets in $A$ that are contained in $U$. In contrast, the right triangular set is contained in $Z$ because it also intersects a set in $A\setminus U$. \label{SubFigLemSetConfiguration} }
	\end{subfigure}
	\begin{subfigure}[t]{0.34\textwidth}
		\centering
		\begin{tikzpicture}[xscale = 1.2,yscale = 1.2,anode/.style = {circle, draw = blue, thick, fill = none, inner sep = 0mm, minimum size = 5mm},bnode/.style = {circle, draw = red!70!black, thick, fill = none, inner sep = 0mm, minimum size = 5mm}]
		\node[anode,fill = blue!10!white] (A2) at (-1.5,1){$2$};
		\node[anode, fill = blue!10!white] (A3) at (-0.5,1){$2$};
		\node[anode] (A4) at (0.5,1){$1$};
		\node[bnode] (B1) at (-2,0){$1$};
		\node[bnode] (B2) at (-1,0){$2$};
		\node[bnode] (B3) at (0,0){$1$};
			\draw[thick] (A2) to node[midway, left] {$1$}(B1);
		\draw[thick] (A2) to node[midway, left]{$1$} (B2);
		\draw[thick] (A3) to node[midway, left] {$1$}(B2);
		\draw[thick] (A3) to node[midway, right] {$1$}(B3);
		\draw[thick] (A4) to node[midway, right] {$1$}(B3);
		\end{tikzpicture}
		\caption{Part of the conflict graph and the weight distribution corresponding to the set configuration in \ref{SubFigLemSetConfiguration}.\label{SubFigLemConflictGraph}}
	\end{subfigure}
	\begin{subfigure}[t]{0.65\textwidth}
		\centering
		\begin{tikzpicture}[scale = 1.2,elem/.style = {circle, draw = black, fill, inner sep = 0.5mm, minimum size = 2mm}]
			\node[elem] (A) at (0,1) {};
		\node[elem] (B) at (1,1) {};
		\node[elem] (C) at (2,1) {};
		\node[elem] at (3,1) {};
		\node[elem] at (4,1) {};
		\node[elem] (I) at (-1,1) {};
		\node[elem] (J) at (-2,1) {};
		\node[elem] (H) at (-3,1) {};
		\node[elem] (K) at (-3,0) {};
		\node[elem] (D) at (-0.5,0) {};
		\node[elem] (E) at (2.5,0) {};
		\draw[blue, thick, rounded corners, fill = blue, fill opacity = 0.1] (-0.4,0.5) rectangle (2.4,1.5);
		\draw[blue, thick, rounded corners, fill = blue, fill opacity = 0.1] (-3.5,0.5) rectangle (-0.5,1.5);
		\draw[blue, thick, rounded corners] (2.5,0.5) rectangle (4.5,1.5);
		\draw[rounded corners = 5mm, red!70!black, thick] (-1.5,1.44)--(0.5,1.44)--(-0.5,-0.5)--cycle;
		\draw[rounded corners, red!70!black, thick] (-3.4,-0.4) rectangle (-2.6,1.4);
		\node at (-4,1) {\color{blue}$A$};
		\node at (-4,0) {\color{red!70!black}$B$};
		\draw[rounded corners, red!70!black, thick] (1.9-0.2,1.2-0.1)--(1.9+0.2,1.2+0.1)--(2.6+0.2,-0.2+0.1)--(2.6-0.2,-0.2-0.1)--cycle;
		\end{tikzpicture}
		\caption{The set collection $X$ (red) we construct in the proof of Lemma~\ref{LemFirstStep} contains all sets from $Y$. For the right triangular set, which is the unique set in $Z$ in our example, we remove the element in which it intersects a set from $A\setminus U$. Then, we add the resulting set of cardinality $2$ to $X$.\label{SubFigLemX}}
	\end{subfigure}
	\caption{Illustration of the construction in the proof of Lemma~\ref{LemFirstStep}. Figure~\ref{SubFigLemSetConfiguration} shows a collection $U\subseteq A$ of sets (blue, filled, horizontal), the collection $N(U,B_1\cup B_2)$ (red) of sets the sets in $U$ receive weight from in the first step, and further sets from $A$ (blue, not filled, horizontal) the sets in $N(U,B_1\cup B_2)$ send weight to. Figure~\ref{SubFigLemConflictGraph} displays the weight distribution from $N(U,B_1\cup B_2)$ to $A$. Figure~\ref{SubFigLemX} illustrates the construction of the set collection $X$.}\label{FigProofLemFirstStep}
\end{figure}
\begin{proof}
	Let $Y\coloneqq\{v\in B_1\cup B_2: N(v,A)\subseteq U\}$ and $Z\coloneqq\{v\in B_2:|N(v,A)\cap U|=1\}$. Then $Y\dot{\cup}Z\subseteq B$ is a disjoint sub-collection of sets. Obtain $Z'$ from $Z$, by, for each $v\in Z$, removing the element in which it intersects the unique set in $N(v,A)\setminus U$. Let $X\coloneqq Y\dot{\cup} Z'$. Then $N(X,A)\subseteq U$ and every set in $X$ intersects at least one set in $U$. Our construction further implies that (\ref{FirstStepProp2}) holds because $Y\dot{\cup} Z\subseteq B_1\cup B_2$. In particular, $X$ consists of pairwise disjoint sets each intersecting a set in $U$, which yields $|X|\leq |\bigcup U|\leq 3\cdot |U|$. Finally, $w(X)=w(Y)+|Z'|=w(Y\cap B_1)+2\cdot |Y\cap B_2|+|Z|$, which equals the amount of weight that $U$ receives in the first step. See Figure~\ref{FigProofLemFirstStep} for an illustration of the proof.
\end{proof}
\begin{corollary}
	No set in $A$ receives more than its own weight in the first step.\label{CorNoMoreThanOwnWeight}
\end{corollary}
\begin{proof}
	Assume towards a contradiction that $u\in A$ receives more than $w(u)$ in the first step. Apply Lemma~\ref{LemFirstStep} with $U=\{u\}$ to obtain a collection $X\subseteq \mathcal{S}$ subject to (\ref{FirstStepProp1})-(\ref{FirstStepProp3}). Then $X$ is a collection of pairwise disjoint sets with $|X|\leq 3$ and such that $w(X)>w(u)=w(N(X,A))$. Thus, $X$ constitutes a local improvement of size at most $3$, contradicting the termination criterion of our algorithm.
\end{proof}\begin{definition}
	Let $C$ consist of those sets from $A$ that receive exactly their own weight in the first step (see Figure~\ref{FigC} for an illustration).
\end{definition}
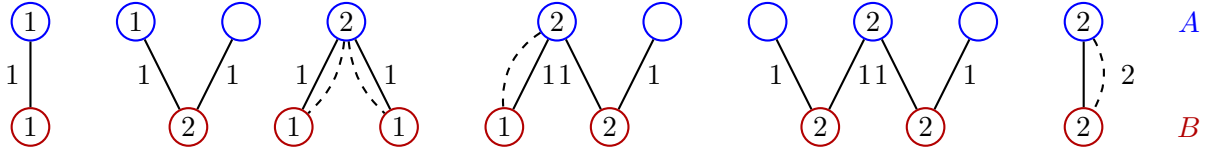
\begin{figure}
	\centering
	\begin{tikzpicture}[scale = 1.4,anode/.style = {circle, draw = blue, thick, fill = none, inner sep = 0mm, minimum size = 5mm},bnode/.style = {circle, draw = red!70!black, thick, fill = none, inner sep = 0mm, minimum size = 5mm}]
	\node[anode] (A1) at (0,1){$1$};
	\node[anode] (A2) at (10,1){$2$};
	\node[bnode] (B1) at (0,0){$1$};
	\node[bnode] (B2) at (10,0){$2$};
	\draw[thick] (A1) to node[midway, left] {$1$}(B1);
	\draw[thick, dashed] (A2) to[bend left = 30] (B2);
	\draw[thick] (A2) to node[midway, right=10pt] {$2$}(B2);	
	\node[anode] (A3) at (1,1){$1$};
	\node[anode] (A4) at (2,1){};
	\node[bnode] (B3) at (1.5,0){$2$};
	\draw[thick] (A3) to node[midway, left] {$1$}(B3);
	\draw[thick] (A4) to node[midway, right] {$1$}(B3);
	\begin{scope}[shift = {(1,0)}]
	\node[anode] (A6) at (4,1){$2$};
	\node[anode] (A7) at (5,1){};
	\node[bnode] (B4) at (3.5,0){$1$};
	\node[bnode] (B5) at (4.5,0){$2$};
	\draw[thick, dashed] (A6) to [bend right = 30] (B4);
	\draw[thick] (A6) to node[midway, right] {$1$}(B4);
	\draw[thick] (A6) to node[midway, left] {$1$}(B5);
	\draw[thick] (A7) to node[midway, right] {$1$}(B5);
	\end{scope}
	\begin{scope}[shift = {(4,0)}]
	\node[anode] (A5) at (3,1){};
	\node[anode] (A6) at (4,1){$2$};
	\node[anode] (A7) at (5,1){};
	\node[bnode] (B4) at (3.5,0){$2$};
	\node[bnode] (B5) at (4.5,0){$2$};
	\draw[thick] (A5) to node[midway, left] {$1$}(B4);
	\draw[thick] (A6) to node[midway, right] {$1$}(B4);
	\draw[thick] (A6) to node[midway, left] {$1$}(B5);
	\draw[thick] (A7) to node[midway, right] {$1$}(B5);
	\end{scope}
	\node[anode] (A6) at (3,1){$2$};
	\node[bnode] (B4) at (2.5,0){$1$};
	\node[bnode] (B5) at (3.5,0){$1$};
	\draw[thick, dashed] (A6) to [bend left = 20](B4);
	\draw[thick, dashed] (A6) to [bend right = 20](B5);
	\draw[thick] (A6) to node[midway, left] {$1$}(B4);
	\draw[thick] (A6) to node[midway, right] {$1$}(B5);
	\node at (11,1) {\color{blue}$A$};
	\node at (11,0) {\color{red!70!black}$B$};
	\end{tikzpicture}
	\caption{There are six different types of sets in $C$, which are displayed from left to right: Sets of weight $1$ receiving one unit of weight from a set of weight $1$ in $B_1$, sets of weight $1$ receiving one unit of weight from a set in $B_2$, sets of weight $2$ receiving one unit of weight from each of two sets in $B_1$ of weight $1$, sets of weight $2$ receiving one unit of weight from a set in $B_1$ and one from a set in $B_2$, sets of weight $2$ receiving one unit of weight from each of two sets in $B_2$, and sets of weight $2$ receiving the full weight of their neighbor in $B_1$.\label{FigC}}
\end{figure}
The intuitive idea behind our analysis is that we now remove the sets in $C$ from our current solution $A$ and the sets in $B_1\cup B_2$ from our optimum solution $B$. If we can find a local improvement in the remaining instance, we will use Lemma~\ref{LemFirstStep} to transform it into a local improvement in the original instance, leading to a contradiction. Lemma~\ref{LemLocalImprovementWithoutC} formalizes this idea. But under the assumption that no local improvement in the remaining instance exists, we can design the second step of the weight distribution in such a way that overall, no set in $A$ receives more than $\frac{4}{3}$ times its own weight.
\begin{lemma}
	There is no collection $Y$ of sets with the following properties:
	\begin{enumerate}[(i)]
		\item  \label{ImprovementWithoutC1} $|Y|+2\cdot |N(Y,C)|\leq 10$
		\item \label{ImprovementWithoutC2}$w(Y) > w(N(Y,A\setminus C))$
		\item \label{ImprovementWithoutC3}There is an injective map $f:Y\rightarrow B\setminus(B_1\cup B_2)$ such that $y\subseteq f(y)$ for all $y\in Y$.
	\end{enumerate}\label{LemLocalImprovementWithoutC}
\end{lemma}
\begin{proof}
	Assume towards a contradiction that such a collection $Y$ would exist and let $U\coloneqq N(Y,C)$. Apply Lemma~\ref{LemFirstStep} to obtain $X$ subject to (\ref{FirstStepProp1})-(\ref{FirstStepProp3}). By Lemma~\ref{LemFirstStep}~(\ref{FirstStepProp1})-(\ref{FirstStepProp2}), property (\ref{ImprovementWithoutC3}) of $Y$ and as each set in $U$ can intersect at most two sets other than the sets contained in $Y$, we get $|X|\leq 2\cdot |U|$. Thus, $Z\coloneqq X\dot{\cup} Y$ contains at most $10$ sets by (\ref{ImprovementWithoutC1}). By Lemma~\ref{LemFirstStep}~(\ref{FirstStepProp2}) and property (\ref{ImprovementWithoutC3}) of $Y$, $X\dot{\cup}Y$ is a disjoint collection of sets. By Lemma~\ref{LemFirstStep}~(\ref{FirstStepProp1}), we know that $N(X,A)\subseteq U$, so \[w(Z)=w(X)+w(Y)\overset{Lem.~\ref{LemFirstStep}}{\underset{(\ref{FirstStepProp3})}{=}} w(U)+w(Y)\overset{(\ref{ImprovementWithoutC2})}{>}\underbrace{w(U)}_{=N(X,A)\cup N(Y,C)}+w(N(Y,A\setminus C))=w(N(Z,A)).\] Hence, $Z$ constitutes a local improvement of size at most $10$, a contradiction to the termination criterion of our algorithm.
\end{proof}
\begin{corollary}
	There is no $v\in B\setminus(B_1\cup B_2)$ with $N(v,A)\subseteq C$.\label{CorNeighborhoodNotContainedinC}
\end{corollary}
\begin{proof}
	Assume towards a contradiction that such a set $v$ would exist and let $Y\coloneqq \{v\}$. As $v$ can intersect at most $3$ sets in $C$, we get $|Y|+2\cdot |N(Y,C)|\leq 7$. Moreover, $w(Y)=w(v)>0=w(N(Y,A\setminus C))$. Thus, $Y$ satisfies Lemma~\ref{LemLocalImprovementWithoutC}~(\ref{ImprovementWithoutC1})-(\ref{ImprovementWithoutC3}), where for the last point, we may just map $v$ to itself. This results in the desired contradiction.
\end{proof}
We now explain how to distribute the weight of the sets in $B\setminus (B_1\cup B_2)$. See Figure~\ref{FigSecondStep} for an illustration.  

First, we consider $v\in B\setminus (B_1\cup B_2)$ with $w(v)=1$. Then $v$ must have two distinct neighbors in $A$ at least one of which is contained in $A\setminus C$ by Corollary~\ref{CorNeighborhoodNotContainedinC}. 
\begin{enumerate}[(a)]
	\item \label{SituationA}	If $v$ has a neighbor in $C$, then this neighbor receives $\frac{1}{3}$ and the neighbor in $A\setminus C$ receives $\frac{2}{3}$. \item \label{SituationB} Otherwise, both neighbors in $A\setminus C$ receive $\frac{1}{2}$.
\end{enumerate}

Now, let $v\in B\setminus(B_1\cup B_2)$ with $w(v)=2$. Then $v$ must have degree $3$ to $A$ and at least one neighbor in $A\setminus C$, again by Corollary~\ref{CorNeighborhoodNotContainedinC}.
\begin{enumerate}[(a),resume]
	\item \label{SituationC} If $v$ has degree $2$ to $C$, then $v$ sends $\frac{1}{3}$ along each edge to $C$ and $\frac{4}{3}$ to the neighbor in $A\setminus C$.
	\item \label{SituationD}If $v$ has degree $1$ to $C$, $v$ sends $1$ along each edge to a vertex in $A\setminus C$ of weight $2$, $\frac{2}{3}$ along each edge to a vertex in $A\setminus C$ of weight $1$, and the remaining amount to the neighbor in $C$.
	\item \label{SituationE}If all three incident edges of $v$ connect to $A\setminus C$, then $v$ sends $\frac{2}{3}$ along each of these edges.
\end{enumerate} 
\begin{figure}
	\begin{subfigure}{0.19\textwidth}
		\centering
		\begin{tikzpicture}[scale = 1.4,anode/.style = {circle, draw = blue, thick, fill = none, inner sep = 0mm, minimum size = 5mm},bnode/.style = {circle, draw = red!70!black, thick, fill = none, inner sep = 0mm, minimum size = 5mm}]
		\node[anode, dashed] (C) at (0,1){};
		\node[anode] (A) at (1,1){};
		\node[bnode] (B) at (0.5,0){$1$};
		\draw[thick] (B) to node[midway, left]{$\frac{1}{3}$} (C);
		\draw[thick] (B) to node[midway, right]{$\frac{2}{3}$} (A);
		\end{tikzpicture}
		\caption*{\ref{SituationA}}
	\end{subfigure}
	\begin{subfigure}{0.19\textwidth}
		\centering
		\begin{tikzpicture}[scale = 1.4,anode/.style = {circle, draw = blue, thick, fill = none, inner sep = 0mm, minimum size = 5mm},bnode/.style = {circle, draw = red!70!black, thick, fill = none, inner sep = 0mm, minimum size = 5mm}]
		\node[anode] (C) at (0,1){};
		\node[anode] (A) at (1,1){};
		\node[bnode] (B) at (0.5,0){$1$};
		\draw[thick] (B) to node[midway, left]{$\frac{1}{2}$} (C);
		\draw[thick] (B) to node[midway, right]{$\frac{1}{2}$} (A);
		\end{tikzpicture}
		\caption*{\ref{SituationB}}
	\end{subfigure}
	\begin{subfigure}{0.3\textwidth}
		\centering
		\begin{tikzpicture}[scale = 1.4,anode/.style = {circle, draw = blue, thick, fill = none, inner sep = 0mm, minimum size = 5mm},bnode/.style = {circle, draw = red!70!black, thick, fill = none, inner sep = 0mm, minimum size = 5mm}]
		\node[anode, dashed] (C1) at (0,1){};
		\node[anode, dashed] (C2) at (-1,1){};
		\node[anode] (A) at (1,1){};
		\node[bnode] (B) at (0.5,0){$2$};
		\draw[thick] (B) to node[midway, left]{$\frac{1}{3}$} (C1);
		\draw[thick] (B) to node[midway, left, below = 0.5pt]{$\frac{1}{3}$} (C2);
		\draw[thick] (B) to node[midway, right]{$\frac{4}{3}$} (A);
		\end{tikzpicture}
		\caption*{\ref{SituationC}}
	\end{subfigure}
	\begin{subfigure}{0.3\textwidth}
		\centering
		\begin{tikzpicture}[scale = 1.4,anode/.style = {circle, draw = blue, thick, fill = none, inner sep = 0mm, minimum size = 5mm},bnode/.style = {circle, draw = red!70!black, thick, fill = none, inner sep = 0mm, minimum size = 5mm}]
		\node[anode] (A1) at (0,1){$1$};
		\node[anode, dashed] (C) at (-1,1){};
		\node[anode] (A2) at (1,1){$1$};
		\node[bnode] (B) at (0.5,0){$2$};
		\draw[thick] (B) to node[midway, left]{$\frac{2}{3}$} (A1);
		\draw[thick] (B) to node[midway, left, below = 0.5pt]{$\frac{2}{3}$} (C);
		\draw[thick] (B) to node[midway, right]{$\frac{2}{3}$} (A2);
		\end{tikzpicture}
		\caption*{\ref{SituationD}}
	\end{subfigure}
	\begin{subfigure}{0.3\textwidth}
		\centering
		\begin{tikzpicture}[scale = 1.4,anode/.style = {circle, draw = blue, thick, fill = none, inner sep = 0mm, minimum size = 5mm},bnode/.style = {circle, draw = red!70!black, thick, fill = none, inner sep = 0mm, minimum size = 5mm}]
		\node[anode] (A1) at (0,1){$1$};
		\node[anode, dashed] (C) at (-1,1){};
		\node[anode] (A2) at (1,1){$2$};
		\node[bnode] (B) at (0.5,0){$2$};
		\draw[thick] (B) to node[midway, left]{$\frac{2}{3}$} (A1);
		\draw[thick] (B) to node[midway, left, below = 0.5pt]{$\frac{1}{3}$} (C);
		\draw[thick] (B) to node[midway, right]{$1$} (A2);
		\end{tikzpicture}
		\caption*{\ref{SituationD}}
	\end{subfigure}
	\begin{subfigure}{0.3\textwidth}
		\centering
		\begin{tikzpicture}[scale = 1.4,anode/.style = {circle, draw = blue, thick, fill = none, inner sep = 0mm, minimum size = 5mm},bnode/.style = {circle, draw = red!70!black, thick, fill = none, inner sep = 0mm, minimum size = 5mm}]
		\node[anode] (A1) at (0,1){$2$};
		\node[anode, dashed] (C) at (-1,1){};
		\node[anode] (A2) at (1,1){$2$};
		\node[bnode] (B) at (0.5,0){$2$};
		\draw[thick] (B) to node[midway, left]{$1$} (A1);
		\draw[thick] (B) to node[midway, left, below = 0.5pt]{$0$} (C);
		\draw[thick] (B) to node[midway, right]{$1$} (A2);
		\end{tikzpicture}
		\caption*{\ref{SituationD}}
	\end{subfigure}
	\begin{subfigure}{0.3\textwidth}
		\centering
		\begin{tikzpicture}[scale = 1.4,anode/.style = {circle, draw = blue, thick, fill = none, inner sep = 0mm, minimum size = 5mm},bnode/.style = {circle, draw = red!70!black, thick, fill = none, inner sep = 0mm, minimum size = 5mm}]
		\node[anode] (A1) at (0,1){};
		\node[anode] (A3) at (-1,1){};
		\node[anode] (A2) at (1,1){};
		\node[bnode] (B) at (0,0){$2$};
		\draw[thick] (B) to node[midway, left]{$\frac{2}{3}$} (A1);
		\draw[thick] (B) to node[midway, left, below = 0.5pt]{$\frac{2}{3}$} (A3);
		\draw[thick] (B) to node[midway, right, below = 0.5pt]{$\frac{2}{3}$} (A2);
		\end{tikzpicture}
		\caption*{\ref{SituationE}}
	\end{subfigure}
	\caption{Illustration of the second step of the weight distribution. Blue circles in the top row indicate sets from $A$, if they are dashed, the corresponding set is contained in $C$. Red circles in the bottom row indicate sets from $B\setminus (B_1\cup B_2)$. The number within a circle indicates the weight of the corresponding set in case it is relevant. Even though drawn as individual circles, the endpoints in $A$ of the incident edges of set $v\in B\setminus(B_1\cup B_2)$ need not be distinct. For example, in \ref{SituationB}, the sets represented by the blue circles may agree, in which case the corresponding set receives $1$ unit of weight.\label{FigSecondStep}}
\end{figure}
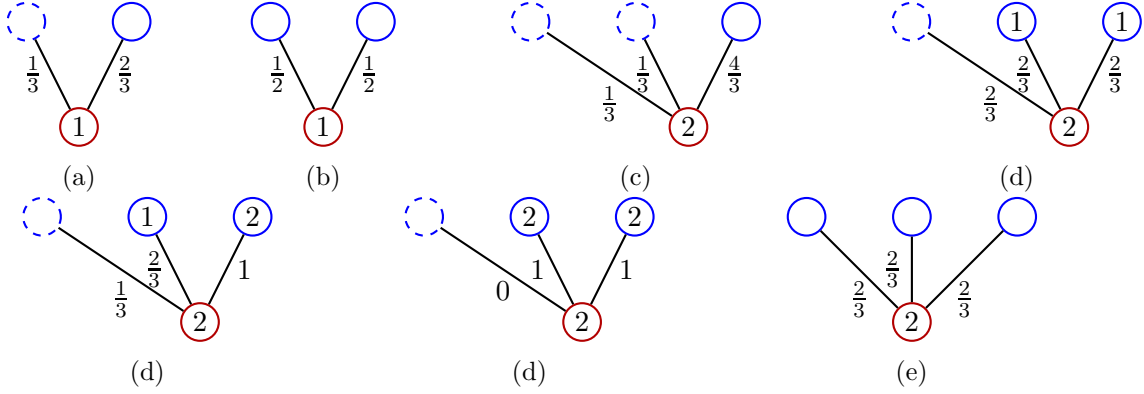
\begin{lemma}
	In situation~\ref{SituationC}, the neighbor of $v$ in $A\setminus C$ has weight $2$.\label{LemSituationC}
\end{lemma} 
\begin{proof}
	Assume towards a contradiction that the neighbor $u$ of $v$ in $A\setminus C$ has weight $1$ and let $Y\coloneqq \{v\}$. Then $|N(Y,C)|\leq 2$, so $|Y|+2\cdot|N(Y,C)|\leq 5$. Moreover, $w(Y)=2>1=w(N(Y,A\setminus C))$. Thus, $Y$ satisfies Lemma~\ref{LemLocalImprovementWithoutC}~(\ref{ImprovementWithoutC1})-(\ref{ImprovementWithoutC3}), where for the last point, we may just map $v$ to itself. This results in the desired contradiction.
\end{proof}
\begin{lemma}
	Let $u\in A\setminus C$ with $w(u)=2$, let $W=N(u,B_1\cup B_2)$, let $R$ consist of all $v\in N(u,B\setminus(B_1\cup B_2))$ to which situation \ref{SituationC} applies, let $Z_1$ consist of all $v\in N(u,B\setminus(B_1\cup B_2))$ to which situation \ref{SituationD} applies that are connected to $u$ by a single edge, and let $Z_2$ consist of all $v\in N(u,B\setminus(B_1\cup B_2))$ to which situation \ref{SituationD} applies that are connected to $u$ by two parallel edges.
	Then $|W|+2|R|+|Z_1|+2 |Z_2|\leq 2$.\label{LemBoundHighCharges}
\end{lemma}
\begin{proof}
	Assume towards a contradiction that $|W|+2|R|+|Z_1|+2 |Z_2|\geq 3$. Note that $|W|\leq 1$ because $u\not\in C$ and $u$ receives at least one unit of weight per neighbor in $B_1\cup B_2$. Pick an inclusion-wise minimal set $\bar{Y}\subseteq R\cup Z_1\cup Z_2$ such that $|W|+2|R\cap\bar{Y}|+|Z_1\cap\bar{Y}|+2 |Z_2\cap\bar{Y}|\geq 3$. Then \begin{align}
	|W|+2|R\cap\bar{Y}|+|Z_1\cap\bar{Y}|+2 |Z_2\cap\bar{Y}|&= 3\text{, or}\label{EqCase1}\\Z_1\cap\bar{Y}=\emptyset\text{ and }3\leq |W|+2|R\cap\bar{Y}|+2 |Z_2\cap\bar{Y}|&\leq 4.\label{EqCase2}
	\end{align} 
	We construct a set collection $Y$ as follows: First, we add all sets in $(R\cup Z_2)\cap\bar{Y}$ to $Y$. Second, for each $v\in Z_1\cap\bar{Y}$, let $v'$ be the set of cardinality $2$ containing the element in which $v$ intersects a set from $C$, and the element in which $v$ intersects $u$. Add $v'$ to $Y$.
	Then $N(Y,A)\subseteq C\cup\{u\}$ and \[w(Y)=2|R\cap\bar{Y}|+|Z_1\cap\bar{Y}|+2 |Z_2\cap\bar{Y}|.\]
	Moreover, by construction,
	\[|N(Y,C)|\leq 2|R\cap\bar{Y}|+|Z_1\cap\bar{Y}|+|Z_2\cap\bar{Y}|.\]
	Let $U\coloneqq N(Y,C)\cup\{u\}$. Apply Lemma~\ref{LemFirstStep} to obtain $X$ subject to (\ref{FirstStepProp1})-(\ref{FirstStepProp3}). Then \[w(X)\geq w(N(Y,C))+|N(u,B_1\cup B_2)|=w(N(Y,C))+|W|\] because each set in $N(Y,C)$ receives its weight in the first step, and $u$ receives at least one per neighbor in $B_1\cup B_2$. By (\ref{FirstStepProp2}), $X$ and $Y$ are disjoint. We would like to show that $X\cup Y$ yields a local improvement of size at most $10$.
	We have 
	\begin{align*}
	w(X\cup Y)&=w(X)+w(Y)\geq |W|+2|R\cap\bar{Y}|+|Z_1\cap\bar{Y}|+2 |Z_2\cap\bar{Y}|+w(N(Y,C)) \\&\geq 3+w(N(Y,C))> w(u)+w(N(Y,C))\geq w(N(X\cup Y,A))
	\end{align*}
	because $N(X,A)\subseteq U = N(Y,C)\cup\{u\}$ and $N(Y,A)\subseteq C\cup\{u\}$ by construction.
	Thus, it remains to show that $|X\cup Y|\leq 10$.
	First of all, each set in $X$ needs to intersect a set in $U$. By (\ref{FirstStepProp2}), the number of sets from $X$ that intersect $u$ can be bounded by $|N(u,B_1\cup B_2)|=|W|$. Moreover, each set from $N(Y,C)$ can intersect at most $2$ sets in $X$ because it also intersects a set in $Y$. 
	Hence, we obtain \begin{align}|Y|+|X|&\leq |Y|+|N(u,B_1\cup B_2)|+2\cdot|N(Y,C)|\notag\\&\leq |W|+|R\cap\bar{Y}|+|Z_1\cap\bar{Y}|+|Z_2\cap\bar{Y}|+2\cdot(2|R\cap\bar{Y}|+|Z_1\cap\bar{Y}|+|Z_2\cap\bar{Y}|).\notag\end{align}  If \eqref{EqCase1} holds, we get 
	\[|Y|+|X|\leq 3 + 2\cdot 3=9<10.\]
	In case \eqref{EqCase2} is satisfied, we obtain
	\[|Y|+|X|\leq  |W|+5|R\cap\bar{Y}|+3|Z_2\cap\bar{Y}|\leq \frac{5}{2}\cdot (|W|+2|R\cap\bar{Y}|+2|Z_2\cap\bar{Y}|)\leq 10.\] Hence, $|X\cup Y|\leq 10$. Thus, we have found a local improvement of size at most $10$, a contradiction.
	
\end{proof}
\begin{lemma}
	If in situation \ref{SituationD}, the neighbor $u$ of $v$ in $C$ receives more than $\frac{1}{3}$, then $w(u)=2$ and $u$ has two incident edges to $B_1\cup B_2$.\label{LemMoreThanOneThird}
\end{lemma}
\begin{proof}
	Assume towards a contradiction that this is not the case. We first observe that $u$ must have exactly one incident edge to $B_1\cup B_2$. First of all, there is at least one such edge because $u\in C$. If further $w(u)=1$, then $u$ can have at most one incident edge to $B_1\cup B_2$ because $u$ is also incident to $v$. On the other hand, if $w(u)=2$, then $u$ can have at most two incident edges to $B_1\cup B_2$. As we have excluded the case of $2$ such edges, $u$ has indeed exactly one incident edge to $B_1\cup B_2$.
	
	In particular, $|u\cap(\bigcup B_1\cup \bigcup B_2)|=1$. By Lemma~\ref{LemFirstStep}, pick a disjoint collection of sets $X$ with $\bigcup X\subseteq \bigcup B_1\cup\bigcup B_2$, $N(X,A)=\{u\}$ and $w(X)=w(u)$. As every $x\in X$ must intersect a set in $A$, and, thus, $u$,  $|u\cap(\bigcup B_1\cup \bigcup B_2)|=1$ and $\bigcup X\subseteq \bigcup B_1\cup\bigcup B_2$ yield $|X|=1$. Let $X=\{x\}$. Then $w(x)=w(u)$. Let $Y\coloneqq \{x,v\}$. By our assumption that $u$ receives more than $\frac{1}{3}$ from $v$, $N(Y,A)=N(x,A)\cup N(v,A)$ must consist of $u$ and one or two sets of weight $1$ from $A\setminus C$. Thus, $w(Y)\geq w(N(Y,A))$ and the number of sets of weight $2$ in $Y$ is larger than the number of sets of weight $2$ in $N(Y,A)$. Thus, $Y$ constitutes a local improvement of size $2$, a contradiction.
\end{proof}
\begin{lemma}
	Each set in $C$ receives at most $\frac{4}{3}$ times its own weight during our weight distribution.
\end{lemma}
\begin{proof}
	First, let $u\in C$ with $w(u)=1$. Then $u$ receives $1$ in the first step and has at most one incident edge to $B\setminus (B_1\cup B_2)$. By Lemma~\ref{LemMoreThanOneThird}, $u$ receives at most $\frac{1}{3}$ via this edge.
	Next, let $u\in C$ with $w(u)=2$. Then $u$ receives $2$ in the first step and $u$ has at most two incident edges to $B\setminus(B_1\cup B_2)$. If $u$ has two incident edges to $B\setminus(B_1\cup B_2)$, then $u$ has at most one incident edge to $B_1\cup B_2$ and Lemma~\ref{LemMoreThanOneThird} implies that $u$ can receive at most $\frac{1}{3}$ via each of the edges to $B\setminus(B_1\cup B_2)$. Thus, $u$ receives at most $\frac{8}{3}=\frac{4}{3}\cdot w(u)$ in total. If $u$ has one incident edge to $B\setminus (B_1\cup B_2)$, then the maximum amount $u$ can receive via this edge is $\frac{2}{3}$ (if in \ref{SituationD}, the endpoints of both incident edges of $v$ to $A\setminus C$ have weight $1$). Again, $u$ receives at most $\frac{8}{3}$ in total.
\end{proof}
\begin{lemma}
	Each set $u\in A\setminus C$ receives at most $\frac{4}{3}$ times its own weight during our weight distribution.
\end{lemma}
\begin{proof}
	If $w(u)=1$, then $u$ cannot receive any weight in the first step because otherwise, it would receive at least $1$ and be contained in $C$. Moreover, $u$ has at most two incident edges and receives at most $\frac{2}{3}$ via either of them in the second step by Lemma~\ref{LemSituationC}.
	
	Next, consider the case where $w(u)=2$. If there exists an edge via which $u$ receives $\frac{4}{3}$ as in situation \ref{SituationC}, then by Lemma~\ref{LemBoundHighCharges}, there is no further edge via which $u$ receives $\frac{4}{3}$ as in situation \ref{SituationC}, or $1$ as in situation \ref{SituationD}. Moreover, $|N(u,B_1\cup B_2)|=\emptyset$, so $u$ does not receive anything in the first step. As $u$ receives at most $\frac{2}{3}$ per edge in all remaining cases, $u$ receives at most $\frac{4}{3}+2\cdot\frac{2}{3}=\frac{8}{3}=\frac{4}{3}\cdot w(u)$.
	Finally, assume that $u$ does not receive $\frac{4}{3}$ as in situation \ref{SituationC}. In the first step, $u$ can receive at most $1$ in total (otherwise, $u\in C$) and this can only happen if $u$ has a neighbor in $B_1\cup B_2$. The maximum amount $u$ can receive through one edge in the second step is $1$, and this can only happen in situation \ref{SituationD}. By Lemma~\ref{LemBoundHighCharges}, there are at most $2$ edges via which $u$ receives $1$. Moreover, $u$ can receive at most $\frac{2}{3}$ via the remaining edges. Again, we obtain an upper bound of $1+1+\frac{2}{3}=\frac{8}{3}$ on the total weight received.	
\end{proof}
This concludes the proof of Theorem~\ref{TheoMainHereditary}.

\bibliography{arborescences_many_leaves}

\begin{thebibliography}{10}

\bibitem{SpanningDirectedTreesWithManyLeaves}
Noga Alon, Fedor~V. Fomin, Gregory Gutin, Michael Krivelevich, and Saket
  Saurabh.
\newblock Spanning directed trees with many leaves.
\newblock {\em SIAM Journal on Discrete Mathematics}, 23(1):466--476, 2009.
\newblock \href {https://doi.org/10.1137/070710494}
  {\path{doi:10.1137/070710494}}.

\bibitem{ColorCoding}
Noga Alon, Raphael Yuster, and Uri Zwick.
\newblock Color-coding.
\newblock {\em J. ACM}, 42(4):844–856, July 1995.
\newblock \href {https://doi.org/10.1145/210332.210337}
  {\path{doi:10.1145/210332.210337}}.

\bibitem{ArkinHassin}
Esther~M. Arkin and Refael Hassin.
\newblock On local search for weighted $k$-set packing.
\newblock {\em Mathematics of Operations Research}, 23(3):640--648, 1998.
\newblock \href {https://doi.org/10.1287/moor.23.3.640}
  {\path{doi:10.1287/moor.23.3.640}}.

\bibitem{Berman}
Piotr Berman.
\newblock A $d/2$ {A}pproximation for {M}aximum {W}eight {I}ndependent {S}et in
  $d$-{C}law {F}ree {G}raphs.
\newblock In {\em Scandinavian Workshop on Algorithm Theory}, pages 214--219.
  Springer, 2000.
\newblock \href {https://doi.org/10.1007/3-540-44985-X_19}
  {\path{doi:10.1007/3-540-44985-X_19}}.

\bibitem{berman1994approximating}
Piotr Berman and Martin F{\"u}rer.
\newblock {Approximating Maximum Independent Set in Bounded Degree Graphs}.
\newblock In {\em Proceedings of the fifth annual ACM-SIAM Symposium on
  Discrete Algorithms}, pages 365--371, 1994.
\newblock URL: \url{https://dl.acm.org/doi/pdf/10.5555/314464.314570}.

\bibitem{KernelsForProblemsWithNoKernel}
Daniel Binkele-Raible, Henning Fernau, Fedor~V. Fomin, Daniel Lokshtanov, Saket
  Saurabh, and Yngve Villanger.
\newblock Kernel(s) for problems with no kernel: On out-trees with many leaves.
\newblock {\em ACM Trans. Algorithms}, 8(4), 2012.
\newblock \href {https://doi.org/10.1145/2344422.2344428}
  {\path{doi:10.1145/2344422.2344428}}.

\bibitem{BONSMA201214}
Paul Bonsma.
\newblock Max-leaves spanning tree is {APX}-hard for cubic graphs.
\newblock {\em Journal of Discrete Algorithms}, 12:14--23, 2012.
\newblock \href {https://doi.org/10.1016/j.jda.2011.06.005}
  {\path{doi:10.1016/j.jda.2011.06.005}}.

\bibitem{ChandraHalldorsson}
Barun Chandra and Magn\'{u}s~M. Halld{\'o}rsson.
\newblock Greedy {L}ocal {I}mprovement and {W}eighted {S}et {P}acking
  {A}pproximation.
\newblock {\em Journal of Algorithms}, 39(2):223--240, 2001.
\newblock \href {https://doi.org/10.1006/jagm.2000.1155}
  {\path{doi:10.1006/jagm.2000.1155}}.

\bibitem{Cygan}
Marek Cygan.
\newblock {I}mproved {A}pproximation for 3-{D}imensional {M}atching via
  {B}ounded {P}athwidth {L}ocal {S}earch.
\newblock In {\em 54th Annual {IEEE} Symposium on Foundations of Computer
  Science, {FOCS} 2013, 26-29 October, 2013, Berkeley, CA, {USA}}, pages
  509--518. {IEEE} Computer Society, 2013.
\newblock \href {https://doi.org/10.1109/FOCS.2013.61}
  {\path{doi:10.1109/FOCS.2013.61}}.

\bibitem{CyganGrandoniMastrolilli}
Marek Cygan, Fabrizio Grandoni, and Monaldo Mastrolilli.
\newblock How to {S}ell {H}yperedges: {T}he {H}ypermatching {A}ssignment
  {P}roblem.
\newblock In {\em Proceedings of the 2013 Annual ACM-SIAM Symposium on Discrete
  Algorithms}, pages 342--351. SIAM, 2013.
\newblock \href {https://doi.org/10.1137/1.9781611973105.25}
  {\path{doi:10.1137/1.9781611973105.25}}.

\bibitem{DaligaultThomasse}
Jean Daligault and St{\'e}phan Thomass{\'e}.
\newblock On finding directed trees with many leaves.
\newblock In Jianer Chen and Fedor~V. Fomin, editors, {\em Parameterized and
  Exact Computation}, pages 86--97. Springer Berlin Heidelberg, 2009.
\newblock \href {https://doi.org/10.1007/978-3-642-11269-0_7}
  {\path{doi:10.1007/978-3-642-11269-0_7}}.

\bibitem{Drescher2010AnAA}
Matthew Drescher and Adrian Vetta.
\newblock An approximation algorithm for the maximum leaf spanning arborescence
  problem.
\newblock {\em ACM Trans. Algorithms}, 6(3), 2010.
\newblock \href {https://doi.org/10.1145/1798596.1798599}
  {\path{doi:10.1145/1798596.1798599}}.

\bibitem{edmonds1965maximum}
Jack Edmonds.
\newblock Maximum matching and a polyhedron with 0,1-vertices.
\newblock {\em Journal of Research of the National Bureau of Standards Section
  B Mathematics and Mathematical Physics}, 69B:125--130, 1965.
\newblock \href {https://doi.org/10.6028/jres.069b.013}
  {\path{doi:10.6028/jres.069b.013}}.

\bibitem{FERNANDES2022217}
Cristina~G. Fernandes and Carla~N. Lintzmayer.
\newblock Leafy spanning arborescences in dags.
\newblock {\em Discrete Applied Mathematics}, 323:217--227, 2022.
\newblock \href {https://doi.org/10.1016/j.dam.2021.06.018}
  {\path{doi:10.1016/j.dam.2021.06.018}}.

\bibitem{FernandesLintzmayer}
Cristina~G. Fernandes and Carla~N. Lintzmayer.
\newblock How heavy independent sets help to find arborescences with many
  leaves in dags.
\newblock {\em Journal of Computer and System Sciences}, 135:158--174, 2023.
\newblock \href {https://doi.org/https://doi.org/10.1016/j.jcss.2023.02.006}
  {\path{doi:https://doi.org/10.1016/j.jcss.2023.02.006}}.

\bibitem{FurerYu}
Martin F{\"{u}}rer and Huiwen Yu.
\newblock Approximating the $k$-{S}et {P}acking {P}roblem by {L}ocal
  {I}mprovements.
\newblock In {\em International Symposium on Combinatorial Optimization}, pages
  408--420. Springer, 2014.
\newblock \href {https://doi.org/10.1007/978-3-319-09174-7_35}
  {\path{doi:10.1007/978-3-319-09174-7_35}}.

\bibitem{GALBIATI199445}
G.~Galbiati, F.~Maffioli, and A.~Morzenti.
\newblock A short note on the approximability of the maximum leaves spanning
  tree problem.
\newblock {\em Information Processing Letters}, 52(1):45--49, 1994.
\newblock \href {https://doi.org/10.1016/0020-0190(94)90139-2}
  {\path{doi:10.1016/0020-0190(94)90139-2}}.

\bibitem{GareyJohnson}
Michael~R. Garey and David~S. Johnson.
\newblock {\em Computers and Intractability; A Guide to the Theory of
  NP-Completeness}.
\newblock W. H. Freeman \& Co., USA, 1990.

\bibitem{Halldorsson}
Magn\'{u}s~M. Halld\'{o}rsson.
\newblock Approximating {D}iscrete {C}ollections via {L}ocal {I}mprovements.
\newblock In {\em Proceedings of the Sixth Annual ACM-SIAM Symposium on
  Discrete Algorithms}, page 160–169, USA, 1995. Society for Industrial and
  Applied Mathematics.
\newblock URL: \url{https://dl.acm.org/doi/10.5555/313651.313687}.

\bibitem{LowerBoundKSetPacking}
Elad Hazan, Shmuel Safra, and Oded Schwartz.
\newblock On the complexity of approximating $k$-{S}et {P}acking.
\newblock {\em Computational Complexity}, 15:20--39, 2006.
\newblock \href {https://doi.org/10.1007/s00037-006-0205-6}
  {\path{doi:10.1007/s00037-006-0205-6}}.

\bibitem{HurkensSchrijver}
C.~A.~J. Hurkens and A.~Schrijver.
\newblock On the size of systems of sets every $t$ of which have an {S}{D}{R},
  with an application to the worst-case ratio of heuristics for packing
  problems.
\newblock {\em SIAM Journal on Discrete Mathematics}, 2(1):68--72, 1989.
\newblock \href {https://doi.org/10.1137/0402008} {\path{doi:10.1137/0402008}}.

\bibitem{karp1972reducibility}
Richard~M. Karp.
\newblock Reducibility among combinatorial problems.
\newblock In Raymond~E. Miller, James~W. Thatcher, and Jean~D. Bohlinger,
  editors, {\em Complexity of Computer Computations: Proceedings of a symposium
  on the Complexity of Computer Computations}. Plenum Press, 1972.
\newblock \href {https://doi.org/10.1007/978-1-4684-2001-2_9}
  {\path{doi:10.1007/978-1-4684-2001-2_9}}.

\bibitem{OnSyntacticVersusComputationalViewsOfApproximability}
Sanjeev Khanna, Rajeev Motwani, Madhu Sudan, and Umesh Vazirani.
\newblock On syntactic versus computational views of approximability.
\newblock {\em SIAM Journal on Computing}, 28(1):164--191, 1998.
\newblock \href {https://doi.org/10.1137/S0097539795286612}
  {\path{doi:10.1137/S0097539795286612}}.

\bibitem{NeuwohnerLipics}
Meike Neuwohner.
\newblock {An Improved Approximation Algorithm for the Maximum Weight
  Independent Set Problem in d-Claw Free Graphs}.
\newblock In {\em 38th International Symposium on Theoretical Aspects of
  Computer Science (STACS 2021)}, volume 187 of {\em Leibniz International
  Proceedings in Informatics (LIPIcs)}, pages 53:1--53:20, 2021.
\newblock \href {https://doi.org/10.4230/LIPIcs.STACS.2021.53}
  {\path{doi:10.4230/LIPIcs.STACS.2021.53}}.

\bibitem{Neuwohner22}
Meike Neuwohner.
\newblock The limits of local search for weighted k-set packing.
\newblock In Karen Aardal and Laura Sanit{\`{a}}, editors, {\em Integer
  Programming and Combinatorial Optimization - 23rd International Conference,
  {IPCO} 2022, Eindhoven, The Netherlands, June 27-29, 2022, Proceedings},
  volume 13265 of {\em Lecture Notes in Computer Science}, pages 415--428.
  Springer, 2022.
\newblock \href {https://doi.org/10.1007/978-3-031-06901-7\_31}
  {\path{doi:10.1007/978-3-031-06901-7\_31}}.

\bibitem{Neuwohner23}
Meike Neuwohner.
\newblock Passing the limits of pure local search for weighted k-set packing.
\newblock In {\em Proceedings of the 2023 Annual ACM-SIAM Symposium on Discrete
  Algorithms (SODA)}, pages 1090--1137. Society for Industrial and Applied
  Mathematics, 2023.
\newblock \href {https://doi.org/10.1137/1.9781611977554.ch41}
  {\path{doi:10.1137/1.9781611977554.ch41}}.

\bibitem{SchmidtSiegel}
Jeanette~P. Schmidt and Alan Siegel.
\newblock The spatial complexity of oblivious k-probe hash functions.
\newblock {\em SIAM Journal on Computing}, 19(5):775--786, 1990.
\newblock \href {https://doi.org/10.1137/0219054} {\path{doi:10.1137/0219054}}.

\bibitem{SchwartgesSpoerhaseWolff}
Nadine Schwartges, Joachim Spoerhase, and Alexander Wolff.
\newblock Approximation algorithms for the maximum leaf spanning tree problem
  on acyclic digraphs.
\newblock In Roberto Solis-Oba and Giuseppe Persiano, editors, {\em
  Approximation and Online Algorithms}, pages 77--88. Springer Berlin
  Heidelberg, 2012.
\newblock \href {https://doi.org/10.1007/978-3-642-29116-6_7}
  {\path{doi:10.1007/978-3-642-29116-6_7}}.

\bibitem{SolisOba2015A2A}
Roberto Solis-Oba, Paul~S. Bonsma, and Stefanie Lowski.
\newblock A 2-approximation algorithm for finding a spanning tree with maximum
  number of leaves.
\newblock {\em Algorithmica}, 77:374--388, 2015.
\newblock \href {https://doi.org/10.1007/s00453-015-0080-0}
  {\path{doi:10.1007/s00453-015-0080-0}}.

\bibitem{SviridenkoWard}
Maxim Sviridenko and Justin Ward.
\newblock {L}arge {N}eighborhood {L}ocal {S}earch for the {M}aximum {S}et
  {P}acking {P}roblem.
\newblock In {\em International Colloquium on Automata, Languages, and
  Programming}, pages 792--803. Springer, 2013.
\newblock \href {https://doi.org/10.1007/978-3-642-39206-1_67}
  {\path{doi:10.1007/978-3-642-39206-1_67}}.

\bibitem{ThieryWard}
Theophile Thiery and Justin Ward.
\newblock An improved approximation for maximum weighted k-set packing.
\newblock In {\em Proceedings of the 2023 Annual ACM-SIAM Symposium on Discrete
  Algorithms (SODA)}, pages 1138--1162. Society for Industrial and Applied
  Mathematics, 2023.
\newblock \href {https://doi.org/10.1137/1.9781611977554.ch42}
  {\path{doi:10.1137/1.9781611977554.ch42}}.

\bibitem{InapproxIndependentSet}
David Zuckerman.
\newblock {Linear Degree Extractors and the Inapproximability of Max Clique and
  Chromatic Number}.
\newblock {\em Theory of Computing}, 3(6):103--128, 2007.
\newblock \href {https://doi.org/10.4086/toc.2007.v003a006}
  {\path{doi:10.4086/toc.2007.v003a006}}.

\end{thebibliography}
\appendix
\section{The structure of minimal binoculars\label{appendix:binoculars}}
\begin{lemma}
	Let $G$ be a graph and let $B$ a minimal binocular in $G$. Then one of the following statements holds:
	\begin{itemize}
		\item There are two not necessarily distinct vertices $u$ and $v$, cycles $C_u$ containing $u$ and $C_v$ containing $v$ of positive lengths and a $u$-$v$-path $P$ (if $u=v$, $P$ is a path of length $0$) such that $E(B)=E(C_1)\dot{\cup} E(C_2)\dot{\cup} E(P)$.  
		\item There are two distinct vertices $u$ and $v$ such that $E(B)$ is the union of the edge sets of three edge-disjoint $u$-$v$-paths.
	\end{itemize}\label{LemStructureBinoculars}
\end{lemma}
\begin{proof}
	We first prove the following claim:
	\begin{claim}
	Every minimal binocular $B$ is connected and every vertex has degree at least $2$. Moreover, $|E(B)|=|V(B)|+1$.
	\end{claim}
\begin{proof}[Proof of the claim]
Let $B$ be a minimal binocular. Then $B$ is connected because otherwise, one of the connected components of $B$ must contain more edges than vertices and thus yields a smaller binocular. In particular, $B$ cannot contain an isolated vertex because $B$ needs to contain at least one edge, which would then be in a different connected component than the isolated vertex. Moreover, $B$ cannot contain a vertex with at most one incident edge because removing the vertex and its incident edge would yield a smaller binocular. Finally, $|E(B)|=|V(B)|+1$ because if $|E(B)|\geq |V(B)|+2$, we could remove an edge and obtain a smaller binocular.	
\end{proof}
Now, assume towards a contradiction that the statement of the lemma was false and pick a minimal binocular $B$ that is a counterexample with a minimum number of edges. As $B$ is a minimal binocular, $B$ must be connected and every vertex of $B$ has at least two incident edges. On the other hand, we have \[\sum_{v\in V(B)} \mathrm{deg}(v)= 2\cdot |E(B)|=2\cdot |V(B)|+2,\] where $\mathrm{deg}(v)$ denotes the degree of a vertex, which is the number of incident two-vertex edges plus twice the number of incident loops.
\begin{claim}
Every vertex of $B$ has degree at least $3$.
\end{claim}
\begin{proof}[Proof of the claim] By the previous claim, it suffices to show that no vertex of $B$ has degree $2$. Assume towards a contradiction that $B$ contains a vertex $v$ of degree $2$. If $v$ has an incident loop, then since $B$ is connected, this must be the only vertex and the loop is the only edge, a contradiction. Thus, $v$ has exactly two incident two-vertex edges $\{v,u\}$ and $\{v,w\}$, where $u$ and $w$ need not be distinct. We obtain $B'$ from $B$ by contracting the edge $\{u,v\}$. Denote the vertex arising from the contraction by $u'$.  We claim that $B'$ is a minimal binocular. If not, there is a strict sub-graph $B''$ of $B'$  that constitutes a binocular. In case $B''$ contains $u'$, we un-contract $\{u,v\}$ again. This results in a strict sub-graph of $B$ that is a binocular, a contradiction. So $B'$ is indeed a minimal binocular. By our assumption on $B$, $B'$ is not a counterexample to the lemma. But un-contracting $\{u,v\}$ again preserves this property, a contradiction.
\end{proof}
As every vertex of $B$ has degree at least $3$ and $\sum_{v\in V(B)} \mathrm{deg}(v)=2|V(B)|+2$, this means that $V(B)$ either consists of one vertex of degree $4$, or of two vertices of degree $3$. In the first case, $B$ consists of two loops incident to the same vertex and is, thus, of the first type.
In the second case, $B$ either contains one edge between the two vertices, and a loop attached to each of them, or three parallel edges between the two vertices. In either case, $B$ is not a counterexample to the lemma.
\end{proof}
\section{Formal proof of Theorem~\ref{TheoNormalization}\label{appendix:Normalization}}
In order to abbreviate the proof of Theorem~\ref{TheoNormalization}, we introduce the following notation:
\begin{definition}
A \emph{nice graph} is a pair $(G,w)$, where $G$ is a $4$-claw free graph, $w:V(G)\rightarrow\{1,2\}$ and every $3$-claw in $G$ is centered at a vertex of weight $2$.
\end{definition}
Observe that Proposition~\ref{PropStructureConflictGraph} tells us that for an instance $(\mathcal{S},w)$ of the $2$-$3$-Set Packing problem, $(G_{\mathcal{S}},w)$ is a nice graph. Moreover, the first condition in Definition~\ref{DefNormalized} simply requires that $(G,w)$ is nice.

We show that Theorem~\ref{TheoNormalization} holds in a slightly more general setting, where $G$ is not necessarily the conflict graph of a $2$-$3$-Set Packing instance, but simply has the property that $(G,w)$ is nice.

We first state the following proposition, which is a direct consequence of the fact that every vertex in $A'$ has at most two neighbors in the independent set $B'$ and vice versa.
\begin{proposition}
	Let $(G,w)$ be nice and let $A,B\subseteq V(G)$ be two independent sets. Then every connected component of $G[A'\cup B']$ is an $A'$-$B'$-alternating path or cycle.\label{PropPathOrCycle}
\end{proposition}
Next, we show that we can discard vertices in $V(G)\setminus (A\cup B)$, $A\cap B$, vertices from cycle components of $G[A'\cup B']$ and vertices from long path components of $G[A'\cup B']$ that are not both contained in $A'$ and connected to $B''$.
\begin{lemma}
	Let $(G,w)$ be a nice graph and let $A$ and $B$ be two independent sets in $G$. Let $D\subseteq V(G)$ such that $N(V(G)\setminus D,A)\subseteq A\setminus D$. Then:
	\begin{itemize}
		\item Every local improvement of $A\setminus D$ in $(G[V(G)\setminus D],w\upharpoonright_{V(G)\setminus D})$ also constitutes a local improvement of $A$ in $(G,w)$.
		\item For every $\tau>0$, $S_\tau(G[V(G)\setminus D],w\upharpoonright_{V(G)\setminus D},A\setminus D)$ is a sub-graph of $S_\tau(G,w,A)$. In particular, every improving minimal binocular in $S_\tau(G[V(G)\setminus D],w\upharpoonright_{V(G)\setminus D},A\setminus D)$ also constitutes an improving minimal binocular in $S_\tau(G,w,A)$.  
	\end{itemize}\label{LemDeletableSet}
\end{lemma}
\begin{proof}[Proof of Lemma~\ref{LemDeletableSet}]
	Both statements follow from the fact that for any $X\subseteq V(G)\setminus D$, we have $N(X,A)=N(X,A)\cap N(V(G)\setminus D,A)=N(X,A)\cap (A\setminus D) = N(X,A\setminus D)$.
\end{proof}

\begin{definition}
	We call a vertex $d\in V(G)$ \emph{deletable} if one of the following properties holds:
	\begin{enumerate}
		\item $d\in V(G)\setminus (A\cup B)$.
		\item $d\in A\cap B$.
		\item $d\in A'\cup B'$ and the connected component of $G[A'\cup B']$ containing $d$ is a cycle or a path with an even number of vertices that is also a connected component of $G[A\cup B]$.
		\item $d\in A'\cup B'$ and the connected component of $G[A'\cup B']$ containing $d$ is a path $P$ with $|\mathrm{in}(P)\cap A'|\geq 3$ (where $\mathrm{in}(P)$ denotes the set of inner vertices of $P$), and $d$ does not have any neighbors in $B\setminus V(P)$. Observe that this applies to all vertices of the path except for possibly endpoints from $A'$.
	\end{enumerate}
	Denote the set of deletable vertices by $D$.
\end{definition}
\begin{proposition}
	\begin{itemize}
		\item[]
		\item $N(V(G)\setminus D, A)\subseteq A\setminus D$.
		\item Let $\alpha\geq\frac{4}{3}$ such that $w(B\setminus D)\leq \alpha\cdot w(A\setminus D)$. Then $w(B)\leq \alpha\cdot w(A)$.
	\end{itemize}\label{PropPropertiesDeletableSet}\end{proposition}
\begin{proof}
For the first item, we need to show that no vertex in $A\cap D$ can have a neighbor in $V(G)\setminus D$. If $d\in A\cap B$, then $d$ cannot have any neighbor in $V(G)\setminus D\subseteq A\Delta B$ because $A$ and $B$ are independent. If $d\in A'\cup B'$ is contained in a cycle component of $G[A'\cup B']$ or in a path component of $G[A'\cup B']$ that is also a connected component of $G[A\cup B]$, then all neighbors of $d$ in $A\cup B\supseteq V(G)\setminus D$ are contained in the same connected component of $G[A'\cup B']$, and, thus, in $D$. Finally, if $d\in A'$ is contained in a path component $P$ such that $N(d,B\setminus V(P))=\emptyset$, then $d$ does not have any neighbors in $V(G)\setminus D\subseteq (A\cup B)\setminus (V(P)\cap B)$ since $A$ is independent.

For the second item, it suffices to show that $w(B\cap D)\leq \frac{4}{3}\cdot w(A\cap D)$. The vertices in $A\cap B$ are counted once on each side. Moreover, by independence of $A$ and $B$, they are isolated in $G[A\cup B]$. For a cycle component or a path component with an even number of vertices, we remove the same number of vertices from $A'$ and $B'$. For a path component $P$ in $G[A'\cup B']$ with $|\mathrm{in}(P)\cap A'|\geq 3$, we remove at least $|\mathrm{in}(P)\cap A'|$ vertices from $A'$, and at most $|\mathrm{in}(P)\cap A'|+1$ vertices from $B'$. Thus, $w(V(P)\cap D\cap B')\leq \frac{4}{3}\cdot w(V(P)\cap D\cap A')$.
\end{proof}

Combining Lemma~\ref{LemDeletableSet} and Proposition~\ref{PropPropertiesDeletableSet} allows us to restrict ourselves to the following setting (note that induced sub-graphs of nice graphs are nice):
\begin{itemize}
	\item $(G,w)$ is nice and bipartite with bipartitions $A$ and $B$.
	\item Every connected component of $G[A'\cup B']$ is a path $P$ with at most $2$ inner vertices from $A'$. Moreover, if $|V(P)|$ is even, then $N(V(P),A''\cup B'')\neq \emptyset$. Note that the case where $|V(P)|$ is odd applies to the isolated vertices that may remain as non-deleted endpoints of paths.
\end{itemize}

We now define a family of paths $\mathcal{P}$ in $G[A'\cup B']$ as follows:
\begin{itemize}
	\item We add each connected component of $G[A'\cup B']$ that is a path with an even number of vertices.
	\item For every connected component of $G[A'\cup B']$ that is a path with an odd number $\geq 3$ of vertices, we remove one of the two endpoints from the path and add the resulting path.
\end{itemize}   As each path in $\mathcal{P}$ comes from a different connected component of $G[A'\cup B']$, the paths are pairwise non-adjacent (i.e.\ no vertex of one path is adjacent to a vertex of another path). Moreover, each path contains an even number of vertices and features at most two inner vertices from $A'$, which implies that it contains at most $3$ vertices from $A$ and $B$ each. Finally, by construction, each of the paths $P$ has at least one neighbor in $(A\cup B)\setminus V(P)$. On the other hand, as vertices in $A'\cup B'$ can have degree at most $2$ in $G=G[A\cup B]$ and one endpoint of the path comes from $A'$ and the other one comes from $B'$, each one of the paths $P$ has at most one neighbor in $A\setminus V(P)$ and $B\setminus V(P)$, each. Thus, we can partition $\mathcal{P}=\mathcal{P}_1\dot{\cup}\mathcal{P}_2\dot{\cup}\mathcal{P}_3$, where $N(V(P),(A\cup B)\setminus V(P))$ consists of
\begin{itemize}
	\item exactly one vertex from $A$ ($P\in\mathcal{P}_1$),
	\item exactly one vertex from $B$ ($P\in\mathcal{P}_2$),
	\item exactly one vertex from $A$ and one vertex from $B$ ($P\in\mathcal{P}_3$).
\end{itemize}
\begin{lemma}
	Let $(\bar{G}, \bar{w})$, as well as $\bar{A}$ and $\bar{B}$ arise from $(G,w)$ and $A$ and $B$ by deleting $\bigcup_{P\in\mathcal{P}} V(P)$ and moreover, for each $P\in\mathcal{P}_3$, connecting the two vertices in $N(V(P),(A\cup B)\setminus V(P))$ by an edge. Note that these vertices are not removed since the paths are non-adjacent. Let $\tau\coloneqq10\cdot\bar{\tau}$. Then the following properties hold:
	\begin{enumerate}[(a)]
		\item $(\bar{G},\bar{w})$ is nice and bipartite with bipartitions $\bar{A}$ and $\bar{B}$. Moreover, $\bar{A}'\cup\bar{B}'$ constitutes an independent set.
		\item If there exists a local improvement $X$ of $\bar{A}$ in $(\bar{G},\bar{w})$ of size at most $\bar{\tau}$, then there exists a local improvement of size at most $\tau$ in $(G,w)$.
		\item If there exists an improving minimal binocular in $S_{\bar{\tau}}(\bar{G},\bar{w},\bar{A})$ of size at most $\bar{\tau}\cdot\log|V(\bar{G})|$, then there exists an improving minimal binocular in $S_\tau(G,w,A)$ of size at most $\tau\cdot\log|V(G)|$.
		\item Let $\alpha\geq\frac{4}{3}$ such that $w(\bar{B})\leq \alpha\cdot w(\bar{A})$. Then $w(B)\leq \alpha\cdot w(A)$.
	\end{enumerate}\label{LemDeletePaths}
\end{lemma}
\begin{proof}[Proof of Lemma~\ref{LemDeletePaths}]
	\begin{proof}[Proof of (a)]
		As $(G,w)$ is nice with bipartitions $A$ and $B$, every vertex in $A''\cup B''$ has degree $\leq 3$ in $G$, and every vertex in $A'\cup B'$ has degree $\leq 2$ in $G$. $\bar{G}$ is bipartite with bipartitions $\bar{A}$ and $\bar{B}$ because neither deleting vertices nor adding an edges between $A$ and $B$ may destroy this property. Moreover, whenever we add a new edge to a vertex from $A$ or $B$, we first remove an incident edge by deleting the respective path from $\mathcal{P}_3$. Thus, every vertex in $\bar{A}''\cup \bar{B}''$ has degree $\leq 3$ in $\bar{G}$, and every vertex in $\bar{A}'\cup \bar{B}'$ has degree $\leq 2$ in $\bar{G}$, so $(\bar{G},\bar{w})$ is nice. 
	\end{proof}
	\begin{proof}[Proof of (b)]
		Let $\bar{X}$ be a local improvement of $\bar{A}$ in $(\bar{G},\bar{w})$ of size at most $\bar{\tau}$. We may assume $\bar{X}\subseteq \bar{B}$ because we can replace $\bar{X}$ by $\bar{X}\cap\bar{B}$ otherwise. Let $\mathcal{P}'\subseteq\mathcal{P}_2\cup \mathcal{P}_3$ be the set of paths $P$ for which $N(V(P),\bar{X})\neq \emptyset$. Obtain $X$ from $\bar{X}$, by, for each such path $P\in\mathcal{P}'$, adding $V(P)\cap B$ to $\bar{X}$. As $G$ is $4$-claw free, we get $|\mathcal{P}'|\leq |N(\bar{X},A)|\leq  3\cdot|\bar{X}|$. Thus, \[|X| = \sum_{P\in\mathcal{P}'} |V(P)\cap B| +|\bar{X}|\leq 3\cdot |\mathcal{P}'|+|\bar{X}|\leq (3\cdot 3+1)\cdot |\bar{X}|\leq 10\cdot \bar{\tau}\leq \tau.\] We claim that $N(X,A)\subseteq N(\bar{X},\bar{A})\cup\bigcup_{P\in\mathcal{P}'} V(P)\cap A$. 
		To see this, let first $x\in\bar{X}$. Then \[N(x,A)=N(x,\bar{A})\cup N\left(x,\bigcup_{P\in \mathcal{P}} V(P)\cap A\right)\subseteq N(\bar{X},A)\cup \bigcup_{P\in\mathcal{P}'} V(P)\cap A.\] Next, let $v\in V(P)\cap B'$ with $P\in\mathcal{P}'$. If $v$ has a neighbor $u\in A\setminus V(P)$, then since $P\in\mathcal{P}'$, there also is $x\in N(V(P),\bar{X})\subseteq N(V(P),B\setminus V(P))$. Thus, $P\in\mathcal{P}_3$ and $\{u,x\}\in E(\bar{G})$. Hence, $u\in N(\bar{X},\bar{A})$, which proves the claim.
		Finally, we observe that
		\begin{align*}
		w(N(X,A))&\leq w(N(\bar{X},\bar{A}))+\sum_{P\in\mathcal{P}'} w(V(P)\cap A)= w(N(\bar{X},\bar{A}))+\sum_{P\in \mathcal{P}'}  w(V(P)\cap B) \\
		w(X)&= w(\bar{X})+\sum_{P\in \mathcal{P}'}  w(V(P)\cap B),\\
		|N(X,A)\cap A''|&= |N(\bar{X},\bar{A})\cap \bar{A}''| \text{ and }\\
		|X\cap A''|&=|\bar{X}\cap\bar{A}''|.
		\end{align*}
		Hence, if $\bar{X}$ constitutes a local improvement, then so does $X$.
	\end{proof}
	\begin{proof}[Proof of (c)]
		Let $\bar{\mathcal{B}}$ be an improving minimal binocular in $S_{\bar{\tau}}(\bar{G},\bar{w},\bar{A})$. We construct an improving minimal binocular $\mathcal{B}$ in $S_\tau(G,w,A)$ as follows: 
		
		For an edge $\bar{e}\in E(\bar{\mathcal{B}})$, let $\mathcal{P}(\bar{e})\coloneqq\{P\in\mathcal{P}: N(V(P),W(\bar{e}))\neq \emptyset\}$. \[\text{Define }U\coloneqq U(\bar{e})\cup\bigcup_{P\in\mathcal{P}(\bar{e})} V(P)\cap A \text{ and }W\coloneqq W(\bar{e})\cup\bigcup_{P\in\mathcal{P}(\bar{e})} V(P)\cap B.\] We add the edge $e(U,W)$ to $E(\mathcal{B})$.
		\begin{claim}
			$(U,W)$ is an edge-inducing pair and $N(W,A\setminus U)=N(\bar{W},\bar{A}\setminus\bar{U})$ (i.e.\ $(U,W)$ induces the same edge as $(\bar{U},\bar{W})$).
		\end{claim}
		\begin{proof}
			We check the properties of an edge-inducing pair one by one.
			\begin{enumerate}[(i)]
				\item  We have $U\subseteq A$ and $W\subseteq B=V\setminus A$ is independent.
				\item We have \[w(W)=w(W(\bar{e}))+\sum_{P\in\mathcal{P}(\bar{e})} w(V(P)\cap B) = w(U(\bar{e}))+2+\sum_{P\in\mathcal{P}(\bar{e})} w(V(P)\cap A)=w(U)+2.\] As $G$ is $4$-claw free and for every $P\in\mathcal{P}(\bar{e})$, $V(P)\cap A$ is adjacent to $W(\bar{e})$, we can infer that \mbox{$|\mathcal{P}(\bar{e})|\leq 3\cdot |W(\bar{e})|\leq 3\cdot \max\{|U(\bar{e})|,|W(\bar{e})|\}$}. This implies \[\max\{|U|,|W|\}\leq \max\{|U(\bar{e})|,|W(\bar{e})|\}+3\cdot |\mathcal{P}(\bar{e})|\leq 10\cdot \max\{|U(\bar{e})|,|W(\bar{e})|\}\leq 10\cdot\bar{\tau}=\tau.\]
				\item It suffices to prove the stronger statement $N(W,A\setminus U)=N(W(\bar{e}),\bar{A}\setminus U(\bar{e})).$
				
				 As $U\setminus U(\bar{e})\subseteq A\setminus\bar{A}$ by construction, we obtain
				\[N(W,A)\setminus U\supseteq N(W(\bar{e}),\bar{A})\setminus U=N(W(\bar{e}),\bar{A})\setminus U(\bar{e}).\] Hence, it remains to verify the other inclusion. We have \begin{align*}
				&\quad N(W,A)=N(W(\bar{e}),\bar{A})\cup N(W(\bar{e}),A\setminus\bar{A})\cup N(W\setminus W(\bar{e}),A)\\
				&=N(W(\bar{e}),\bar{A}) \cup \underbrace{N\left(W(\bar{e}),\bigcup_{P\in\mathcal{P}} V(P)\cap A\right)}_{\subseteq U\setminus U(\bar{e})}\cup N\left(\bigcup_{P\in\mathcal{P}(\bar{e})} V(P)\cap B, A\right).
				\end{align*}Thus, it suffices show that for each $P\in\mathcal{P}(\bar{e})$, we have $N(V(P)\cap B, A)\subseteq U\cup N(W(\bar{e}),\bar{A})$. Pick a path $P\in\mathcal{P}(\bar{e})$. Then $N(V(P),B\setminus V(P))\supseteq N(V(P),W(\bar{e}))\neq \emptyset$, so $P\in\mathcal{P}_2\cup\mathcal{P}_3$. If $P\in\mathcal{P}_2$, then $N(V(P)\cap B,A)\subseteq V(P)\cap A\subseteq U$. If $P\in\mathcal{P}_3$, let $u$ be the unique vertex in $N(V(P),A\setminus V(P))$ and let $v$ be the unique vertex in $N(V(P),B\setminus V(P))$. Then $v\in W(\bar{e})$ and $\bar{G}$ contains the edge $\{u,v\}$, so $u\in N(W(\bar{e}),\bar{A})$. Thus, $N(V(P)\cap B,A)\subseteq \{u\}\cup V(P)\subseteq N(W(\bar{e}),\bar{A})\cup U$.
			\end{enumerate}
		\end{proof}
		\begin{claim}
			$\mathcal{B}$ is an improving minimal binocular in the search graph of $(G,w)$.
		\end{claim}
		\begin{proof}We have $|E(\mathcal{B})|=|E(\bar{\mathcal{B}})|\leq \bar{\tau}\cdot\log|V(\bar{G})|\leq \tau\cdot\log|V(G)|.$
			The previous claim implies that $\mathcal{B}$ is a minimal binocular since $E(\mathcal{B})$ contains the same underlying edges as $E(\bar{\mathcal{B}})$. We further know that edges of size two in $E(\bar{\mathcal{B}})$ result in edges of size two in $E(\mathcal{B})$, while loops in $E(\bar{\mathcal{B}})$ produce loops in $E(\mathcal{B})$. Let $E(\bar{\mathcal{B}})=\bar{E}_1\dot{\cup}\bar{E}_2$ be the partition into loops and edges of size two and let $E(\mathcal{B})=E_1\dot{\cup} E_2$ be the respective partition for $\mathcal{B}$. We now check the remaining three criteria from Def.~\ref{DefImprovingBinocular}.
			\begin{enumerate}[(i)]
				\item As $\bar{\mathcal{B}}$ is improving, the sets $W(\bar{e}),\bar{e}\in\bar{E}_2$ are pairwise disjoint. As every path in $\mathcal{P}$ has at most one endpoint in $B$ and the paths in $\mathcal{P}$ are pairwise vertex-disjoint, the sets $W(e),e\in E_2$ must be pairwise disjoint as well.
				\item As $\bar{\mathcal{B}}$ is improving, we have
				\[w\left(\bigcup_{\bar{e}\in\bar{E}_1} W(\bar{e})\setminus \bigcup_{\bar{e}\in\bar{E}_2} W(\bar{e})\right)\geq w\left(\bigcup_{\bar{e}\in\bar{E}_1} U(\bar{e})\setminus \bigcup_{\bar{e}\in\bar{E}_2} U(\bar{e})\right)+2|\bar{E}_1|.\]
				Moreover, $|E_1|=|\bar{E}_1|$. Let $\tilde{\mathcal{P}}\coloneqq \bigcup_{\bar{e}\in\bar{E}_1} \mathcal{P}(\bar{e})\setminus \bigcup_{\bar{e}\in\bar{E}_2} \mathcal{P}(\bar{e})$.
				Going from $\bar{E}_1$ and $\bar{E}_2$ to $E_1$ and $E_2$, the left-hand-side increases by $\sum_{P\in\tilde{\mathcal{P}}} w(V(P)\cap B),$ whereas the right-hand-side increases by $\sum_{P\in\tilde{\mathcal{P}}} w(V(P)\cap A).$ As all paths in $\mathcal{P}$ are contained in $G[A'\cup B']$ and feature the same amount of vertices from $A'$ and $B'$, respectively, these sums are equal.
				\item By construction, $W(\mathcal{B})$ is a subset of $B$, and, thus, independent.
			\end{enumerate}
		\end{proof}
	\end{proof}
	\begin{proof}[Proof of (d)]
		\begin{align*}\text{We have } w(A)&=w(\bar{A})+\sum_{P\in\mathcal{P}} |V(P)\cap A|=\bar{w}(\bar{A})+\sum_{P\in\mathcal{P}} |V(P)\cap B|\text{ and }\\w(B)&=w(\bar{B})+\sum_{P\in\mathcal{P}} |V(P)\cap B|=\bar{w}(\bar{B})+\sum_{P\in\mathcal{P}} |V(P)\cap B|.\end{align*}	
	\end{proof}
\end{proof}
Removing the vertices in $\bigcup_{P\in\mathcal{P}} V(P)$ and applying Lemma~\ref{LemDeletePaths} concludes the proof of Theorem~\ref{TheoNormalization}.

\end{document}